\documentclass[pra,aps,amsmath,amssymb,twocolumn,floatfix,superscriptaddress]{revtex4-1}
\usepackage{amsmath}
\usepackage{graphicx}
\usepackage{color}
\usepackage{amssymb}
\usepackage[utf8]{inputenc}
\usepackage[T1]{fontenc}

\usepackage[normalem]{ulem}
\usepackage{hyperref}
\usepackage{array}

\begin{document}
\title{Emergence of unstable avoided crossing in the collective excitations of spin-1 spin-orbit coupled  Bose-Einstein condensates}

\author{Sanu Kumar Gangwar}
\affiliation{Department of Physics, Indian Institute of Technology, Guwahati 781039, Assam, India} 

\author{Rajamanickam Ravisankar}
\email{Corresponding author: ravicpc2012@gmail.com}
\affiliation{Department of Physics, Zhejiang Normal University, Jinhua 321004, China}

\affiliation{Zhejiang Institute of Photoelectronics \& Zhejiang Institute for Advanced Light Source, Zhejiang Normal University, Jinhua, Zhejiang 321004, China}

\author{Henrique Fabrelli}
\affiliation{Centro Brasileiro de Pesquisas Físicas, 22290-180 Rio de Janeiro, RJ, Brazil}

\author{Paulsamy Muruganandam}
\affiliation{Department of Physics, Bharathidasan University, Tiruchirappalli 620024, Tamilnadu, India}

\author{Pankaj Kumar Mishra}
\email{Corresponding author: pankaj.mishra@iitg.ac.in}
\affiliation{Department of Physics, Indian Institute of Technology, Guwahati 781039, Assam, India}

\date{\today}
\begin{abstract} 
We present the analytical and numerical results on the collective excitation spectrum of quasi-one-dimensional spin-orbit (SO) coupled spin-1 spinor ferromagnetic Bose-Einstein condensates. The collective excitation spectrum, using Bogoliubov-de-Gennes theory, reveals the existence of a diverse range of phases in the SO and Rabi ($k_L-\Omega$) coupling plane. Based on the nature of the eigenvalue of the excitation spectrum, we categorize the $k_L-\Omega$ plane into three distinct regions, namely I, II, and III. In region I, a stable mode with phonon-like excitations is observed. In region IIa, single and multi-band instabilities are noted with a gapped mode, while multi-band instability accompanied by a gapless mode between low-lying and first excited states is realized in region IIb, which also provides evidence of unstable avoided crossing between low-lying and first excited modes, responsible for the $I_o$ type of oscillatory non-equilibrium dynamical pattern formation. The gap between low-lying and first-excited states increases upon increasing the Rabi coupling while decreases upon increase of SO coupling. Using eigenvector analysis, we confirm the presence of the spin-dipole mode in the spin-like modes in Region II. We corroborate the nature of the collective excitation through real-time dynamical evolution of the ground state perturbed with the quench of the trap using the mean-field Gross-Pitaevskii model. This analysis suggests the presence of dynamical instability leading to the disappearance of the $0$-th component of the condensate. In Region III, mainly encompassing $\Omega \sim 0$ and finite $k_L$, we observe phonon-like excitations in both the first excited and the low-lying state. The eigenvectors in this region reveal alternative in- and out-of-phase behaviours of the spin components. Numerical analysis reveals the presence of a super stripe phase for small Rabi coupling in this region, wherein the eigenvector indicates the presence of more complicated spin-like-density mixed modes. 
\end{abstract}

\maketitle
\section{Introduction}
Ultracold atoms offer a versatile but precise platform for exploring quantum matter in the presence of diverse synthetic fields~\cite{zhang2016properties}. The recent experimental realization of spin-orbit (SO) coupling in spinor ultracold atoms~\cite{Jacob2012, Campbell2016} has sparked great interest among researchers to explore the SO coupling-related physics in large-spin systems, which is challenging to achieve in the electronic materials system of condensed matter physics. Over the last few decades, research on spinor Bose-Einstein condensates (BECs) has captured the community's attention due to its magnanimity in showing the interesting physical phenomena such as vortices~\cite{Kasamatsu2005}, turbulence state triggered by the modulational instabilities as a result of the counterflow motion between the binary component ~\cite{Tsubota2012}, quantum phase transitions~\cite{Yu2016, Martone2016}, modulational instability~\cite{Li2017}, solitons~\cite{Malomed2018}, magnetized vector solitons~\cite{Peng2019}, and supersolid-like behaviour in quasi-2D~\cite{Adhikari2021}. 

In the field of ultracold gases, numerical simulations have become an indispensable tool for exploring numerous complex phases and the underlying detailed mechanisms that appear in those systems as low-lying excitation phases. To this end, the mean-field model Gross-Pitaevskii equations-based models have been widely used to investigate various aspects of SO-coupled spinor BECs, such as phase separation~\cite{Gautam2014}, phase separation of vector solitons~\cite{Adhikari2019}, stability, and dynamics of solitons~\cite{Gautam2015, Adhikari2020, Gautam2021}, spin precession and the separation between the spin component owing to the anomalous spin-dependent velocities~\cite{Mardonov2015, Mardonov2018, Mardonov2019}, etc. Some studies have explored the dynamics of modulation instability~\cite{Mithun2019}, preparation of stripe states~\cite{Cabedo2021}, and condensate flow past an obstacle~\cite{Zhu2020} in SO-coupled BECs.

One of the intriguing features of the SO-coupled BECs is the presence of a modulated ground state that appears as stripe phases for high values of SO couplings~\cite{Lin2011, Ravisankar2021influence}. In recent years, there have been several studies indicating an intimate connection between the stripe phase and the supersolid behaviour of the BECs. For instance, Adhikari~\cite{Adhikari2021} reported supersolid-like states in quasi-two-dimensional trapped SO-coupled spinor BECs. In multicomponent condensates, Goldstein et al. \cite{Goldstein1997} have initiated the study of the Hartree-Bogoliubov theory to determine the resulting quasi-particle frequency spectrum and observed the interferences resulting from cross-coupling among the condensate leading to the reversal of the sign of two-body interaction which triggers the onset of spatial instabilities. There are some studies that suggest domain formation through the dynamical stability analysis~\cite{Zhang2005}. 


 
The investigation of collective excitations, which are low-lying excitations in BECs, has played a pivotal role in comprehending the fundamental characteristics of these quantum degenerate gases. This includes aspects such as the stability of various ground state phases, superconductivity, and superfluidity~\cite{Pethick2008, Pitaevskii2016}. In 1941, Landau coined the term ``excitations'' to elucidate the emergence of superfluids as a combined effect of a weakly interacting mixture of phonons and rotons quasi-particles. Subsequently, Bogoliubov employed Landau's excitation spectrum tools to explicate the superfluidity in BECs~\cite{Bogoliubov1947}. Experimentally, the realization of low-lying excitations has been achieved in dilute gases of rubidium~\cite{Jin1996} and sodium~\cite{Mewes1996} atoms. For trapless spinor condensates, the study of collective excitations involved considering equal Rashba and Dresselhaus couplings. This investigation revealed the presence of maxon-roton excitations~\cite{Martone2012, Yun2013}. Utilizing Bragg spectroscopy, Khamehchi et al. provided experimental evidence of collective excitations by realizing roton-maxon modes in SO coupled spinor systems~\cite{Khamehchi2014}.

Following the experimental realization of the collective excitation mode for SO coupled BECs in the laboratory, a series of theoretical and numerical works have been undertaken in the recent past that reveal the more complex nature of these excitations. For instance, Chen et al.~\cite{Chen2017} demonstrated the collective excitation spectrum of Raman-induced SO coupled spinor BECs confined in a quasi-1D harmonic trap. By tuning the Raman coupling strength, they projected the presence of three distinct phases: stripe (ST), plane-wave (PW), and zero-momentum (ZM). Additionally, they identified the Spin dipole and breathing modes of collective excitations, indicating clear phase boundaries. Numerical confirmation of these features was obtained through quench dynamics. Few studies indicate the presence of the PW, ZM, and ST phases of the condensate as quadratic Zeeman coupling is introduced in the SO coupled spin-1 BECs~\cite{chen2022elementary, he2023stationary}. Ozawa et al.~\cite{Ozawa2013} numerically analyzed the dynamical and energetic instabilities of quasi-one-dimensional SO-coupled BECs. Ravisankar et al.~\cite{Mishra2021} analytically and numerically analyzed the stability of the spin-1/2 binary SO-coupled BECs in quasi-2D using collective excitation spectrum and found the presence of phonon-maxon-roton mode in the spectrum for finite SO and Rabi couplings. Katsimiga et al.~\cite{Katsimiga2021} analyzed the non-linear solitary wave excitation that appear in the form of different combinations of the dark and bright solitons for the trapped spin-1 condensate in the harmonic trap. Subsequently different sorts of vortex-bright type excitations in the 2D harmonically confined spin-1 BECs were also explored ~\cite{Katsimiga2023}. Rajat et al.~\cite{roy2022collective} theoretically and numerically analyzed the collective excitation of SO coupled spin-1 BECs in a cigar-shaped trap both at zero and finite temperature, and demonstrated the presence of density and spin excitation, which exhibits qualitatively different features at finite temperature than those at zero temperature.     





So far, the studies of collective excitation in the spinor BECs are mainly restricted to the specified range of coupling parameters. However, a comprehensive picture of the instabilities arising as a result of the collective excitation has not been understood clearly. Moreover, from the dynamics point of view, it has been demonstrated that in the presence of finite detuning between the two spin states unstable avoided crossing between the low-lying and first excited states appears upon perturbation. Such physical mechanism is responsible for constructing a class of $I_o$ (oscillatory patterns) dynamically oscillatory non-equilibrium pattern formation in the quantum systems~\cite{Cross1993, Bernier2014} which remains to be unexplored in the spin-1 coupled BECs along with spin-orbit and linear Rabi coupling. In this paper, we attempt to address these issues considering the ferromagnetic spin-1 BECs by scanning a wide range of the coupling parameters regime using analytical computation of the eigenspectrum and eigenvectors. Based on the detailed nature of the eigenspectrum, we divide $k_L - \Omega$ plane into three regimes: (i) a stable region characterized by a phonon-like mode, (ii) unstable regimes broadly divided into two parts characterized by the presence of a gap (IIa) and gapless (IIb) behaviour between the low-lying (ll) and first-excited (fe) spectra, and also yields the evidence of unstable avoided crossing modes, and (iii) for coupling values near $\Omega \sim 0$, we find that the system exhibits a phonon-like symmetric maxon-roton mode, which is also a gapless mode between the low-lying and first-excited spectra.  In the presence of a relatively small imaginary frequency, we observe both a weak instability and the emergence of a superstripe phase. In all the regions, we corroborate the analytical observation of the eigenspectrum with the real-time dynamics of the ground state using the GP equations. 

The paper is organized as follows. In Section~\ref{sec:2}, we present our theoretical model for investigating the collective excitations and instabilities of SO-coupled spin-1 ferromagnetic BECs. Section~\ref{sec:3} illustrates the single-particle spectrum, followed by the analysis of the collective excitation spectrum using the Bogoliubov-de-Gennes theory in Section~\ref{sec:4}. In Section~\ref{sec:5}, we analytically and numerically report the collective excitation spectrum, identifying various stability regions. Each subsection within Section~\ref{sec:5} provides the excitation spectrum and corresponding dynamical observations using the GP equations. Finally, Section~\ref{sec:6} summarizes the findings.
\section{Mean-Field Model}
\label{sec:2}
We consider a quasi-1D spin-$1$ spinor SO coupled BECs realized by tight confinement in the transverse direction~\cite{salas2002}. The non-dimensional dynamical equations for quasi-1D SO coupled spin-$1$ BECs are given by \cite{Gautam2015, Katsimiga2021, Adhikariss2021, Ueda2012},
\begin{subequations}
\begin{align}
\mathrm{i} \frac{\partial \psi_{\pm1}}{\partial t} &=  \bigg[- \frac{1}{2 } \frac{\partial^{2}}{\partial x^{2}}+ V+c_{0}\rho\bigg] \psi_{\pm1} \mp \frac{ k_{L}}{\sqrt{2}} \frac{\partial \psi_{0}}{\partial x}  \notag \\ & + c_{2}^{\pm} \psi_{\pm1} + \psi_{0}^{2} \psi_{\mp1}^{*} + \frac{\Omega}{\sqrt{2}}\psi_{0}, \label{gpe1} \\ 
\mathrm{i} \frac{\partial \psi_{0}}{\partial t} &=  \bigg[- \frac{1}{2} \frac{\partial^{2}}{\partial x^{2}}+ V +c_{0}\rho\bigg] \psi_{0} + \frac{ k_{L}}{\sqrt{2}} \bigg[ \frac{\partial \psi_{+1}}{\partial x}  - \frac{\partial \psi_{-1}}{\partial x}\bigg]  \notag \\ &  + c_{2}^{0}\psi_{0} + 2 \psi_{0}^{*} \psi_{+1} \psi_{-1}  +\frac{\Omega}{\sqrt{2}}(\psi_{1}+\psi_{-1}) \label{gpe2}
\end{align}
\end{subequations}
where, $c_2^{\pm} = \left(\rho_{\pm1} + \rho_{0} - \rho_{\mp1}\right) $, $c_2^0 = \left( \rho_{+1} + \rho_{-1}\right)$, $\psi_{j}$; $j= +1,0,-1$, are the spinor condensate wavefunctions that satisfy the normalization condition $\int_{-\infty}^{\infty} \rho d x   = 1$, where $\rho_j = \vert\psi_{j}\vert^{2}$, and $\rho = \vert\psi_{+1}\vert^{2} + \vert\psi_{0}\vert^{2} + \vert\psi_{-1}\vert^{2}$. 
The dimensionless equations (\ref{gpe1})-(\ref{gpe2}) are obtained by non-dimensionalized the time, length, and energy with respect to $\omega^{-1}$, $l_{0} = \sqrt{\hbar / m \omega}$, and $\hslash \omega$ respectively, where $\omega = \omega_{x}$ is the trap frequency along the $x$-axis. The resulting condensate wavefunction takes the form as $\psi_{j} = \sqrt{ \frac{l_{0}}{N}} \tilde{\psi}_{j}$, and $N$ is the total number of atoms.  $V(x) = x^{2} /2$ is the trap potential, $c_{0} = 2 N l_{0}(a_{0} + 2a_{2})/ 3 l_{\perp}^{2}$ is the density-density interaction strength, and $c_{2} = 2 N l_{0}(a_{2} - a_{0})/ 3 l_{\perp}^{2}$, spin-exchange interaction strength with $a_{0}$ and $a_{2}$ are the s-wave scattering length in total spin channels $0$ and $2$, respectively . The nature of interaction strength depends on the sign of $c_{2}$. $c_{2} < 0$ represents the ferromagnetic condensate, while  $c_{2} > 0$ denotes the anti-ferromagnetic condensate~\cite{Ueda2012, Stamper2013}. Here, $l_{\perp} = \sqrt{ \hbar / (m \omega_{\perp})}$ is the harmonic oscillator length in the transverse direction with $\omega_{\perp} = \sqrt{\omega_{y}\omega_{z}}$. The SO and Rabi coupling strengths are given by  $k_{L} = \tilde{k_{L}} / \omega_{x} l_{0}$, $\Omega = \tilde{\Omega} / (\hbar \omega_{x})$, respectively. In the above description, the quantities with tilde represent dimension-full quantities.

One important entity that characterizes the miscibility of different spin components is the magnetization of spin-1 spinor condensates defined by,
\begin{align}\label{magnt}
\mathcal{M} = \int_{-\infty}^{\infty} \bigg\{{\rho}_{+1}({x}) - \rho_{-1}({x}) \bigg\} d {x}
\end{align}
The energy functional corresponding to the coupled GP Eqs.(\ref{gpe1})-(\ref{gpe2}) is given by~\cite{Gautam2014, Ravisankar2021},
\begin{align}\label{eqn5}
E = & \frac{1}{2} \int_{-\infty}^{\infty} dx  \bigg\{\sum_{j} \lvert \partial_{x} \psi_{j} \rvert^{2} + 2 V(x) \rho + c_{0} \rho^{2}  \notag \\  & + c_{2}[ \rho_{+1}^{2} + \rho_{-1}^{2} + 2( \rho_{+1}\rho_{0}+\rho_{-1}\rho_{0} -\rho_{+1}\rho_{-1} \notag \\  & +\psi_{-1}^{*}\psi_{0}^{2}\psi_{+1}^{*}+ \psi_{-1}\psi_{0}^{*2}\psi_{+1})]  + \sqrt{2} \Omega [(\psi_{+1}^{*}+\psi_{-1}^{*})\psi_{0} \notag \\  & +\psi_{0}^{*}(\psi_{+1}+\psi_{-1})]+ \sqrt{2} k_{L}[  (\psi_{-1}^{*}-\psi_{+1}^{*})  \partial_{x} \psi_{0}  \notag \\  & + \psi_{0}^{*}(\partial_{x} \psi_{+1}- \partial_{x}\psi_{-1})]\bigg\} 
\end{align}

In order to make the dimensionless set of parameters used for our simulations viable for an experiment, here, we outline the dimensionful experimental parameters. For the ferromagnetic system, we consider $^{87}$Rb BECs with $N \sim 2\times 10^{4}$ number of atoms. The axial trap frequency is $\omega_x = 2 \pi \times 50$ Hz while the transverse trap frequencies are considered to be $\omega_y = \omega_z = 2 \pi \times 500$ Hz. The spin-dependent and spin-independent interactions can be achieved by controlling the $s$-wave scattering lengths through Feshbach resonance; by varying the magnetic field, we can tune the $s$-wave scattering lengths \cite{Inoye1998, Marte2002, Chin2010}. We can achieve the SO coupling strengths in the range of $k_L = \{ 0.1 - 5 \}$ by changing the laser wavelengths from $\{68.86 \mu$m - $1377.22 \text{nm}\}$. Further, one can also control the Rabi coupling strength in a similar range by changing the frequency of the laser, $2\pi \hbar\times \{5 - 250\}$ Hz. %
\section{Single-particle spectrum}
\label{sec:3}
In this section, we present the single-particle spectrum~\cite{Wen2012, Peng2019, Rajaswathi2023} of the non-interacting spinor condensate for trapless SO and Rabi-coupled spin-1 spinor BECs. Following that, we introduce the collective excitation spectrum for the interacting systems.

For $V(x) =0$, and $ c_{0} =  c_{2} = 0$ as we substitute $\psi_{0,\pm 1} = \phi_{0,\pm 1} e^{\mathrm{i}({q_{x} x -\omega t})}$ in the Eqs.(\ref{gpe1})-(\ref{gpe2})) we obtain
\begin{align}\label{eqn6}
 \mathcal{L}_{sp}\begin{pmatrix}
           \phi_{1}\\
     \phi_{0}\\
     \phi_{-1}
\end{pmatrix} &= \omega\begin{pmatrix}
     \phi_{1}\\
     \phi_{0}\\
     \phi_{-1}
     \end{pmatrix} 
\end{align}
with
\begin{align}
\mathcal{L}_{sp} = \frac{1}{2}
 \begin{pmatrix}
 q_{x}^{2} & L & 0\\
 S & q_{x}^{2} & L \\
 0 & S & q_{x}^{2} \\
 \end{pmatrix},
\end{align}
where $L = - \sqrt{2} (\mathrm{i}k_{L} q_{x} - \Omega), S = \sqrt{2} (\mathrm{i}k_{L} q_{x} + \Omega)$. Upon diagonalizing the Eq.(\ref{eqn6}), we obtain the single-particle spectrum as
\begin{subequations}\label{eqn7}
\begin{align}%
\omega_{0}= &\frac{q_{x}^{2}}{2} \\
\omega_{\pm}=&\frac{1}{2}(q_{x}^{2}\pm 2 \sqrt{\Omega^{2}+ k_{L}^{2} q_{x}^{2}})
\end{align}
\end{subequations}%
From Eq.~\ref{eqn7}(a) and Eq.~\ref{eqn7}(b), it is quite evident that the single-particle spectrum exhibits three branches, namely,  $\omega_{0}$ and $\omega_{\pm}$. Among them, $\omega_{0}$ is independent of SO and Rabi couplings, a constant branch of the single-particle spectrum. However, the detailed nature of the other two branches $\omega_{\pm}$ depends on the SO and Rabi couplings, which have been designated as positive and negative branches of the spectrum, respectively, throughout the paper. 

\begin{figure}[!htp]
\centering\includegraphics[width=0.99\linewidth]{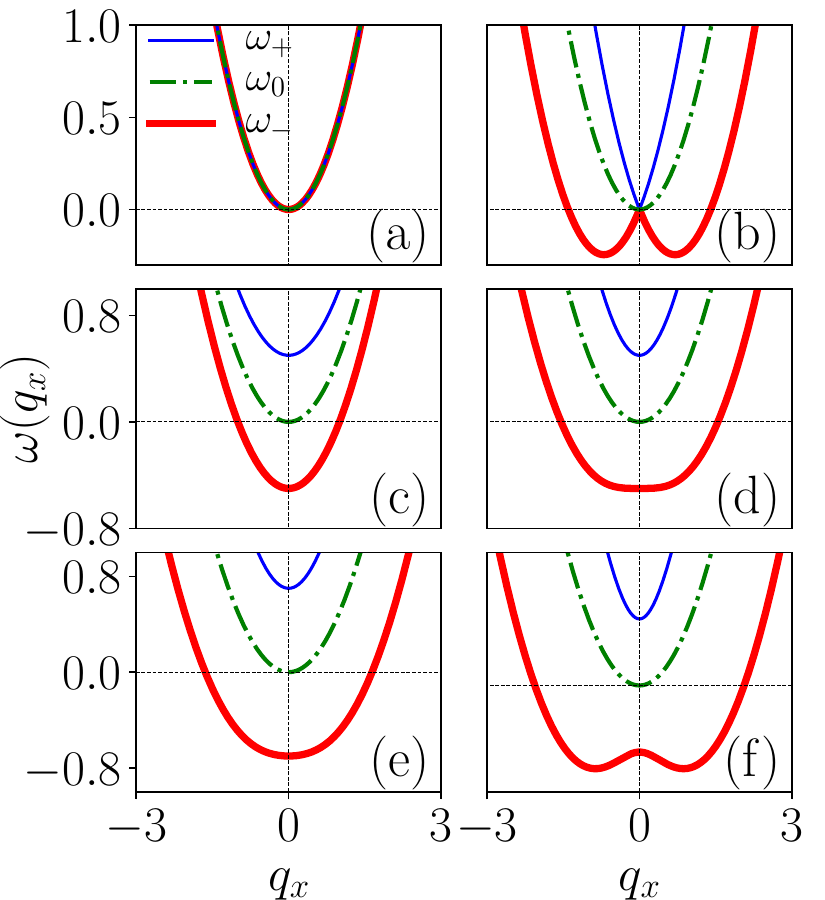}
\caption{Single-particle energy spectrum in momentum space for the different set of SO and Rabi coupling strengths $(k_L,\Omega)$: (a)-(f): $(0, 0)$, $(0.7, 0)$, $(0, 0.5)$, $(0.7, 0.5)$, $(0.7, 0.7)$, $(1.0, 0.5)$. The thick-solid-red, dash-dotted green line and thin-solid-blue line represent the energy eigenspectrum for $\{-1, 0, +1\}$ components of the spin, respectively. Varying the coupling parameters lead to the changes in the eigenspectrum of different components, where $\omega_{-}$, only having the lowest minimum for $\Omega > k_L^2$, whereas $\Omega < k_L^2$, the  $\omega_{-}$ exhibits double minima appearance of plane-wave phase, indicating the presence of stripe phase.}
\label{fig1} 
\end{figure}

To start with, we consider both the SO ($k_{L}$) and the Rabi ($\Omega$) coupling strengths as zero. We obtain the non-degenerate parabolic spectrum of the SO-coupled spin-1 system as couplings are set to zero [see Fig.~\ref{fig1}(a)]. As $\omega_{0}$ of the spectrum is independent of both the SO and Rabi couplings, it does not show any change upon the variation of coupling strengths as shown in Figs.~\ref{fig1}(a)- \ref{fig1}(e) with the dashed green line. To understand the behaviour of the positive branch ($\omega_{+}$) and negative branch ($\omega_{-}$) of the single-particle spectrum, we introduce the finite value of SO coupling in the absence of Rabi coupling and vice-versa. First, we introduce a finite value to SO coupling strength, $k_{L} = 0.7$, in the absence of Rabi coupling strength, $\Omega = 0.0$. For this parameter, the negative branch of the spectrum gets transformed from the parabolic to double minima, which appears at $q_{x} =\pm k_{L}$ as shown in Fig.~\ref{fig1}(b). Here the energy minima of $\omega_{-}= -0.245$ occurs at $q_{x} = \pm 0.7$. As Rabi coupling is increased to a finite value, $\Omega = 0.5$ while keeping  $k_{L} = 0.0$, we find a change from double minima to a single minimum ($\Omega > k_L^2$) and also leading to opening up a gap between the $\omega_{+}-\omega_{0}$ branches. The magnitude of the gap between the $\omega_{+}-\omega_{0}$ branches, and $\omega_{0}-\omega_{-}$ branches, is of the order of $\Omega$. It appears to be of $2 \Omega$ between the $\omega_{+}-\omega_{-}$ branches as shown in Fig.~\ref{fig1}(c) which exhibits a single minimum with $\omega_{-}= -0.5$ at $q_{x} = 0$. Further, we analyze the spectrum by introducing both couplings strengths,i.e., $k_{L}  =0.7$, $\Omega = 0.5$ for which we obtain the global minima in the spectrum, also satisfies the condition, $\Omega \approx k_L^2$. Upon increasing the Rabi coupling value from $\Omega = 0.5 \to \Omega = 0.7$, we obtain a similar kind of spectrum, but the gap between the branches increases with an increase in Rabi coupling as shown in Fig.~\ref{fig1}(d) and Fig.~\ref{fig1}(e), with energies $\omega_{-}= -0.5$, $\omega_{-}= -0.7$ at $q_{x} \approx 0.0$.  On the other hand, upon increasing the SO coupling from $k_{L} = 0.7$  to $k_{L} = 1.0$, we obtain the transition of single minimum nature of $\omega_{-}$ into a double minimum as shown in Fig.~\ref{fig1}(f). At a later stage, we will see that this feature (occurrence of double minima) in the single particle spectrum is responsible for the appearance of the stripe phase for the situation when $\Omega < k_L^2$. In Fig.~\ref{fig1}(f), the lowest energy minima is $\omega_{-}= -0.625$ at $q_{x} \approx \pm 0.9$. From the observation we find that, when $\Omega > k_L^2$, the energy depends on only the Rabi coupling, while if $\Omega < k_L^2$ energy depends on SO coupling strength. Overall, we find that, upon variation of SO coupling, the positive branch exhibits a single minimum upon fixing the Rabi coupling at a finite value. However, the negative branch exhibits a transition from the single minimum to a double minimum upon varying the SO coupling, keeping the Rabi coupling fixed. This typical feature of the negative branch of the spectrum appears to indicate the change in the phases of the condensate upon varying the coupling parameters, and therefore, for later parts of our discussion, we will focus on analyzing the typical features of the $\omega_{-}$ branch.

In the next section, we discuss the effect of both coupling terms (SO and Rabi) on the collective excitation spectrum for the interacting SO and Rabi-coupled spin-1 BECs.%

\section{Collective Excitation spectrum: Bogoliubov-de-Gennes Analysis}
\label{sec:4}
In the previous section, we have discussed the single-particle spectrum of SO-coupled spin-1 ferromagnetic BECs. In this section, we shall present the collective excitation spectrum of the condensate using the Bogoliubov-de-Gennes (BdG) analysis. The excitation wave function in terms of perturbation term $\delta \phi_{j}$ and ground state wave function $\phi_{j}$ can be represented as~\cite{Zhu2020, Goldstein1997},
\begin{align}\label{excwave}
\psi_{j}(x,t) = e^{-i \mu_{j} t}[\phi_{j} +  \delta \phi_{j}(x, t) ]
\end{align}    
where,
\begin{align}\label{pertwave}
\delta \phi_{j}(x,t) = u_{j} e^{i( q x - \omega t)} + v_{j}^{*} e^{- i(q x - \omega^{*} t)}
\end{align}%
represents the ground state wave function with $\phi_{j} = (1/2, -1/\sqrt{2}, 1/2)^{T}$. Here, $\mu_{j}$ denotes the chemical potential, $u_{j}$,  $v_{j}$ are the Bogoliubov amplitudes, with $j =$  $+1$, $0$ $-1$ corresponding to the three component of spinor density of condensates. As we substitute Eq. (\ref{excwave}) in the dynamical equations (Eqs. (\ref{gpe1})-(\ref{gpe2})) we  obtain,
\begin{widetext}
 \begin{align}\label{bdgeigprbm}
    \mathcal{L} \begin{pmatrix}
            u_{+1} & v_{+1} & u_{0} & v_{0} & u_{-1} & v_{-1}\\
          \end{pmatrix}^{T}
     &=\omega \begin{pmatrix}
     u_{+1} & v_{+1} & u_{0} & v_{0} & u_{-1} & v_{-1}\\
     \end{pmatrix}^{T} 
   \end{align}
where $T$ represents the transpose of the matrix and  $\mathcal{L}$ is $6 \times 6$ matrix given by,
\begin{align} \label{bdgmatrix}
 \mathcal{L} = 
 \begin{pmatrix}
 H_{+}-\mu_{+} & \mathcal{L}_{12} & \mathcal{L}_{13}
 & \mathcal{L}_{14} & \mathcal{L}_{15} &
 \mathcal{L}_{16}\\
 \mathcal{L}_{21} & -H_{+}+\mu_{+} & \mathcal{L}_{23} & \mathcal{L}_{24}
& \mathcal{L}_{25} & \mathcal{L}_{26}\\
 \mathcal{L}_{31} & \mathcal{L}_{32} & H_{0}-\mu_{0} & \mathcal{L}_{34} & \mathcal{L}_{35} & \mathcal{L}_{36}\\
 \mathcal{L}_{41} & \mathcal{L}_{42} & \mathcal{L}_{43} & -H_{0}+\mu_{0} & \mathcal{L}_{45} & \mathcal{L}_{46}\\
 \mathcal{L}_{51} & \mathcal{L}_{52} & \mathcal{L}_{53} & \mathcal{L}_{54} & H_{-} -\mu_{-} & \mathcal{L}_{56}\\
 \mathcal{L}_{61} & \mathcal{L}_{62} & \mathcal{L}_{63} & \mathcal{L}_{64} & \mathcal{L}_{65} & -H_{-}+\mu_{-}
 \end{pmatrix}
 \end{align}
 \end{widetext}
The matrix elements of $\mathcal{L}$ are given in appendix~\ref{matrx:BdG}. Bogoliubov coefficients follows the normalization condition,
\begin{align}\label{normbdg}
 \int (\lvert u_{j} \rvert^{2} - \lvert v_{j}^{*} \rvert^{2}) dx = 1
\end{align}
The simplified form of the BdG excitation spectrum is obtained by calculating the determinant of the matrix $\mathcal{L}$ and equate it with zero, i.e., $det~\mathcal{L}=0$. The characteristic equation can be written as,
\begin{align}\label{bdgex}
\omega^{6}+b \omega^{4}+ c \omega^{2} + d =0
\end{align}
where the coefficient of $b$, $c$, and $d$ are given in the appendix~\ref{matrx:BdG}.%

\begin{figure}[!ht]
\centering\includegraphics[width=0.99\linewidth]{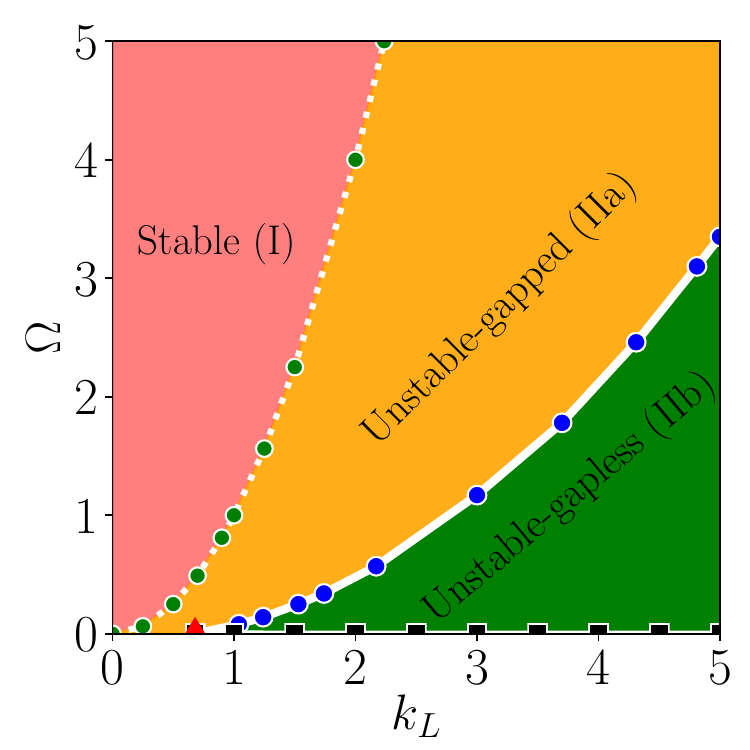}
\caption{Stability phase diagram in the $k_{L}$- $\Omega$ plane for the interaction parameter $c_{0} = 0.5$, $c_{2} = -0.1$ which is being considered for all the simulation runs. Based on the different characteristics of the eigenspectrum and ground state, the phase diagram is divided into regions I, IIa, IIb, and III. While Region I is stable, regions IIa, IIb, and III are unstable. The dotted white line along with green dots represent the $\Omega=k_L^2$ curv that separates region I from IIa. Blue dots with a white solid line representing $\Omega \approx 0.1365 k_{L}^{2} - 0.0686$ separates the regions IIa and IIb. The horizontal line with a filled half square for $\Omega\sim 0$ denotes the region III after the cutoff, $k_L^c = 0.68$, which is a tricritical point for regions IIa-IIb-III, indicated as a red triangle.}
\label{fig8}
\end{figure}%

\section{Effect of Rabi and SO  coupling on the excitation spectrum}%
\label{sec:5}
To better understand the stability of the ground state phases of the system, we plot the stability diagram of the system in $k_{L}$ - $\Omega$ plane with interaction parameters $c_{0} = 0.5$, and $c_{2} = -0.1$ ~\cite{Wang2010}. We obtain the stability phase plot by solving the collective excitation spectrum. The real frequencies of the BdG matrix imply the dynamically stable phase while the imaginary frequency indicates a dynamically unstable phase~\cite{Ueda2012, Goldstein1997, Mishra2021, Ozawa2013, Zhu2012}. In Fig.~\ref{fig8}, we show the stability diagram in the $k_L - \Omega$ plane. Based upon the eigenvalue, the plane is divided mainly into three regions, viz., stable region I, unstable regions II, and III. White dots with the dotted line separate the stable and unstable regions, which hold the phase transition boundary, i.e., $ \Omega=k_L^{2}$~\cite{Mishra2021}. Further, we divide the unstable region into two separate parts:  regions IIa and IIb. Blue dots with white solid line separate the regions IIa and IIb. The phase transition critical points for region-IIa to region-IIb can be obtained by a polynomial fit given by $\Omega \approx 0.1365 k_{L}^{2} - 0.0686$. %

\begin{figure*}[] 
\begin{centering}
\centering\includegraphics[width=0.95\linewidth]{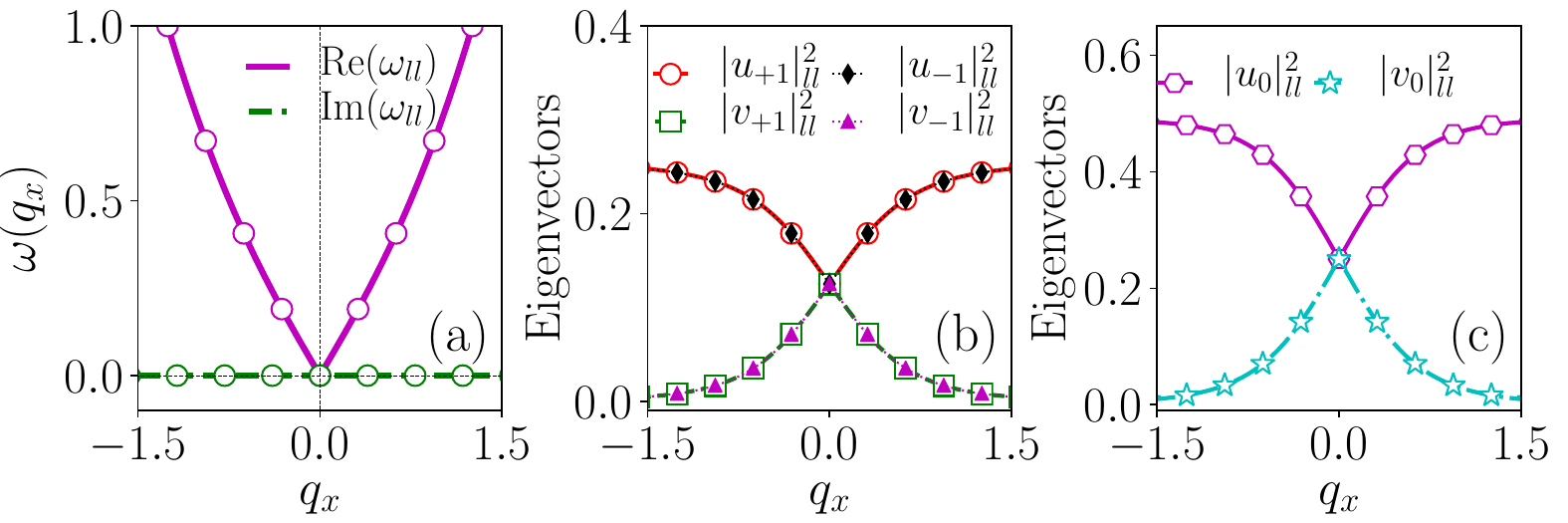}
\caption{Excitation spectrum (a) and eigenvectors (b) and (c) for the coupling parameters $k_{L} = 0.5$, $\Omega = 2.0$. (a) solid magenta line represents $\mathrm{Re}(\omega_{-})$, and dashed-dotted green line represents the $\vert{\mathrm{Im}(\omega_{-})}\vert$, here solid and dash-dotted lines are the analytical results of BdG Eq.~\ref{bdgex} and open circles are numerical results obtained by solving Eq.~\ref{bdgeigprbm}. In the eigenvectors, $\vert u_{+1}\vert^{2}$(red open circles), $\vert u_{-1}\vert^{2}$ (black diamonds), $\vert u_{0}\vert^{2}$ (magenta open hexagons), $\vert v_{0}\vert^{2}$ (cyan open stars ),  $\vert v_{+1}\vert^{2}$ (green open squares), and $\vert v_{-1}\vert^{2}$ (magenta triangles), (b-c) the eigenvectors corresponding to eigenspectrum has shown, which we obtained by solving Eq.~\ref{bdgeigprbm} numerically. The first column of the figure shows the presence of phonon mode in the eigenspectrum. Also, it has no negative or complex eigenfrequency thus, it is energetically and dynamically stable.} %
\label{fig9}
\end{centering}
\end{figure*}
In region IIa, we obtain a clear gap between the low-lying (negative branch) and first-excited (positive branch) branches of the spectrum. Also, for a fixed value of Rabi coupling strength ($\Omega$), upon increasing the SO coupling strength ($k_{L}$), the number of instability bands changes from one to two or vice versa. In region IIb, we obtain overlap among the low-lying and first-excited branches of the eigenspectrum.  Thus, the line delineates regions IIa and IIb, encompassing points where instability is present by means of multiple instability bands and gap openings between low and first excited states. Furthermore, in region IIb, we observe a spectrum characterized by multiple instability bands. In this region for a fixed value, Rabi strength ($\Omega$), upon increasing the SO coupling strength $k_{L}$, we achieve gapless mode between the low-lying and first-excited branches of the spectrum for some range of quasi-momentum. By increasing the Rabi coupling strength with fixed SO coupling, we can again achieve the gapped mode between the low-lying and first excited branches. %

The transition from the region-IIa to IIb can be obtained after a certain critical coupling point. We obtain the gapless behaviour of eigenspectrum either increasing SO coupling ($k_L > k_L^c$) for a fixed Rabi coupling ($\Omega > \Omega^c$) or decreasing Rabi coupling ($\Omega^c < \Omega$) for a fixed SO coupling ($k_L > k_L^c$), with the critical coupling parameters as $k_{L}^c = 0.68$, and $\Omega^c = 0.001$. This point is the origin of the line which separates regions IIa and IIb. The horizontal line at which Rabi coupling strength is approximately zero, ($\Omega \approx 0.0$), we realize the presence of phonon mode in the low-lying as well as in the first-excited branch of the spectrum, along with the presence of instability band. These features present in IIa and IIb are completely different compared to those obtained for the region I. The cutoff value of SO coupling strength to obtain this behaviour along the line is $k_{L}^{c} = 0.68$. We analyze the behaviour of the system along the horizontal line separately under Region III. Here, $k_{L}^{c}$ is the tricritical point for regions IIa, IIb, and III, which is indicated by the red triangle in the phase plot [see Fig.~\ref{fig8}]. Based on the above observations, we divide the phase plot into three different regions: regions I, IIa, IIb, and III. In Appendix C, we show a similar kind of phase diagram for two other different sets of interaction strengths $c_{0} = 5.0$, $c_{2} = -0.1$, and $c_{0} = 885.72$, $c_{2} = 4.09$. With these interactions, we also obtain similar kinds of phase plots ~[see Appendix.~\ref{app:phase2}] as those in Fig.~\ref{fig8}.%

As we divide the phase diagram of the $k_L - \Omega$ plane into three regions, we discuss their behaviour in each region individually. The detailed structure of the presentation in different regions is as follows. We perform collective excitation spectrum calculations to understand the stability of the system, both dynamically and energetically. As the real but negative eigenfrequencies of the BdG matrix suggests energetically unstable nature of the condensate while complex eigenfrequencies indicates dynamically unstable nature of the condensate~\cite{Ueda2012, Ozawa2013, Ravisankar2021}. To validate these claims, we perform numerical simulations using the GP equations. We provide the eigenvalue spectrum, eigenvectors, and numerical simulation in every subsection.

Following this, we corroborate the analytical results for the excitation spectrum by numerically solving the BdG equations~(\ref{bdgmatrix}) from which we also obtain the eigenvectors as a function of $q_x$. First, we consider a $[-1000:1000]$ grid in real space with step size $h_x=0.2$. Then, we use the Fourier collocation method, where we numerically compute the Fourier transform of the BdG equations and obtain a truncated reduced BdG matrix, which is subsequently diagonalized using the LAPACK package~\cite{Anderson1999}. In momentum space, we consider $[-700:700]$ modes in the $q_x$ direction, with a grid step size of $h_{q_x}=0.0157$. %

Further we provide the real-time dynamics analysis corresponding to the collective excitation spectrum by solving coupled GP equations (Eqs. (\ref{gpe1}) - (\ref{gpe2})) numerically. We use the imaginary-time propagation method to obtain the ground state. Then, we evolve the ground state in time using the real-time propagation method. For this purpose, we adopt the split-step Crank-Nicholson scheme outlined in Refs.~\cite{Muruganandam2009, Luis2016, Ravisankar2021}. We have considered a grid of $1280$ space points with space step $dx = 0.05$, time step $dt = 0.00025$ for imaginary-time-propagation, and $dt = 0.0005$ for real-time-propagation. Initially, we obtain the ground states according to the regions in the phase plot [see Fig.~\ref{fig8}] by using imaginary-time propagation. After calculating the ground state, we evolve it using real-time propagation by quenching the trap strength.
\begin{figure*}[]
\begin{centering}
\centering\includegraphics[width=0.99\linewidth]{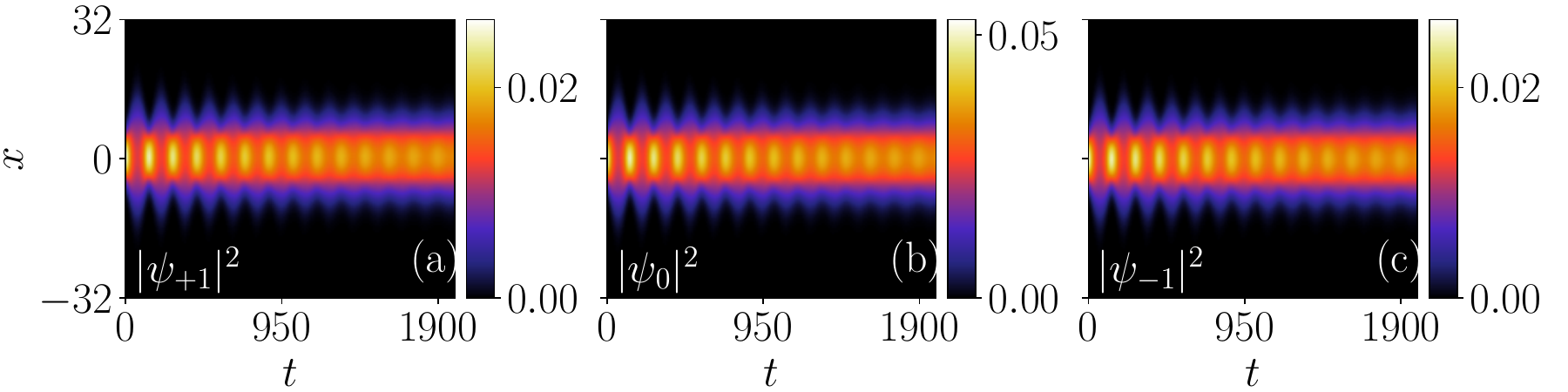}  
\caption{Time evolution of ground state density profiles illustrating dynamical stability of (a) $+1$, (b) $0$, and (c) $-1$ components of the condensate, where the parameters are similar to Fig.~\ref{fig9}. All three components $\vert\psi_{+1}\vert^{2}$, $\vert\psi_{0}\vert^{2}$, and $\vert\psi_{-1}\vert^{2}$ depict stable breather-like dynamics, which validate the dynamical stability of the condensate.}
\label{fig11a}
\end{centering}
\end{figure*}
\subsection{Excitation spectrum in region I}

\paragraph*{Excitation spectrum:} We present the collective excitation spectrum for the stable case and corresponding eigenvectors in Fig.~\ref{fig9}. Here, we consider the coupling strengths, $k_{L} = 0.5$ and $\Omega = 2.0$. Fig.~\ref{fig9}(a) shows only the presence of $\mathrm{Re}(\omega)$ in the eigenspectrum. The above suggests the presence of phonon mode~\cite{Pethick2008}. The absence of negative and imaginary eigenfrequency in the spectrum indicates that the phase is energetically and dynamically stable throughout the region-I. 

We present the eigenvectors corresponding to the eigenspectrum in Fig.\ref{fig9} (b), (c). We have three sets of eigenvectors: $\lvert u_{+1}\rvert^{2}$, $\lvert u_{-1}\rvert^{2}$, $\lvert u_{0}\rvert^{2}$, $\lvert v_{0}\rvert^{2}$, $\lvert v_{+1}\rvert^{2}$, and $\lvert v_{-1}\rvert^{2}$. The eigenvectors exhibit in-phase behaviour in all components for all wavenumber, which means that $\lvert u_{+1}(q_x)\rvert ^2=\lvert u_{-1}(q_x)\rvert^2$ and $\lvert v_{+1}(q_x)\rvert^2=\lvert v_{-1}(q_x)\rvert^2$, indicating the presence of a density-like (in-phase) excitations. [see Figs. \ref{fig9}(b) and \ref{fig9}(c)]. We demonstrate that both $u$ and $v$ eigenvectors are in-phase. For $q_{x} \approx 0$, the eigenvector components approach each other, and at $q_{x} = 0$, they have equal values, signifying the presence of a phonon mode, as mentioned earlier. As anticipated, the eigenspectrum and eigenvectors exhibit the same behaviour for other sets of parameters ($k_L$, $\Omega$) in the same region (region I of Fig.~\ref{fig8}) of the $k_L$-$\Omega$ plane. The critical change in the eigenspectrum is that with increasing coupling strengths, the real eigenspectrum widens similarly to how the eigenvectors also widen after the junction at $q_x = 0$.
\begin{figure}[!ht]
\begin{centering}
\centering\includegraphics[width=0.8\linewidth]{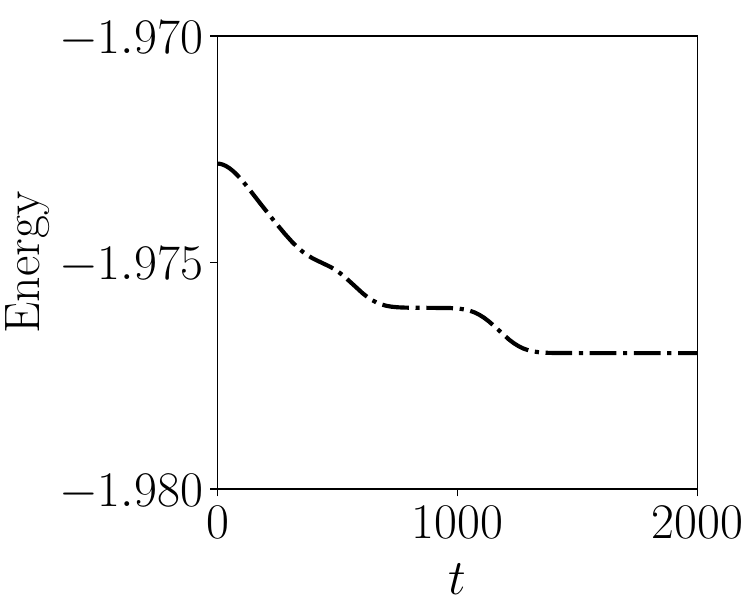}
\caption{Energy variation in the time evolution for the stable region I. The parameter values considered in this figure are similar to Fig.~\ref{fig9}. During time evolution, the energy of the condensate decreases in the beginning and further shows stable behaviour, which signifies that the condensate is energetically stable due to the lack of negative eigenfrequency.}
\label{fig11b}
\end{centering}
\end{figure}

\paragraph*{Dynamical stability:}
We have considered the same coupling parameters, $(k_{L}, \Omega) = (0.5, 2)$, to examine the dynamical behaviour of region-I, and the interaction strengths are $c_{0} = 0.5$, and $c_{2} = -0.1$ by solving the GP equations (Eqs. (\ref{gpe1}) - (\ref{gpe2})). Here, we obtain the plane-wave phase as a ground state of the condensate. To understand the stability of the ground state, quench the trap strength to half and further evolve the condensate. The density of the condensate shows stable breather-like motion throughout the dynamical evolution [See Fig.~\ref{fig11a}(a)-(c)]. In addition, the system density profile is symmetric in nature and doesn't show any oscillations, as shown in Figs.~\ref{fig11a}(a)-(c). Thus, it is clear that the system is not showing polarization behaviour as we expected from the in-phase nature of eigenvectors shown in Figs.~\ref{fig9}(b)-(c).%
  
In Fig.~\ref{fig11b}, we plot the total energy of the condensate. In real-time evolution, we quench the trap strength to one-half its original value. Hence, in the beginning, the energy of the condensate starts decreasing. After a while, the energy gets stabilized and constant throughout the dynamical evolution. Therefore, we find that the ground state of the condensate is energetically and dynamically stable, as shown in the stability phase plot Fig.~\ref{fig8} corresponding to the excitation spectrum. We also investigate the dynamical spin texture of this region, which does not change its behaviour during the dynamical evolution.
\begin{figure*}
\centering\includegraphics[width=0.95\linewidth]{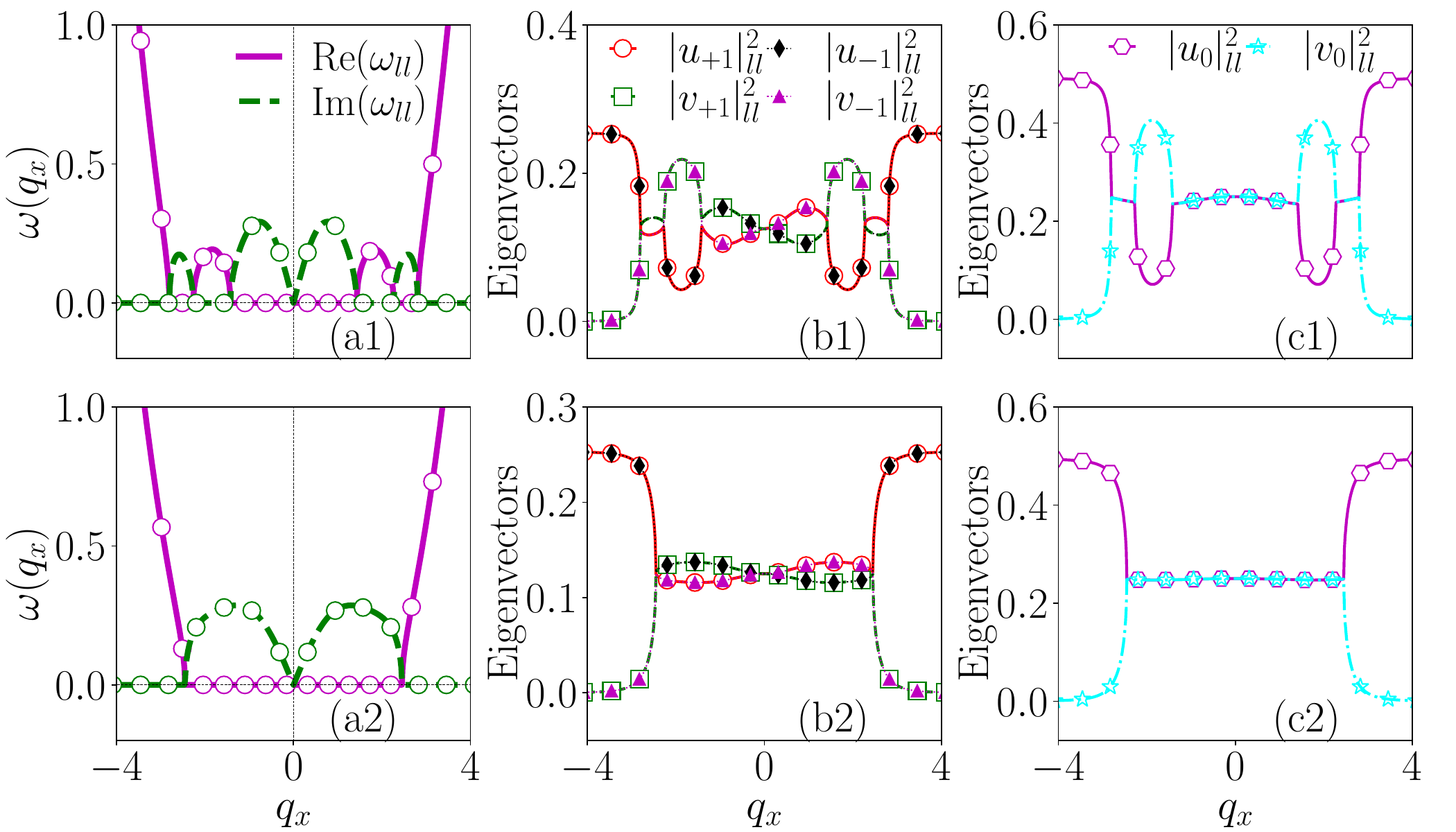}
\caption{
Stability excitation spectrum eigenvalues and eigenvectors: The SO and Rabi coupling strength are $(k_{L},\Omega)$ $= (2.0,2.0)$ (top row), and $(2.35, 4.0)$ (bottom row), respectively. (a1), (a2) shows eigenenergy, (b1), (c1), and (b2), (c2 ) corresponding eigenvectors, figure labels are as like Fig.~\ref{fig9}. The eigenenergy spectrum shows the presence of complex eigenfrequencies in terms of two (one) bands in the top (bottom) row, which indicates the system is dynamically unstable. Whereas the eigenvectors display the spin-dipole along with spin-like mode along the $q_x$ momentum direction, also symmetric along the axis.
}
\label{fig19}
\end{figure*}
\subsection{Excitation spectrum in region II{a}}
\paragraph*{Excitation spectrum:} 
To investigate region II, we divide it into two sub-parts, region IIa and IIb. In the first sub-part, we consider two points with fixed SO and Rabi coupling strengths: $(k_{L}, \Omega) = (2, 2)$,  and $(k_{L}, \Omega) = (2.35, 4)$, respectively. We plot the dispersion relation of collective excitations and corresponding eigenvectors at these values of the coupling parameters in Fig.~\ref{fig19}. Fig.~\ref{fig19}(a1) shows the presence of the imaginary part of the eigenspectrum at larger values of quasi-momentum $q_{x}$ having two imaginary bands. As we increase the value of Rabi coupling to larger extents than SO coupling strength, the double imaginary eigen-bands change into a single band [see Fig.~\ref{fig19}(a2)]. It is also shown in Fig.~\ref{fig8}, where the second point falls on the boundary of regions I and IIa. Therefore, we can state that the second point is approaching towards the stable regime.

The eigenvector corresponding to the eigenenergy is presented in the subsequent column of Fig.~\ref{fig19}. Unlike the behaviour observed in Region I, we find complex behaviour in the eigenvector components here. The eigenvectors shows in momentum direction the transition from density-like (in-phase) excitations, where $\lvert u_{+1}(q_x)\rvert^2=\lvert u_{-1}(q_x)\rvert^2$ and $\lvert v_{+1}(q_x)\rvert^2=\lvert v_{-1}(q_x)\rvert^2$, to the spin-like (out-of-phase) excitations, where $\lvert u_{+1}(q_x)\rvert^2=\lvert v_{-1}(q_x)\rvert^2$ and $\lvert v_{+1}(q_x)\rvert^2=\lvert u_{-1}(q_x)\rvert^2$. In Fig.~\ref{fig19}(a1), the eigenspectrum has double band instability; corresponding to this, dual transitions from density-like mode to the spin-like mode for $\pm 1$ components of eigenvectors. Still, the $0$-th component exists only in density-like mode. Also, there is an amplitude difference among all three components of the eigenvector; the $\pm 1$ components have half of the amplitude value than the $0$-th component [see Figs.~\ref{fig19}(b1) and ~\ref{fig19}(c1)]. 
\begin{figure*}
\begin{center}
\includegraphics[width=0.95\linewidth]{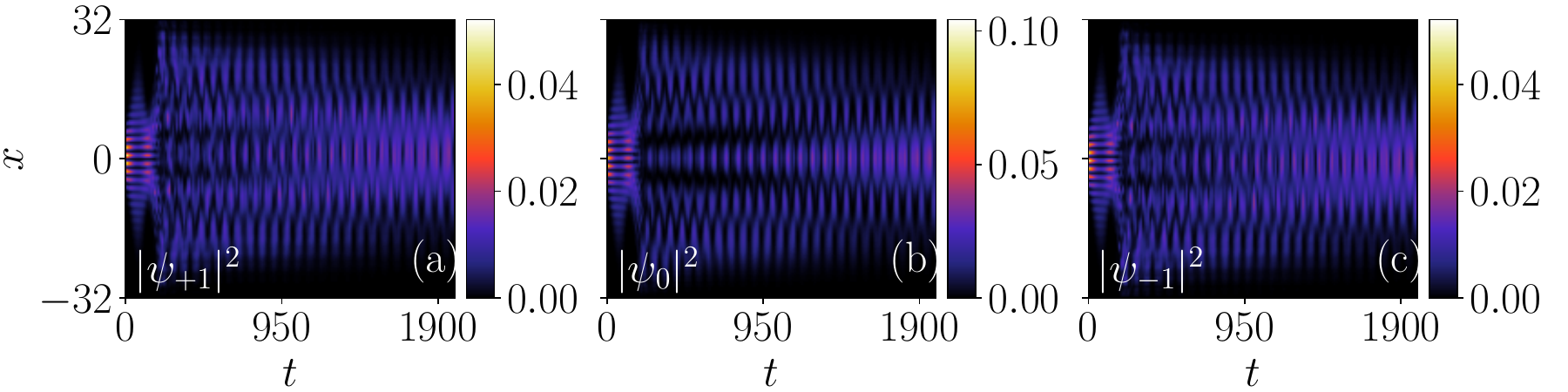}
\includegraphics[width=0.95\linewidth]{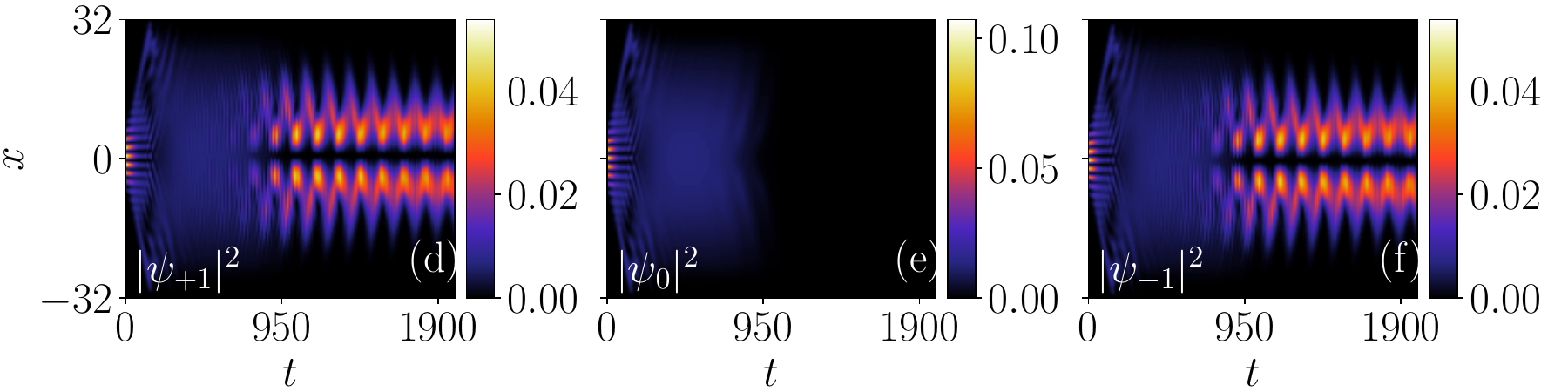}
\end{center}
\caption{Dynamics: (a-c) and (d-f) depict temporal evolution of ground state density profile of $\vert\psi_{+1}\vert^{2}$, $\vert\psi_{0}\vert^{2}$, and $\vert\psi_{-1}\vert^{2}$ for $(k_{L},\Omega)$: top row $(2.0, 2.0)$ and bottom row $(2.35, 4.0)$, respectively. The density profile holds stripe wave behavior in both cases for $t < 150$, and $t < 180$ for the top and bottom row, respectively. Further, it gets fragmented into several small domains. This behavior signifies the change in shape and amplitude of the density profile, where the bottom row $0$-th component gets diminished, whereas the $\pm 1$ components gained the density.}
\label{fig21a}
\end{figure*} 
For the second point in the region [see Fig.~\ref{fig19}(a2)], the eigenspectrum has only single-band instability. And there is a constant transition from the density-like mode to the spin-like mode of $\pm 1$ components of eigenvectors [see Fig.~\ref{fig19}(b2)]. The amplitude of $\pm 1$ and $0$ appears the same as for the previous point. It is observed from the above discussion whenever the $\mathrm{Im}(\omega_{-})$ appears, the density-like mode (where we have $\mathrm{Re}(\omega_{-}) = 0$) changes into the spin-like mode. From the transition point of view, when comparing these two points in this region-IIa, one can notice that the multi-band instability has mixed density-like and spin-like mode behaviour in the eigenvectors. In detail, by comparing Figs.~\ref{fig19}(b1) and ~\ref{fig19}(b2), the second only has the density-like mode in the major portion of the momentum space. But first one has a mixed type of mode in the same range of momentum space, where imaginary eigenvalues appeared at $0< q_x <2$, it has spin-like mode when $2< q_x <3$ eigenvalues are real, then it exhibits density-like mode, for $3< q_x <3.5$ again spin-like mode, $ q_x >3.5$ density-like mode (behaviour is symmetric about the momentum axis). Such complex-mixed-mode has not been explored in the spinor SO coupled system. We highlight that this study will be helpful for the experimental researcher to achieve the stable ferromagnetic spinor SO coupled BEC experimentally.
 
\begin{figure}[!htp] 
  \begin{centering}
  \centering\includegraphics[width=0.99\linewidth]{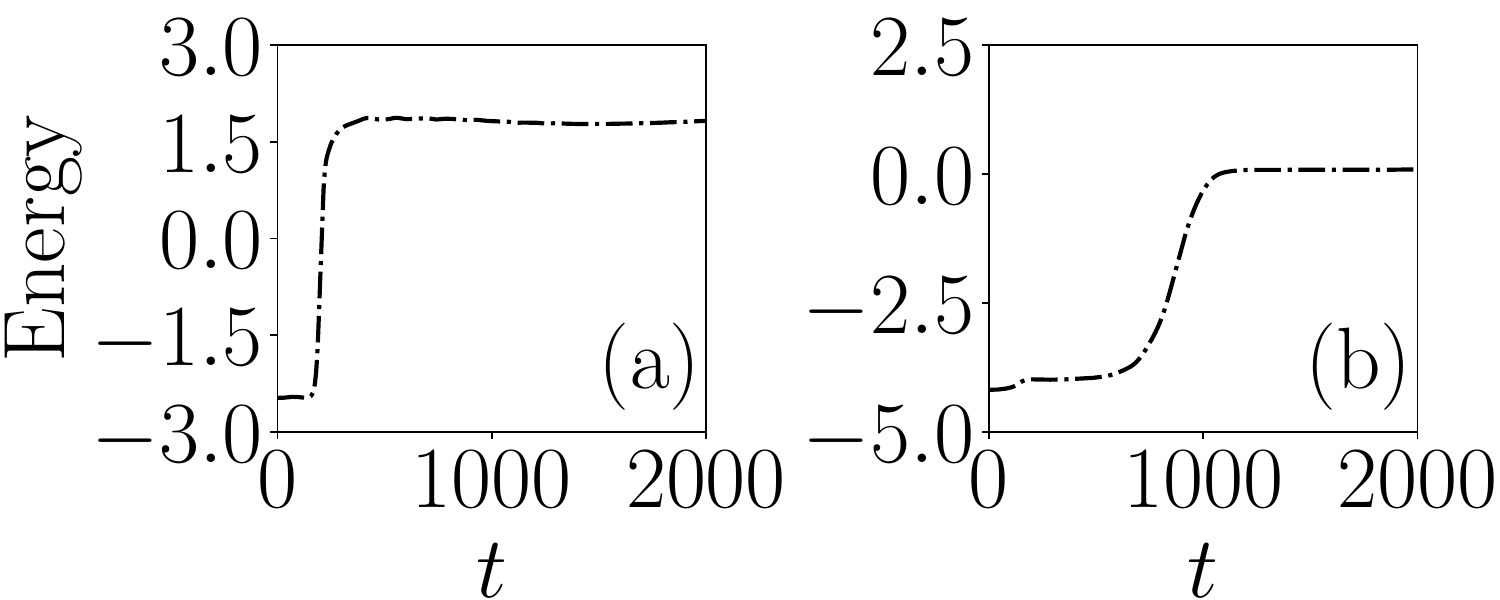}
  \caption{(a) and (b) represent the total energy of the condensate for the density evolution in Fig.~\ref{fig21a}. In both cases, the energy of the condensate increases at the beginning of dynamical evolution, but finally shows well-settled stable behavior, such kind of sudden change happened due to the external perturbation. This behavior validates the energetic stability of the condensate, however, the system is dynamically unstable.}
  \label{fig21b}
  \end{centering}
  \end{figure}

\paragraph*{Dynamical stability:} 
We divide region II into two subregions: IIa and IIb. We focus on the system's dynamics in region IIa, specifically at two different points within the region IIa. For each point, we analyzed the dynamic behaviour of the system separately. We discuss the dynamic behaviour of the system for each point separately. We choose $(k_{L}, \Omega) = (2, 2)$. We obtain the ground state using imaginary-time propagation, which is the stripe-wave and evolve the ground state using real-time propagation. The dynamical evolution of the state shows the stripe wave phase holds up to $t = 150$. Furthermore, the density profile of all three components fragmented into non-periodic domains [see Figs.~\ref{fig21a}(a)-(c)]~\cite{Mithun2019}. If the system has a nonvanishing $0$-th component in the dynamical evolution, then the system tends to be less subjected to instability. Conversely, if the $0$-th component vanishes, the system becomes more prone to instability~\cite{Tasgal2015}. In this case, the appearance of the $0$-th component in the dynamical evolution confirms that the chosen coupling point has a lower risk of instability. We plot the total energy of the condensate in Fig.~\ref{fig21b}(a). The energy of the condensate starts with a value of $-2.471$. It changes sign from negative to positive at $t = 205$, due to the quench similar kind of agitation happens, and further, it shows stable behaviour near $t = 365$. For this point, as shown in Fig.~\ref{fig19}(a1), the BdG excitation spectrum has complex eigenfrequencies, and the eigenvector shows the out-of-phase behaviour for the certain range of quasi-momentum. The above indicates that the system is dynamically unstable and generates non-linear wave patterns, which matches well with numerical simulation results.
We choose another point in the region IIa at which we consider the value of SO coupling strength $k_{L} = 2.35$ and $\Omega = 4.0$. We achieve the ground state, which is stripe-wave-phase. The excitation spectrum has complex eigenfrequencies, as shown in Fig.~\ref{fig19}(a2). The presence of complex eigenfrequencies indicates that the system is dynamically unstable. To verify this claim numerically, we evolve the ground state of the system using the real-time-propagation method as illustrated in Figs.~\ref{fig21a}(d)-(f). The stripe wave behaviour of the condensate holds its shape for a while. Then, it gets fragmented into several small domains~\cite{Mithun2019}. In addition, we observe that with time, the zeroth component of density starts diminishing, and finally, it disappears [see Fig.~\ref{fig21a}(e)]. Other components of the condensate density ($\lvert\psi_{\pm 1}\rvert^2$) boosted the magnitude after the zeroth component of the density disappears [see Figs.~\ref{fig21a}(d), and (f)]. As one considers the quadratic Zeeman effect, then such kind of dynamical disappearance does not take place~\cite{Tasgal2015}. However, in this work, we see such sort of dynamical instability in the excitation of spin-$1$ BECs in the presence of SO coupling alone. So far we find that the density pattern of the system changes its shape and amplitude accompanied with disappearance of the $0$-th component, which is a signature of dynamical instability. Further, we show the total energy of the condensate in Fig.~\ref{fig21b}(b). It starts from $-4.182$ and changes its sign from negative to positive at $t = 1080$, which is higher than the previous point, where the $0$-th component vanishes, and then the particles are shared by the spin-up and down components, there, the energy gets settled. Such kind of $0$-th component vanishing behaviour is not observed for the previous case due to the appearance of spin and density-mixed modes. Upon investigating the dynamical spin texture of this region, we notice that the initial pattern does not maintain its shape as time progresses and changes its spin texture during the dynamical evolution [see Appendix.~\ref{app:mag-vec}].
\begin{figure}
\begin{centering}
\centering\includegraphics[width=0.8\linewidth]{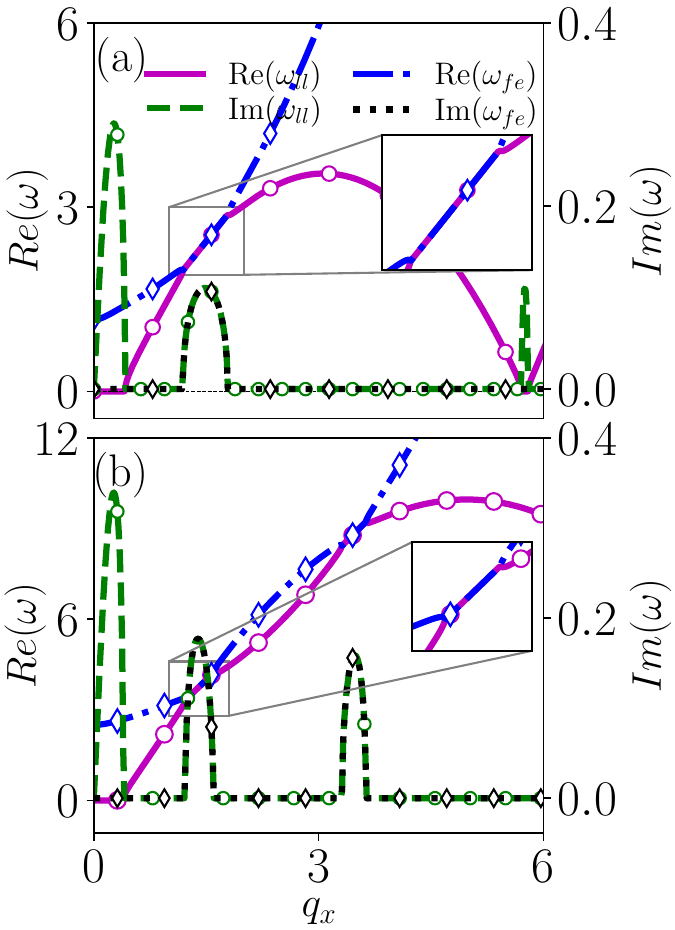}
\caption{Excitation spectrum (a), and (b) for the coupling strengths, $(k_{L}, \Omega) = (3.1, 1.17)$ and $(5.0, 2.5)$ depict the low-lying and first-excited branches of the spectrum. The solid magenta line represents the $\mathrm{Re}(\omega_{-})$ and the dashed green line represents the $\vert\mathrm{Im}(\omega_{-})\vert$ which is low-lying branch of the spectrum, while the blue dash-dotted line represents $\mathrm{Re}(\omega_{+})$ and black dotted line represents $\vert\mathrm{Im}(\omega_{+})\vert$ corresponding to the first-excited branch of the spectrum, which obtained from Eq.~\ref{bdgex}. The eigenspectrum in (a, b) shows multiple instability bands as we obtained in~\ref{fig19}(a1), however, (a) and (b) show one and two gapless modes. Insets are for the better visualization of gapless mode, the secondary $y$-axis scale is for representing the imaginary eigenfrequencies magnitudes.}
 \label{fig29a}
\end{centering}
\end{figure}
\begin{figure}
\begin{centering}
\centering\includegraphics[width=0.99\linewidth]{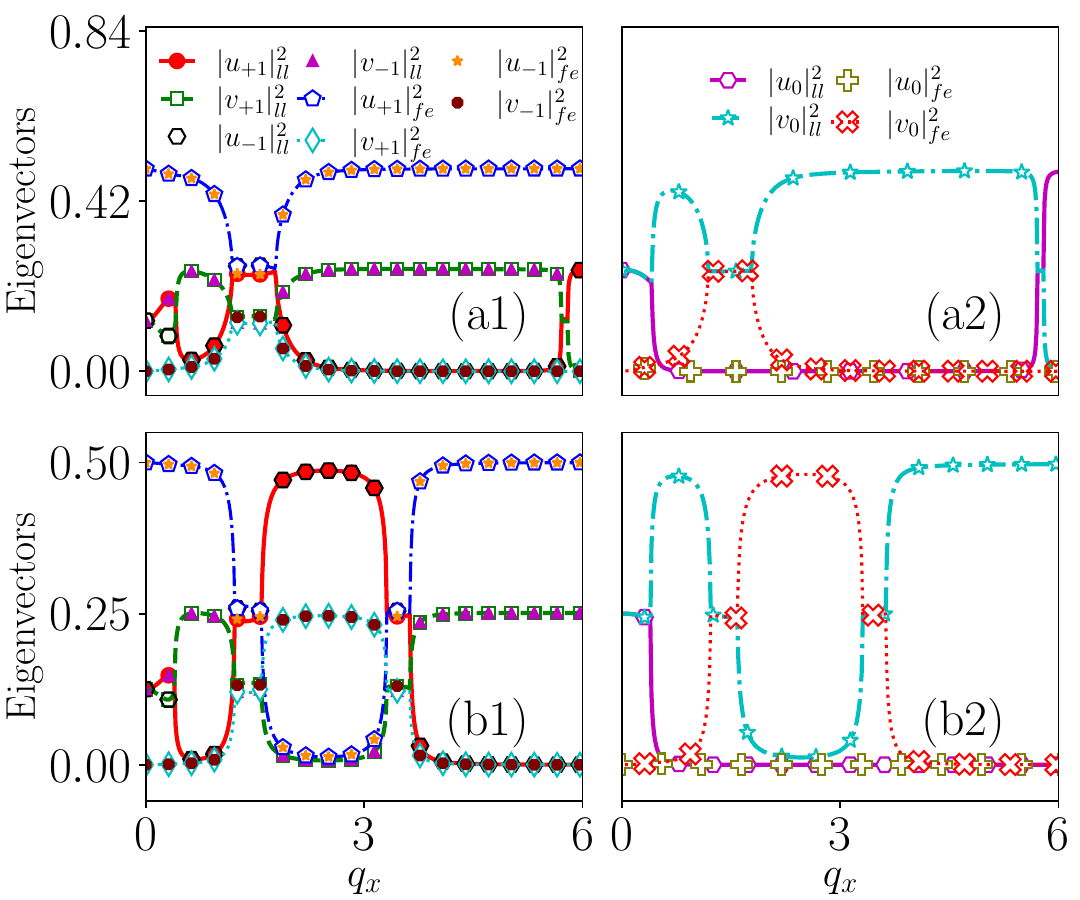}
\caption{The eigenvectors top and bottom rows corresponding to Fig.~\ref{fig29a}(a), and Fig.~\ref{fig29a}(b), respectively.
The eigenvectors represents as $\vert u_{+1}\vert^{2}_{ll}$ (red dot), $\vert u_{-1}\vert^{2}_{ll}$ (black hexagons), $\vert u_{0}\vert^{2}_{ll}$ (magenta hexagons), $\vert v_{0}\vert^{2}_{ll}$ (cyan open stars), $\vert v_{+1}\vert^{2}_{ll}$ (green open squares), and $\vert v_{-1}\vert^{2}_{ll}$ (magenta triangles). Whereas, $\vert u_{+1}\vert^{2}_{fe}$ (blue open pentagon), $\vert u_{-1}\vert^{2}_{fe}$ (orange stars), $\vert u_{0}\vert^{2}_{fe}$ (olive open plus), $\vert v_{0}\vert^{2}_{fe}$ (red open x with the dotted line), $\vert v_{+1}\vert^{2}_{fe}$ (cyan open diamond), and $\vert v_{-1}\vert^{2}_{fe}$ (maroon dots). We depict the $+q_x$, however, the eigenfrequencies and eigenvectors are symmetric about the axis. Here we found that the eigenvectors show the spin-like mode when the complex-frequency exhibits, the density-like mode when it is only real. }
 \label{fig29b}
\end{centering}
\end{figure}
\subsection{Excitation spectrum in region IIb}
\paragraph*{Excitation spectrum:} In this subpart of region II, we choose two different points at which we consider the coupling strengths as $k_{L} = 3.1$, $\Omega = 1.17$, and $k_{L} = 5.0$,  $\Omega = 2.5$, respectively. The collective excitation spectrum of the condensate for these sets of parameters is given in Fig.~\ref{fig29a}. We obtain complex eigen frequencies for both points [see Figs.~\ref{fig29a}(a)- ~\ref{fig29a}(b)]. Still, the number of instability bands along the quasimomentum direction and its amplitude differ for both points. In Fig.~\ref{fig29a}(a), the number of low-lying instability bands is three, and the positions and amplitudes are $ \{ q_x, \omega \}$ = $\{0.27, 0.29\}$, $\{1.48, 0.11 \}$, and $\{5.75, 0.11 \}$, respectively. Further, we show the first excited ($\omega_{fe}$) and low-lying ($\omega_{ll}$) branches of the eigenspectrum, which overlap with each other, where we also find that the complex eigenfrequency is not similar to the eigenspectrum of the previous case region-IIa. This results in gapless modes between the first- and low-lying excited spectrum. However, we find three complex eigenfrequencies in which the overlapping point shows complex eigenfrequencies for both spectrums. The phenomenon that has been observed in the spinor BECs is related to the \textit{unstable avoided crossing} between spin and charge modes by considering $\theta = \pi$, the phase difference between the perturbed wave fields, which also provides the class $I_o$ oscillatory patterns as a result of the dynamical instability~\cite{Cross1993,Bernier2014}. However, in this work, we have realized the same phenomenon only for the symmetric case $\theta = 0$, even though we consider only SO and Rabi couplings without Raman detuning. Due to the $I_o$ dynamical instability, we find out-of-equilibrium dynamics in the density patterns, and it is the route to the nonlinear patterns formation, which we shall discuss later in this subsection. In Fig.~\ref{fig29a}(b), the number of instability bands is four, and the positions and amplitudes are $ \{ q_x, \omega \}$ = $\{0.27, 0.33\}$, $\{1.39, 0.18 \}$, $\{3.48, 0.16 \}$, and $\{9.45, 0.11 \}$ (not shown in figure), respectively. The appearance of instability bands for both points is symmetric about the momentum ($q_x$) axis. In addition to this, we show the first excited state in the same plot.  Here, the first excited and low-lying branches overlap twice, where we observed that double-unstable avoided crossing in SO-coupled spinor ferromagnetic BECs. For this case, the first excited spectrum also has two complex eigenfrequencies, while the unstable-crossing avoided points. %
\begin{figure*} 
\begin{centering}
\centering\includegraphics[width=0.95\linewidth]{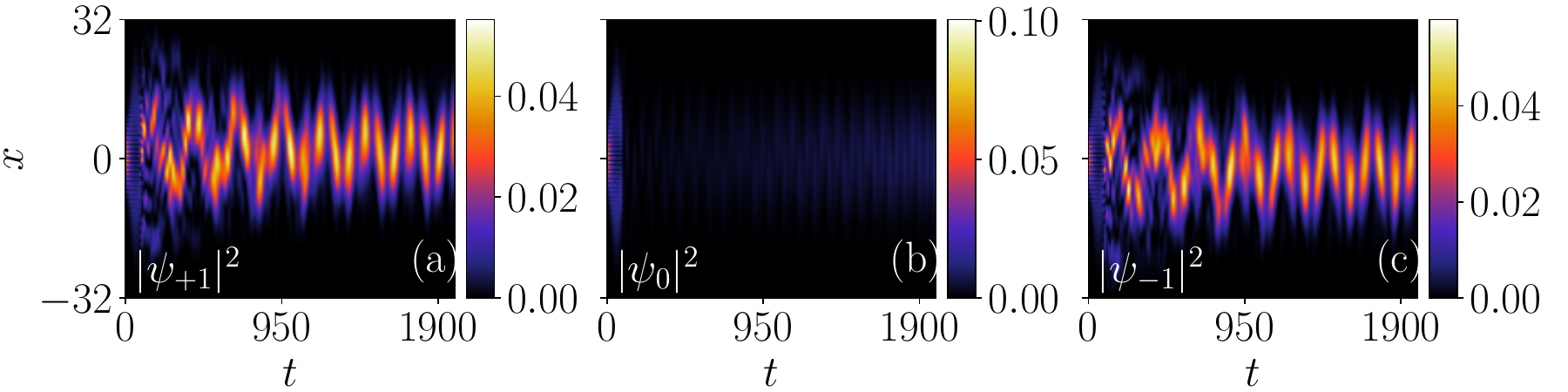}
\centering\includegraphics[width=0.95\linewidth]{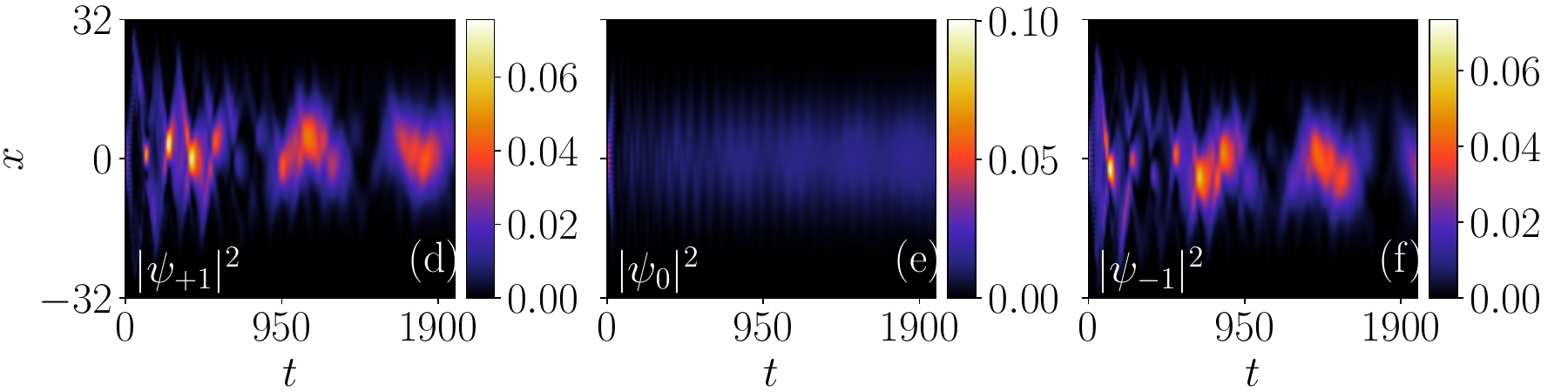}
\caption{Dynamics: (a)-(c) and (d)-(f) represent the dynamics of the ground state density profile of $\vert\psi_{+1}\vert^{2}$, $\vert\psi_{0}\vert^{2}$, and $\vert\psi_{-1}\vert^{2}$  components for $(k_{L}, \Omega)$ top row $(3.1, 1.17)$, and bottom row $(5.0, 2.5)$. The density profile holds stripe wave behavior in both cases for $t < 100$ and $t < 50$ units for the top and bottom row, respectively. Top row: $t> 100$, it shows non-periodic oscillation and non-uniform density in space-time, whereas the $0$-th component gets disappeared. Bottom row: $t>50$, it further gets fragmented into several small domains. This behaviour signifies the change in shape and amplitude of the density profile, along with the disappearance of the $0$-th component, which confirms the dynamical instability.}
\label{fig31a}
\end{centering}
\end{figure*} 

The eigenvector corresponding to the eigenspectrum is shown in the subsequent columns of Fig.~\ref{fig29b}. For the choice of parameters $(k_{L}, \Omega) = (3.1, 1.17)$, we observed no overlap between the low-lying and the first excited spectrum eigenvectors. However, the gap between them closes, and the overlap happens for certain values of quasimomentum $q_x$.  The interval $q_x=(-0.4,0.4)$, where the low-lying spectrum exhibits imaginary eigenvalues, for this case, the eigenvectors correspond to the spin-like mode, as we see from the out-of-phase behaviour between the low-lying eigenvectors. Further, when $q_x > \pm 0.4$ up to the gapless point, we notice that the low-lying eigenvectors display density-like (in-phase) mode. However, the first-excited state only have a density-like mode up to this gapless point. Nevertheless, in the region of the spectra where the low-lying and the first excited branches overlap, we see that the eigenvectors of both the low-lying and first-excited states are out-of-phase among them. When the gapless mode transforms as gapped mode, all the eigenvectors are in phase [see Fig.~\ref{fig29b}(a1)], where the $0$-th component shows only density-like mode [see Fig.~\ref{fig29b}(a2)].%

For the second point of this region, all eigenvector components have the same amplitude value. Choosing the coupling parameters $(k_{L}, \Omega) = (5, 2.5)$ reveals a substantial difference. With these parameters, we observed double overlaps between the low-lying and the first excited spectrum, closing the gap between them. These overlaps occur for certain values of quasimomentum $q_x$. The first two eigenvectors exhibit behaviour similar to that observed at the previous point.

As the gapless mode transforms into a gapped mode, all the eigenvectors become in phase. We also observe similar feature for the next gapless mode and further momentum directions. Additionally, we find that when the real eigenvalues fall to zero and then rise, the eigenvectors are flipped but remain in phase as shown in Fig.~\ref{fig29b}(a1). The obvious change in the spectrum from the gapped to the gapless mode of the eigenvectors demonstrates the transition from a constant density-like mode to a hybridized density-spin-like mode ~\cite{Abad2013} [see Figs.~\ref{fig29b}(b1, b2)].

\paragraph*{Dynamical stability:} 
We consider the points $(k_{L}, \Omega) = (3.1, 1.17)$, with interaction parameters, $c_{0} = 0.5$ and $c_{2}  = -0.1$. The collective excitation spectrum corresponding to this point is shown in Fig.~\ref{fig29a}(a), which has complex eigenfrequencies in terms of multiple bands. The presence of complex eigenfrequencies makes the system dynamically unstable. To verify that the system is dynamically unstable, we perform numerical simulation. We obtain the ground state, which is the stripe-wave phase. We evolve the ground state by quenching the trap, and during dynamical evolution, the density profile holds stripe wave behaviour initially. After a while, the $\pm 1$ components are fragmented into small domains and show non-periodic oscillation in space-time evolution, shown in the density profiles in  Figs.~\ref{fig31a}(a) and~\ref{fig31a}(c)]. On the other hand, the zeroth component of density starts diminishing as time progresses [see Fig.~\ref{fig31a}(b)]. Thus, we state that the density pattern of the condensate changes its shape and amplitude during dynamical evolution, which confirms the dynamical instability as the $0$-th component disappears due to the instability of the system~\cite{Tasgal2015}.  
\begin{figure}[!htp] 
\begin{centering}
\centering\includegraphics[width=0.99\linewidth]{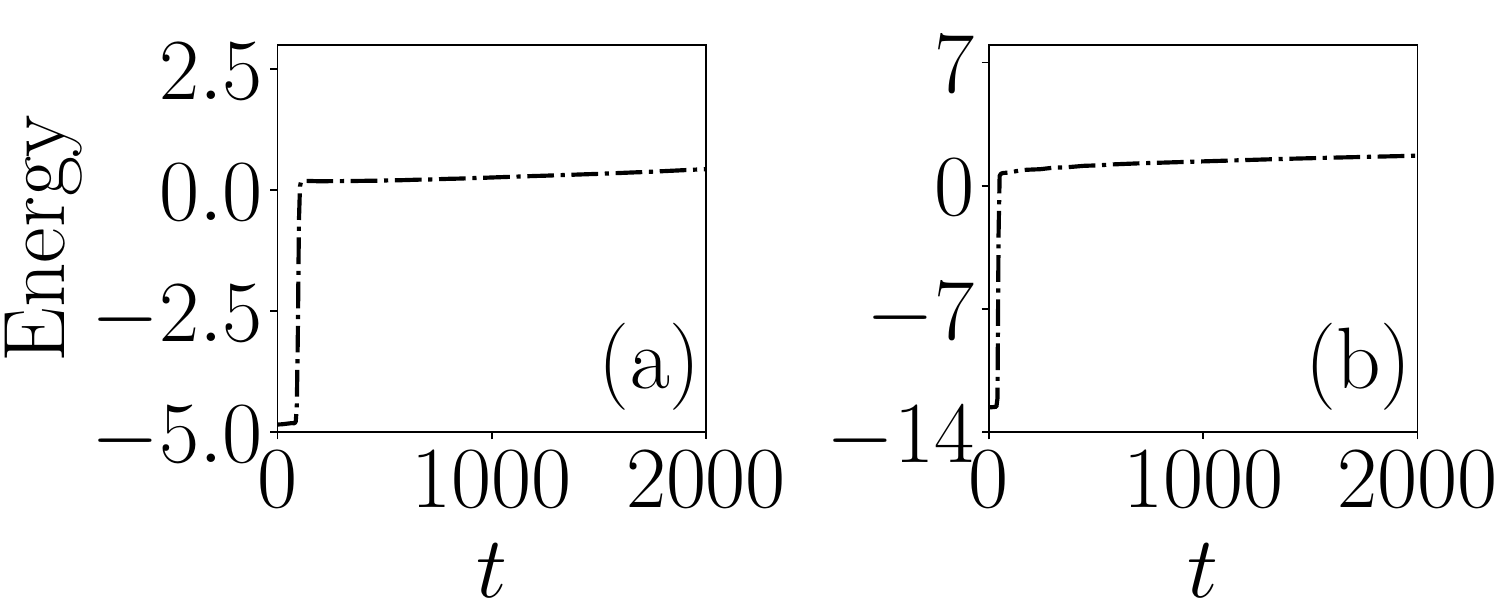}
\caption{(a) and (b) depict the total energy of the condensate for the dynamical evolution for the Fig.~\ref{fig31a}. From the energy saturation, we understood that the system is energetically stable but dynamically unstable. }
\label{fig31b}
\end{centering}
\end{figure} 
We investigate the second point by choosing $(k_{L}, \Omega) = (5, 2.5)$. For this, we obtain the ground state as the stripe wave and evolve the ground state numerically. The dynamical evolution of this phase shows the stripe wave phase holds for a while. Further, the density profile of all three components fragmented into non-periodic domains [see Figs.~\ref{fig31a}(a) and~\ref{fig31a}(c)]~\cite{Mithun2019}. Also, the density of the zeroth component starts diminishing, explicitly showing that the system holds dynamical instability. We also observe that the density pattern of the system changes its shape and amplitude during dynamical evolution. 

We plot the total energy of the condensate in the Fig.~\ref{fig31b}. The ground state energy of the condensate starts with the values (a) $E_0 = -4.845$,  (b) $E_0 = -12.593$. It changes sign from negative to positive, then stable, after the sudden change in the parameter during time evolution. We observed that the second point is more energetic than the first point as well as in the ground state and excited state, i.e., (a) $E_{ex} \approx 0.432$,  (b) $E_{ex} \approx 1.708$. Further, we investigated the dynamical spin texture, where the initial pattern is not observed, and at the final time, they changed their spin-textures in the dynamical evolution [see Appendix.~\ref{app:mag-vec}].
\subsection{Excitation spectrum in region-III}
\begin{figure}[!ht] 
\centering\includegraphics[width=0.8\linewidth]{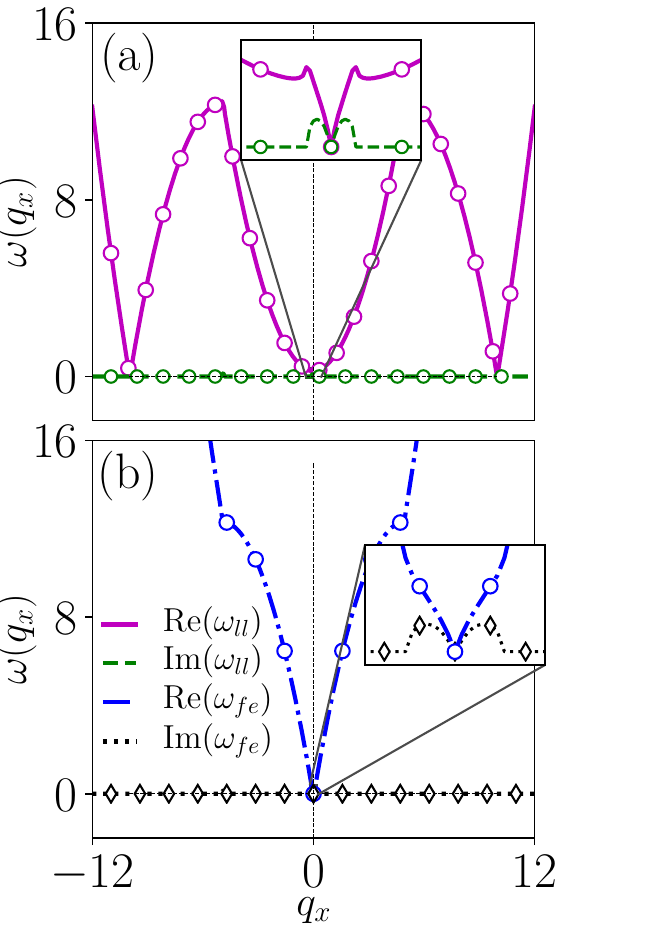}
\caption{Excitation spectrum (a) low-lying and (b) first-excited branches, for the coupling strengths $k_{L} = 5.0$, $\Omega = 0.0$. The label representation is similar to the Fig.~\ref{fig29a}. Insets are maximized for the visualization of the complex frequency mode. Whereas the gapless mode also appeared along with the symmetric roton mode.}
\label{fig59a}
\end{figure}

\paragraph*{Excitation spectrum:} 
In region-III, we choose points along the horizontal axis to understand the system's behaviour in the absence of Rabi coupling strength ($\Omega=0$), and we consider the SO coupling strength as $k_{L} = 5.0$. The collective excitation spectrum for this is demonstrated in Fig.~\ref{fig59a}(a, b), where differently from the previous cases, in region III we observe the presence of roton-maxon-like modes. As demonstrated in previous studies ~\cite{chen2022elementary,Khamehchi2014,Lyu2020} when $\Omega \rightarrow 0$ there is the appearance of roton-maxon modes and the subsequent softening of the roton mode when $\Omega$ approaches to zero. The softening of the roton mode favors the appearance of superstripe structures.  The low-lying excitation has complex eigenfrequencies alongside the phonon and symmetric roton-like mode, indicating that the system is dynamically unstable due to the presence of the imaginary eigenfrequencies. 
Further, the first-excited state also shows similar behaviour, i.e., the presence of the phonon mode along with the complex eigenfrequencies but no symmetric roton-like structure. Although, the unstable avoided-crossing is noticed at $q_x \approx 4.8499$, between the first- and low-lying states. After that point, the real eigenfrequency decreases and reaches zero, where it exhibits the imaginary eigenfrequency. Beyond this point, the real frequency gains the growth. Based on this observation, one has to understand such behaviour using the eigenvectors. Additionally, the complex eigenfrequency presents multiple instability bands along the quasimomentum axis. The positions of the low-lying excitation spectrum instability bands are $q_{x} = 0.06, 4.91, 9.97 $, and the corresponding complex eigenfrequencies are $\omega_{-} = 0.1037, 0.1404, 0.0993$, respectively. The first excited state has two instability bands, and its position and complex eigenfrequency are the same as those of the low-lying state. Notably, the instability bands are symmetric about the $x$-axis.

\begin{figure}
\begin{centering}
\centering\includegraphics[width=0.99\linewidth]{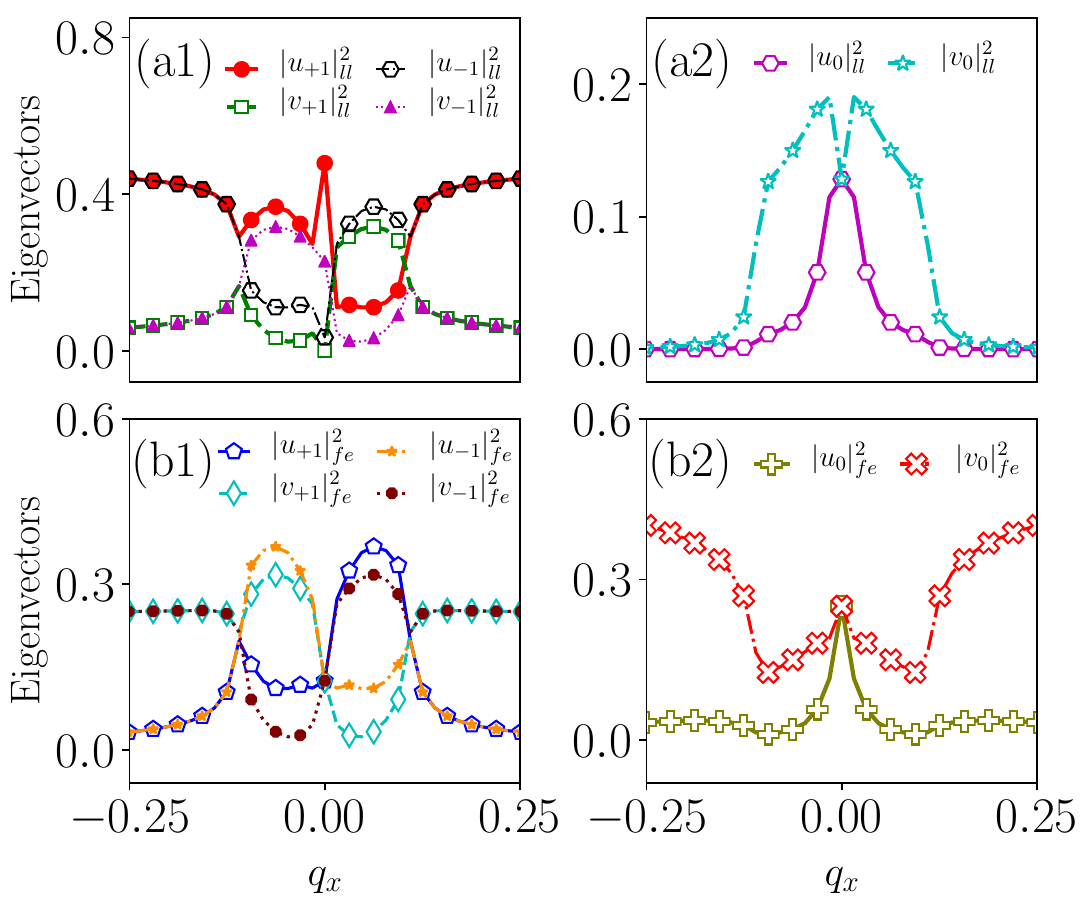}
\caption{Eigenvectors top and bottom rows are corresponding to the Fig.~\ref{fig59a}(a) and, Fig.~\ref{fig59a}(b) respectively. The label representation is similar to the Fig.~\ref{fig29b}. In this region, both low-lying and the first excited state assume the value $\omega=0$ for $q_x=0$ with eigenvectors are in-phase with the low-lying and first-excited states, which indicates the existence of the phonon mode for these branches.}
 \label{fig59b}
\end{centering}
\end{figure}

Fig.~\ref{fig59b} represents the eigenvector components, in this region, the low-lying excitation eigenvectors exhibits in-phase mode between $\vert u_{+1}\vert_{ll}^2$ and $\vert u_{-1}\vert_{ll}^2$ as well as between $\vert v_{+1}\vert_{ll}^2$ and $\vert v_{-1}\vert_{ll}^2$. This feature continues up to the point of real eigenvalues, when the imaginary eigenvalues exhibit out-of-phase between them ($\vert u_{+1}\vert_{ll}^2 - \vert u_{-1}\vert_{ll}^2 \neq 0$). Although at $q_x=0$, the eigenvalues become zero $\omega=0$, a phonon-mode is observed, which is explicitly shown in the form of sharp spike-like-eigenvectors appeared in $\vert u_{+1}\vert_{ll}^2$ and $\vert v_{+1}\vert_{ll}^2$.
However, in this complex eigenvalues regime, we have both phonon-mode as well as complicated spin-like mode. In addition to this, the $0$-th component eigenvectors show a mixed-mode behaviour, and as a result of this, we observed different dynamical phases from the ground state. Also, the first excited state exhibits $\omega=0$ for $q_x=0$, where all the eigenvectors met at a single point as before. The first-excited state behaves like a spin-like mode with an asymmetric nature, indicating the existence of the superstripe phase. Further, the density mode appears while the real eigenfrequency is present, also evident from the $0$-th component eigenvector behaviour. %

\begin{figure*}[] 
\begin{centering}
\centering\includegraphics[width=0.95\linewidth]{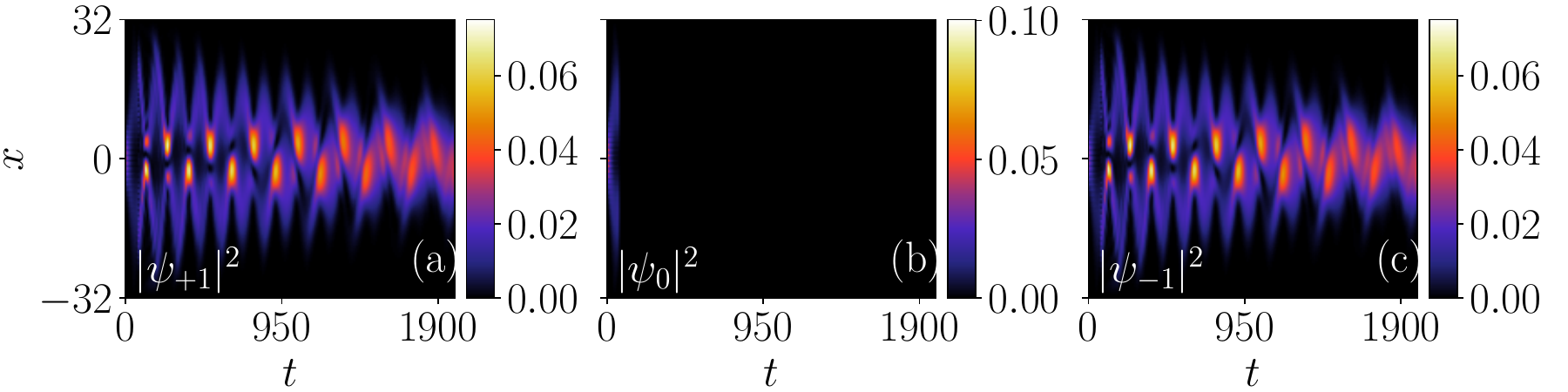}
\caption{Dynamics: (a)-(c) represent the time evolution of super-stripe ground state density profiles of $\vert\psi_{+1}\vert^{2}$, $\vert\psi_{0}\vert^{2}$, and $\vert\psi_{-1}\vert^{2}$ components of the condensate. The coupling strengths are $k_{L} = 5.0$, $\Omega = 0.0$ with interaction parameters $c_{0} = 0.5$ and $c_{2} = -0.1$. During dynamics, the unpolarized super-stripe wave holds its behavior for $t < 30$. Further, it breaks into two domains and oscillated either sides of the condensates of $\pm 1$ components, and the $0$-th component disappears, where $\pm 1$ components densities are obtaining the gain, which shows the change in the density profile confirms the dynamical instability.}
\label{fig61}
\end{centering}
\end{figure*}

\paragraph*{Dynamical stability:} 
We consider a case with SO coupling strength $k_{L} = 5.0$ and zero Rabi coupling strength $\Omega = 0.0$. The corresponding BdG excitation spectrum in Figs.~\ref{fig59a}(a) and (b) reveal the presence of phonon-gapless complex eigenfrequencies indicating the dynamical unstable nature of the condensate. To investigate this numerically, we obtained the ground state of the condensate, which falls within the unpolarized stripe-wave phase. We then evolved this state in time using a real-time propagation method. The density profile initially maintains its unpolarized stripe-wave behaviour however subsequently breaks into two parts at $t = 85$. Following this, the density of $\pm 1$ components oscillates around the trap centre, exhibiting accumulated breather behaviour for a period. Ultimately, the fragmented oscillating waves merge and form several small domains [see Figs.~\ref{fig61}(a) and (c)]~\cite{Mithun2019}, while the zeroth component of the density disappears [see Fig.~\ref{fig61}(b)]. This behaviour signifies that the dynamical instability also changes its shape and size~\cite{Tasgal2015}. We examined the energy during its dynamic evolution, which eventually stabilizes, indicating energetic stability. This finding is consistent with the analysis of the excitation spectrum, suggesting that the numerical simulation accurately reflects the theoretical predictions.
\subsection{Effect of coupling strength on the Band gap}
\begin{figure}[!ht] 
\begin{centering}
\centering\includegraphics[width=0.99\linewidth]{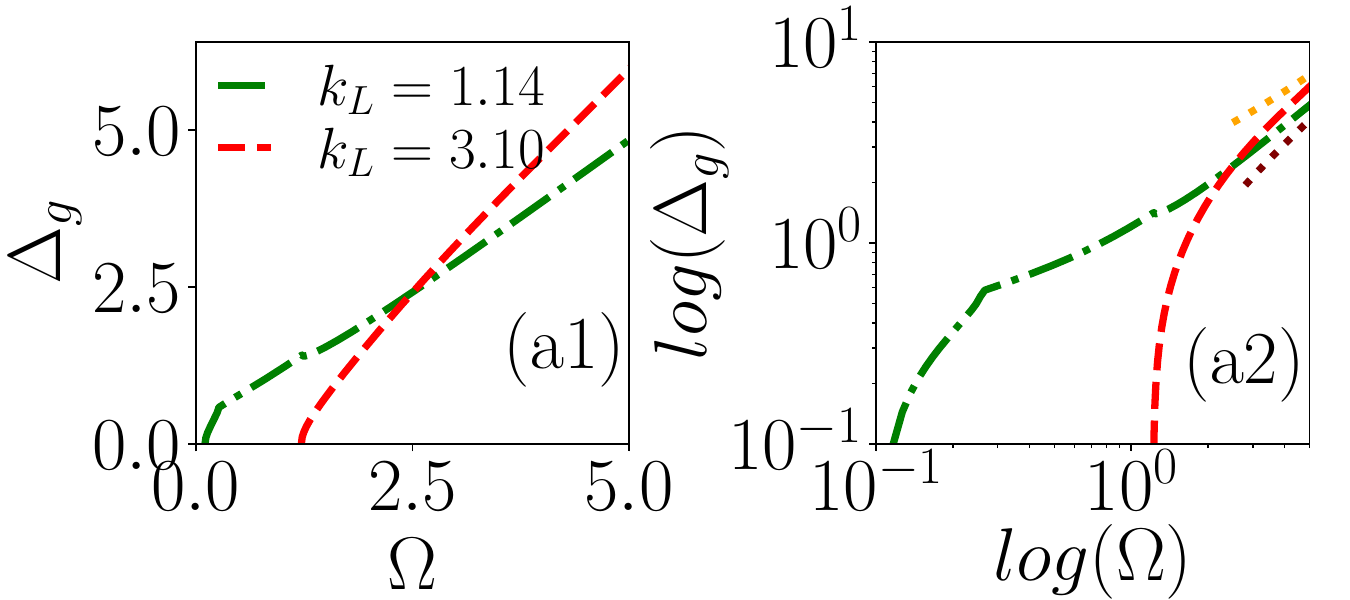}
\centering\includegraphics[width=0.99\linewidth]{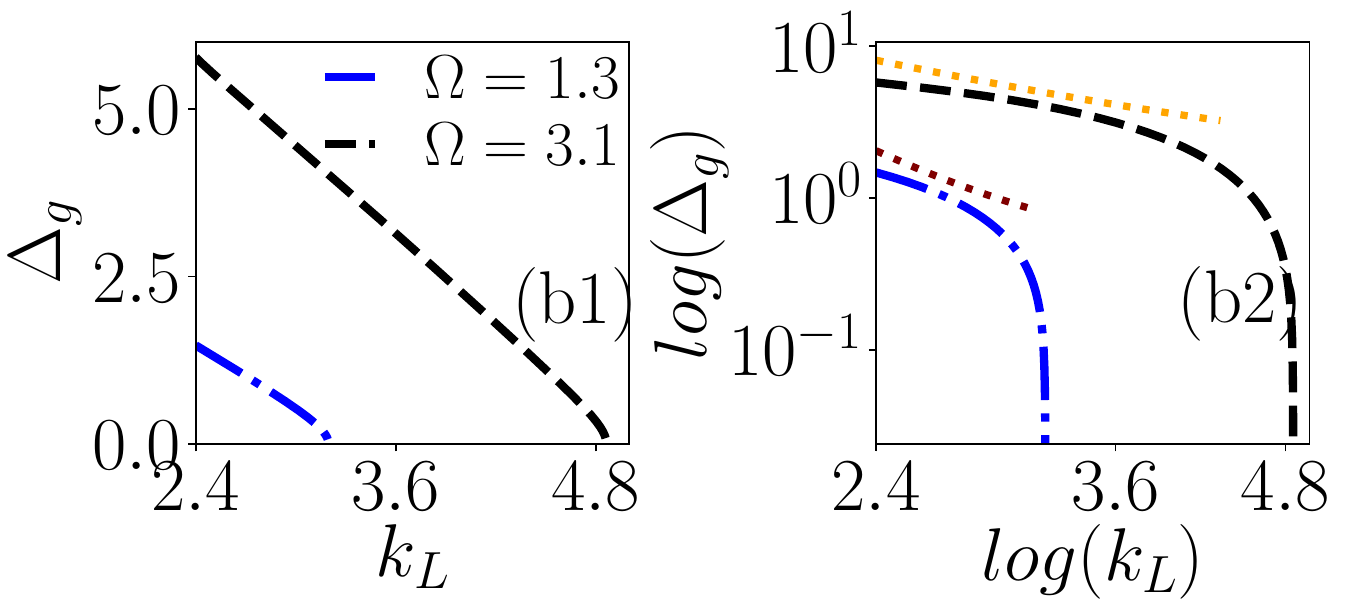}
\caption{Variation of band gap $\Delta_{g}$ between negative and positive branches of the spectrum. (a1) For the fixed value SO coupling strength upon variation of Rabi coupling strength, vice versa in (b1). In (a1), the dash-dotted green and the dashed red lines represent gap calculation for $k_{L} = 1.14$ and $k_{L} = 3.10$, respectively. In (b1), the dash-dotted blue and the dashed black lines represent gap calculation for $\Omega = 1.3$ and $\Omega = 3.1$, respectively.  The interaction parameter strength are $c_{0} = 0.5$, and $c_{2} = -0.1$. (a2) and (b2) are the logarithmic scale plots corresponding to (a1) and (b1), respectively. Further, in (a2) and (b2) the maroon (orange) dotted lines represent the fitting curve $\Delta_{g} = a \Omega^{b}$ with the parameters $ a= 0.9811, b = 0.9974$,  ($ a = 1.0825, b = 1.0258$), and  $a = 126.2992, b = -4.9568$, ($ a = 46.4438, b = -2.2146$), respectively. }
\label{fig63}
\end{centering}
\end{figure}

So far, we have discussed the nature of collective excitations, eigenvectors and further the dynamical behaviour of the condensate in different regions of the $k_L - \Omega$ plane, next we consider the effect of SO and Rabi coupling strengths on the band gap $\Delta_{g}$ between the positive and negative branches of the spectrum, where we define the band gap as $\Delta_{g}$. In Fig.~\ref{fig63}(a1), we show the band gap variation upon varying the Rabi coupling strength ($\Omega$) for a fixed $k_L$. In what follows we discuss it for two distinct SO strengths, for instance, $k_{L} = 1.14$ and $3.10$. The gap closes at a certain value of $\Omega$ for these SO couplings. For the point $k_{L} = 1.14$, $\Delta_{g} = 0$ for $\Omega = 0.107$ and similarly for $k_{L} = 3.10$, it closes at $\Omega = 1.218$ for the quasi-momentum values $q = 0.521$, $1.486$, respectively. Starting from the point $\Delta_{g} = 0$, upon increasing the $\Omega$, we observe that the gap becomes non-zero even for a relatively small increment in the Rabi coupling strength, and the magnitude of the band gap increases with an increase in $\Omega$. Also, we provide the logarithmic scale plot corresponding to Fig.~\ref{fig63}(a1) wherein we show the fitting curve $\Delta_{g} = a \Omega^{b}$ with the maroon and orange dotted line for $k_{L} = 1.14$ and $k_{L} = 3.10$, respectively. The coefficient for the fitting function is $a = 0.981$, $b = 0.997$ and $a = 1.083$, $b = 1.026$ for the SO coupling strength $k_{L} = 1.14$ and $k_{L} = 3.10$, respectively. Here, we wish to highlight that the band gap gets increased upon increasing $\Omega$ for fixed $k_L$. For higher $k_L$, the gap attains sudden maximum even for the lesser $\Omega$ while comparing small $k_L$, evident from the red dashed line crosses the green dashed-dotted line in Fig.~\ref{fig63}(a1). Moreover, for higher $k_L$ the gap closes at higher $\Omega$, compare starting points in Fig.~\ref{fig63}(a1).

Further, we discuss the variation of band gap upon varying the SO coupling for fixed Rabi coupling strength. We choose two different values of Rabi coupling strength $\Omega = 1.3, 3.1$. Since Rabi coupling strength is responsible for opening the gap among the branches of the collective excitation spectrum (as discussed above). As result of this the band gap $\Delta_{g}$ value is higher for a lesser value of $k_L$ at fixed $\Omega$ [see Fig.~\ref{fig63}(b1)]. Upon increasing $k_L$, the $\Delta_{g}$ starts decreasing, and further, it reaches zero at the certain quasi-momentum value $q_x = 1.5325, 2.3381$ with $k_{L} = 3.20, 4.86$ for the Rabi coupling strengths $\Omega = 1.3, 3.1$, respectively. We find that the band gap $\Delta_{g}$ decreases upon increasing the $k_L$ at fixed $\Omega$.  In Fig.~\ref{fig63}(b2), we show the fitting function $\Delta_{g} = c \Omega^{d}$ with the maroon and orange dotted line for $\Omega = 1.3$ and $\omega = 3.1$, respectively. The coefficient of fitting function is $c = 126.30, d = -4.96$ and $c = 46.44, d = - 2.22$ for the SO coupling strength $\Omega = 1.3$ and $k_{L} = 3.1$, respectively. %


\section{Summary and Conclusions }
\label{sec:6}
In this paper, we have investigated the collective excitation spectrum of SO coupled spin-1 ferromagnetic BECs with Rabi coupling in quasi-one dimension using Bogoliubov-de Gennes theory and numerical simulation. Firstly we have analyzed the single-particle spectrum for different sets of SO and Rabi coupling strengths. Upon increasing the strength of coupling parameters, we noticed separation among different negative ($\omega_{-}$) and positive ($\omega_{+}$) branches of the eigenspectrum, with $\omega_{-}$ having the lowest minima. However for $\Omega < k_{L}^{2}$, the negative branch of the spectrum shows double minima indicating the presence of a stripe phases. 

Based upon collective excitation calculation, we have obtained the phase diagram in the $k_L - \Omega$ plane which have been broadly divided into three regions: stable phase (region I) and unstable (region II, III). We have divided the unstable region II into two subparts according the nature of eigenspectrum, which is based upon the appearance of the the gapped and gapless (unstable-avoided crossing) mode. In region I, the excitation spectrum exhibits solely real positive eigenvalues with the presence of a phonon mode. The eigenvectors corresponding to these eigenvalues substantiate the existence of the phonon mode by exhibiting the density like mode in the wavenumber space. The absence of the negative and complex eigenfrequencies in the spectrum confirms dynamically and energetically stable condensate in this region.

As the coupling strengths increase, we find a transition from Region I to the unstable region II, attributed to the relation $\Omega < k_{L}^{2}$. In this region, complex eigenfrequencies emerge throughout the spectrum, accompanied by gapped and gapless modes between the low-lying $(\omega_{-})$ and first-excited $(\omega_{+})$ branches. Based on the presence of gapped and gapless modes region II is divided into two subregions, IIa and IIb,  respectively. The eigenspectrum in region IIa exhibits gapped behaviour across the entire region. Notably, the presence of complex eigenfrequencies is reflected in the corresponding eigenvectors, indicating a transition from spin-mode to density-mode nature of the eigenvectors as the eigenfrequency shifts from complex to real. This phenomenon is evident in the $\pm 1$ components of the eigenspectrum, while the zeroth component exclusively displays a density-like mode. The combined analysis of eigenvectors and dynamic behaviour reveals a mixed-spin-density mode.

Region IIb is separated from Region IIa by a line, the latter of which exhibits a multi-instability band across its entirety. Upon increasing the SO coupling for a point along the separation line, we observe gapless behaviour in the eigenspectrum for a particular range of quasi-momentum among the low-lying and first-excited branches of the spectrum. Similarly, decreasing the Rabi coupling strength for any point on the separation line will also yield gapless behaviour. Similar to the previous region, we also identify the mixed-spin-density eigenvector mode in this region. For the gapless region, the eigenvector of the low-lying and first-excited branches exhibits a spin-like mode for all its components. We also characterize the gapless mode as an unstable avoided crossing between the low-lying and first-excited spectrum. This instability manifests as the emergence of complex eigenfrequencies when the two branches come into contact.
The corresponding dynamical evolution also exhibits an unstable stripe state, which we have examined numerically. We observed that the $0$-th component is non-zero in region IIa but disappears in region IIb. This suggests that a non-zero $0$-th component tends to be less susceptible to dynamical instability, while a vanishing $0$-th component tends to be higher than the previous one.

In addition to these two regions, we have identified the region III which corresponds to the  $\Omega=0$ line. In this region, both the low-lying and the first excited states assume the value $\omega=0$ for $q_x=0$ with eigenvectors in phase with the low-lying state. This indicates the existence of a phonon mode. The low-lying and first-excited spectra also exhibit gapless, unstable-avoided crossing behaviour between them. Further analysis of the eigenvectors reveals not only the phonon character but also the presence of a spin-like mode in both states. Numerical simulations for this case reveal the presence of super stripe ground state, and their respective dynamics exhibit unstable dynamical-nonlinear patterns. Our observations suggest that this work would be a valuable resource for experimental researchers seeking to achieve stable and unstable SO coupled spinor ferromagnetic BECs. Here, we have provided a comprehensive numerical simulation of regions with diverse dynamical behaviours.

In this work we find the absence of the PW phase in the excitation spectrum. In the recent studies it has been shown that in presence of the Zeeman coupling the double degeneracy in the excitation spectrum gets lifted off hence the PW phase appears ~\cite{chen2022elementary, he2023stationary}. We would be interested to extend the present work in presence of the Zeeman field and explore the effect of the field on the nature of overall eigenspectrum and eigenvectors. One of the another interesting features that have emerged in the SO coupled BECs is the appearance of self-bound quantum droplet in the binary BECs~\cite{Cabrera2018,Petrov2015} in which the repulsive nature of the mean-field interaction is balanced by the beyond mean-field quantum fluctuation in one-dimension~\cite{Petrov2015, Gangwar2022}. It would be interesting to extend the present work of spin-1 considering the effect of quantum fluctuation on different phases of the collective excitation modes in the similar line as presented for spin-1/2 SO coupled BECs~\cite{Gangwar2023}.  
    

\acknowledgments %
We gratefully acknowledge our super-computing facility Param-Ishan (IITG), where all the simulation runs were performed. R.R. acknowledges the postdoctoral fellowship supported by Zhejiang Normal University, China, under Grants No. YS304023964. H. F. acknowledges the financial support from \textit{Programa de Capacitação Institucional} (PCI-CBPF) [No. 31.7202/2023-5]. P.M. acknowledges the financial support from MoE RUSA 2.0 (Bharathidasan University - Physical Sciences).

\onecolumngrid
\appendix
\counterwithin{figure}{section}

\section{Relevant terms of the BdG matrix of collective excitations}
\label{matrx:BdG}

In this appendix, we provide an explicit form of the matrix elements of the BdG matrix Eq. ~(\ref{bdgmatrix}) used in Section~\ref{sec:5}. The matrix elements of Eq. ~(\ref{bdgmatrix}) read as
\begin{subequations}
 \begin{align}
 H_{+} = & \frac{q_{x}^{2}}{2} + c_{0}(2\phi_{+1}^{2} + \phi_{0}^{2} +\phi_{-1}^{2}) + c_{2}(2\phi_{+1}^{2} + \phi_{0}^{2} -\phi_{-1}^{2}) \\
 H_{0} = & \frac{q_{x}^{2}}{2} + c_{0}(\phi_{+1}^{2} + 2\phi_{0}^{2} +\phi_{-1}^{2}) + c_{2}(\phi_{+1}^{2} +\phi_{-1}^{2})
\\
H_{-} = & \frac{q_{x}^{2}}{2} + c_{0}(\phi_{+1}^{2} + \phi_{0}^{2} +2\phi_{-1}^{2}) + c_{2}(2\phi_{-1}^{2} + \phi_{0}^{2} -\phi_{+1}^{2})
\\
\mu_{+}\phi_{+1} = & c_{0}(\phi_{+1}^{2} + \phi_{0}^{2} +\phi_{-1}^{2}) \phi_{+1} + c_{2}(\phi_{+1}^{2} + \phi_{0}^{2} -\phi_{-1}^{2}) \phi_{+1} + c_{2} \phi_{0}^{2} \phi_{-1}^{*} +  \frac{\Omega}{\sqrt{2}} \phi_{0}
\\
\mu_{0} \phi_{0}= & c_{0}(\phi_{+1}^{2} + \phi_{0}^{2} +\phi_{-1}^{2}) \phi_{0} + c_{2}(\phi_{+1}^{2} + \phi_{-1}^{2})\phi_{0} + 2 c_{2}\phi_{0}^{*}\phi_{+1}\phi_{-1} +  \frac{\Omega}{\sqrt{2}}(\phi_{+1}+\phi_{-1}) 
\\
\mu_{-} \phi_{-1} = & c_{0}(\phi_{+1}^{2} + \phi_{0}^{2} +\phi_{-1}^{2}) \phi_{-1} + c_{2}(\phi_{-1}^{2} + \phi_{0}^{2} -\phi_{+1}^{2}) \phi_{-1} + c_{2} \phi_{0}^{2} \phi_{+1}^{*} +  \frac{\Omega}{\sqrt{2}} \phi_{0}
\end{align}
\end{subequations}
\begin{align}
  \mathcal{L}_{12} & = C^{+}\phi_{+1}^{2}; \;\; \mathcal{L}_{13}= C^{+}\phi_{0}^{*}\phi_{+1}- \frac{k_{L}}{\sqrt{2}}i q_{x} + 2c_{2}\phi_{0}\phi_{-1}^{*}+ \frac{\Omega}{\sqrt{2}}; \;\; \mathcal{L}_{14}=C^{+}\phi_{0}\phi_{+1}; \;\; \mathcal{L}_{15}=C^{-}\phi_{-1}^{*}\phi_{+1} \notag \\
  \mathcal{L}_{16} & =C^{-}\phi_{-1}\phi_{+1} +c_{2} \phi_{0}^{2}; \;\; \mathcal{L}_{21}=-C^{+}\phi_{+1}^{*2}; \;\; \mathcal{L}_{23}=-C^{+}\phi_{0}^{*}\phi_{+1}^{*}; \;\; \mathcal{L}_{24} = -C^{+}\phi_{0}\phi_{+1}^{*} + \frac{k_{L}}{\sqrt{2}}i q_{x} -2c_{2}\phi_{0}^{*}\phi_{-1}-\frac{\Omega}{\sqrt{2}}; \notag \\
  \mathcal{L}_{25} & =-C^{-}\phi_{-1}^{*}\phi_{+1}{*} - c_{2}\phi_{0}^{*2}; 
  ;\;\ \mathcal{L}_{26}=-C^{-}\phi_{-1}\phi_{+1}^{*}; \;\; \mathcal{L}_{31} = C^{+}\phi_{+1}^{*}\phi_{0} + 2c_{2}\phi_{0}^{*}\phi_{-1} +\frac{k_{L}}{\sqrt{2}}i q_{x} +\frac{\Omega}{\sqrt{2}}; \;\; \mathcal{L}_{32}=C^{+}\phi_{+1}\phi_{0}; \notag \\ \mathcal{L}_{34} & =c_{0}\phi_{0}^{2} +2 c_{2}\phi_{+1}\phi_{-1}; \;\;  \mathcal{L}_{35} = C^{+}\phi_{-1}^{*}\phi_{0} + 2c_{2}\phi_{0}^{*}\phi_{+1} -\frac{k_{L}}{\sqrt{2}}i q_{x} +\frac{\Omega}{\sqrt{2}}; \;\; \mathcal{L}_{36}=C^{+}\phi_{-1}\phi_{0}; \;\; \mathcal{L}_{41}=-C^{+}\phi_{+1}^{*}\phi_{0}^{*} \notag \\
  \mathcal{L}_{42} & = -C^{+}\phi_{+1}\phi_{0}^{*}-\frac{k_{L}}{\sqrt{2}}i q_{x}- 2c_{2}\phi_{0}\phi_{-1}^{*}-\frac{\Omega}{\sqrt{2}}; \;\; \mathcal{L}_{43}=-c_{0}\phi_{0}^{*2}-2 c_{2}\phi_{+1}^{*}\phi_{-1}^{*}; \;\; \mathcal{L}_{45}=-C^{+}\phi_{-1}^{*}\phi_{0}^{*}; \;\; \notag
\end{align}
\begin{align}
  \mathcal{L}_{46} & = -C^{+}\phi_{-1}\phi_{0}^{*} + \frac{k_{L}}{\sqrt{2}} i q_{x}- 2c_{2}\phi_{0}\phi_{+1}^{*}- \frac{\Omega}{\sqrt{2}}; \;\; \mathcal{L}_{51}=C^{-}\phi_{+1}^{*}\phi_{-1}; \;\; \mathcal{L}_{52}=C^{-}\phi_{+1}\phi_{-1}+c_{2}\phi_{0}^{2}; \;\;  \notag \\
  \mathcal{L}_{53} & = C^{+}\phi_{0}^{*}\phi_{-1}+ \frac{k_{L}}{\sqrt{2}}i q_{x} + 2c_{2}\phi_{0}\phi_{+1}^{*} + \frac{\Omega}{\sqrt{2}}; \;\; \mathcal{L}_{54}=C^{+}\phi_{0}\phi_{-1}; \;\; \mathcal{L}_{56}=C^{+}\phi_{-1}^{2}; \;\; \mathcal{L}_{61}= -C^{-}\phi_{+1}^{*}\phi_{-1}^{*}-c_{2}\phi_{0}^{*2}; \notag \\
  \mathcal{L}_{62} & =- C^{-}\phi_{+1}\phi_{-1}^{*}; \;\; \mathcal{L}_{63}=-C^{+}\phi_{0}^{*}\phi_{-1}^{*}; \;\; \mathcal{L}_{64} = -C^{+}\phi_{0}\phi_{-1}^{*}-\frac{k_{L}}{\sqrt{2}}i q_{x}-2c_{2}\phi_{0}^{8}\phi_{+1}-\frac{\Omega}{\sqrt{2}}; \;\; \mathcal{L}_{65}= -C^{+} \phi_{-1}^{*2}. \notag
\end{align}
Also,
\begin{align}
 C^{+}\equiv c_{0} + c_{2}, \;C^{-}\equiv c_{0} - c_{2}   \notag
\end{align}

The coefficients for the BdG characteristic equation~(\ref{bdgex}) are given as follows:
\begin{align}
b= &-5 \Omega^{2}-4 c_{2}^{2}-(2 k_{L}^{2}+3 \Omega + c_{0})q_{x}^{2}-\frac{3}{4}q_{x}^{4}+ c_{2}(8 \Omega + q_{x}^{2}), \\
c= &4 \Omega^{4}+ \Omega \left[(2 \Omega ( k_{L}^{2} + 3 \Omega ) - ( k_{L}^{2} -5 \Omega ) c_{0}\right]q_{x}^{2} + 4 c_{2}^{3} q_{x}^{2}  + \frac{1}{2}\left[2 k_{L}^{4} + 9 \Omega^{2} + ( k_{L}^{2} + 6 \Omega )c_{0} \right] q_{x}^{4} + \frac{1}{2}(3 \Omega + c_{0}) q_{x}^{6}  \notag \\
&+ \frac{3}{16} q_{x}^{8} + 4 c_{2}^{2}\left[\Omega^{2}+ ( k_{L}^{2}-\Omega + c_{0} ) q_{x}^{2} \right] - \frac{1}{2} c_{2} \left[16 \Omega^{3} +2 \Omega ( 7 k_{L}^{2} + 5 \Omega + 8 c_{0} )q_{x}^{2} +( k_{L}^{2} + 6 \Omega +4 c_{0} )q_{x}^{4}+ q_{x}^{6}\right], \\
d=&-\frac{1}{64} q_{x}^{2}\left[ ( 4 k_{L}^{2} - 4 \Omega - q_{x}^{2} )( 2 \Omega + q_{x}^{2} ) + c_{2}( -8 k_{L}^{2} + 8 \Omega + 4 q_{x}^{2} )\right]  \bigg[-4 c_{0}\bigg(8 \Omega^{2} -2 (k_{L}^{2}-3 \Omega)q_{x}^{2} + q_{x}^{4} - 4 c_{2}(2 \Omega + q_{x}^{2})\bigg) \notag \\
&+(2 \Omega + q_{x}^{2})\bigg(-16 \Omega c_{2} + 16 c_{2}^{2}-q_{x}^{2}(-4 k_{L}^{2} + 4 \Omega + q_{x}^{2})\bigg)\bigg].
\end{align}

\twocolumngrid

\section{Phase diagram for different interaction strengths}
\label{app:phase2}
\begin{figure}[!ht]
\centering\includegraphics[width=0.49\linewidth]{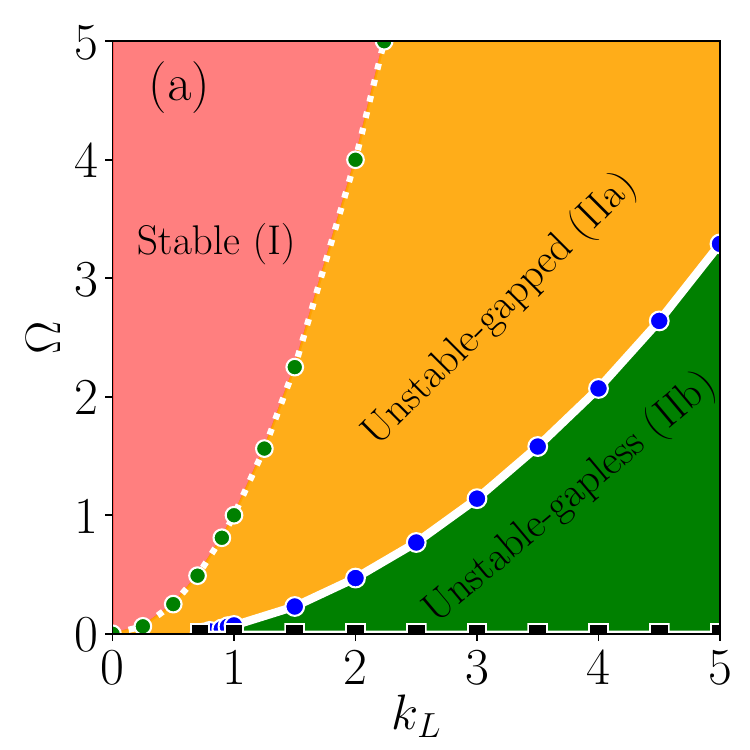}
\centering\includegraphics[width=0.49\linewidth]{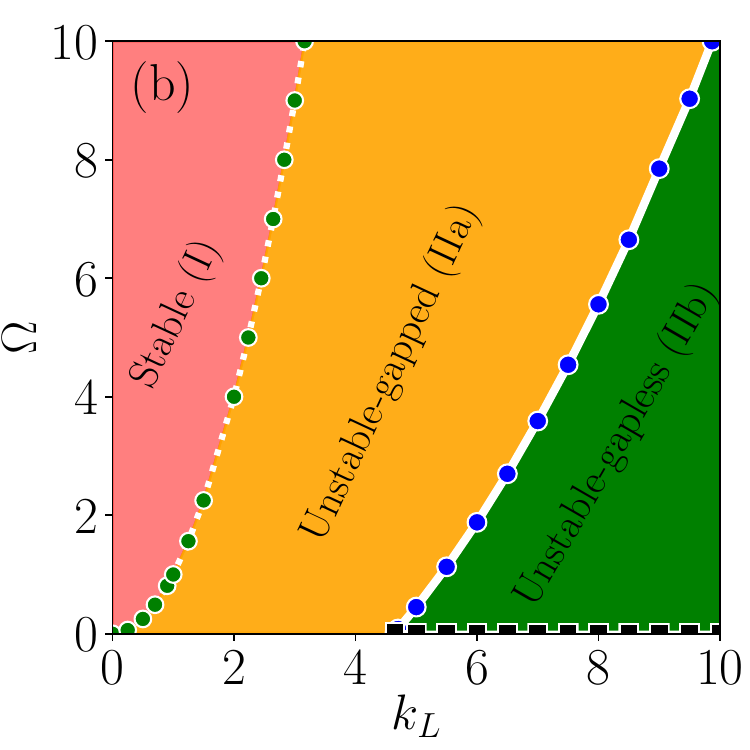}
\caption{Phase diagram in the $k_{L}$- $\Omega$ plane for different interaction parameters:  (a) For $(c_{0}=5.0$ and  $c_{2}= -0.1$, and  (b) $ (c_{0}=885.72$  and  $c_{2}= -4.09)$. Based on the different characteristics of the eigenspectrum and ground state, the phase diagram is divided into regions I, IIa, IIb, and III. Region I is stable, while regions IIa and IIb are unstable. The line $\Omega=k_L^2$ has been drawn, separating region I from II. Blue dots separate the regions IIa and IIb. The horizontal line with $\Omega\sim 0$ denotes the region III. The behaviour of all the regions remain same as those for  $c_{0} = 0.5$, and $c_{2} = -0.1$.}
\label{fig64}
\end{figure}

Here we present a detailed phase diagram in the $k_L - \Omega$ plane for the interaction strengths other than $c_{0} = 0.5$, and $c_{2} = -0.1$ as considered in this work. For  $c_{0} = 0.5$, and $c_{2} = -0.1$ we have mainly observed three different regions, on the basis of the collective excitation spectrum. 
Based on the above observations, here, we show such kind of phase diagrams for two other different sets of interaction strengths $c_{0} = 5.0$, $c_{2} = -0.1$, and $c_{0} = 885.72$, $c_{2} = -4.09$. 
In Fig.~\ref{fig64}(a) we show the stability phase diagram in the $k_{L}$-$\Omega$ plane for the interaction strength $c_{0} = 5$, $c_{2} = -0.1$. The stability phase diagram has been obtained by solving Eq.~(\ref{bdgmatrix}) and analyzing the collective excitation spectrum for these set of parameters. The phase diagram shows the presence of stable region (region I) and unstable region, region II, which is given using the relation $k_{L}^{2} = \Omega$. Region II has been divided into two subparts: region IIa, and region IIb. The transition line that separates region IIa and IIb is given using the relation $\Omega = 0.134k_{L}^{2} - 0.0668$. Where we observed that only the boundary can be changed otherwise three regions are found. %
The gapless behavior emerges after the point $k_{L}^c = 0.72$, $\Omega^c = 0.001$, which is also the origin of the line that separates regions IIa, and IIb. The horizontal line at which $\Omega = 0.0$, we obtain phonon mode in the low-lying as well as in the first-excited branch of the spectrum, along with the presence of instability band, which is different than previous cases. The cutoff value of SO coupling strength to obtain this behavior along the line is $k_{L} = 0.96$. %

Figure \ref{fig64}(b) illustrates a similar stability phase diagram $k_{L}$-$\Omega$ plane, featuring a very high interaction parameter strength for the interaction parameters $c_{0} = 885.72$ and $c_{2} = -4.09$. Both real and complex eigenfrequencies coexist with gapped and gapless modes for this set of interaction strengths. Varying the strengths of the spin-orbit and Rabi couplings allows us to observe the transition from the stable region (region I) to the unstable region (region II) in the phase diagram. The relationship $k_{L}^{2} = \Omega$ defines the boundary between region I and region II.
The transition line that separates regions IIa and IIb is given by the relation $\Omega = 0.1317 k_{L}^{2} - 2.8512$. The origin point of the line is $k_{L}^c = 4.65$, $\Omega^c = 0.02$. The horizontal line along which Rabi coupling strength is zero, we treat it as region III, where the cut-off value of SO coupling to achieve this behaviour is $k_{L} = 8.1$. 

\section{Spin-magnetization density vectors}
\label{app:mag-vec}

The spatial distribution (or orientation) of spin magnetization density vectors can be referred to as spin texture~\cite{Sadler2006}. Comparing the spin textures at the initial and finite times we have characterized the dynamical behaviour of the stable (region I) and unstable regions (IIa, IIb, and III) in the systemic way. The relevant quantities for the characterization, the spin magnetization density vectors, are defined as, %
\begin{subequations}%
\begin{align}
m_{x}  &= \mathrm{Re} \left[\sqrt{2} (\psi_{+1} + \psi_{-1})^{*} \psi_{0} \right],\label{eq:maga}\\
m_{y}  & = \mathrm{Im}\left[\sqrt{2} (\psi_{+1} - \psi_{-1})^{*}  \psi_{0} \right],\label{eq:magb}  \\
m_{z} & = \vert \psi_{+1}\vert^{2} - \vert \psi_{-1}\vert^{2}.
\label{eq:magc}
\end{align}
\end{subequations}%
\begin{figure*}[!ht]
\centering\includegraphics[width=0.425\linewidth]{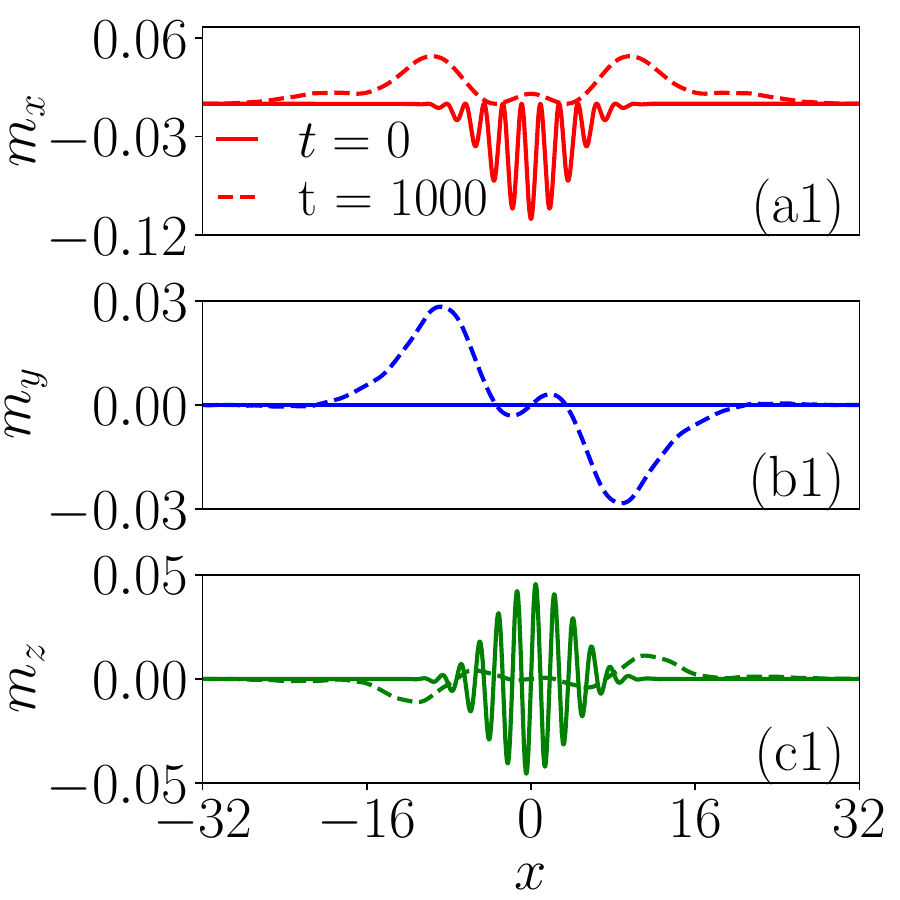}
\centering\includegraphics[width=0.425\linewidth]{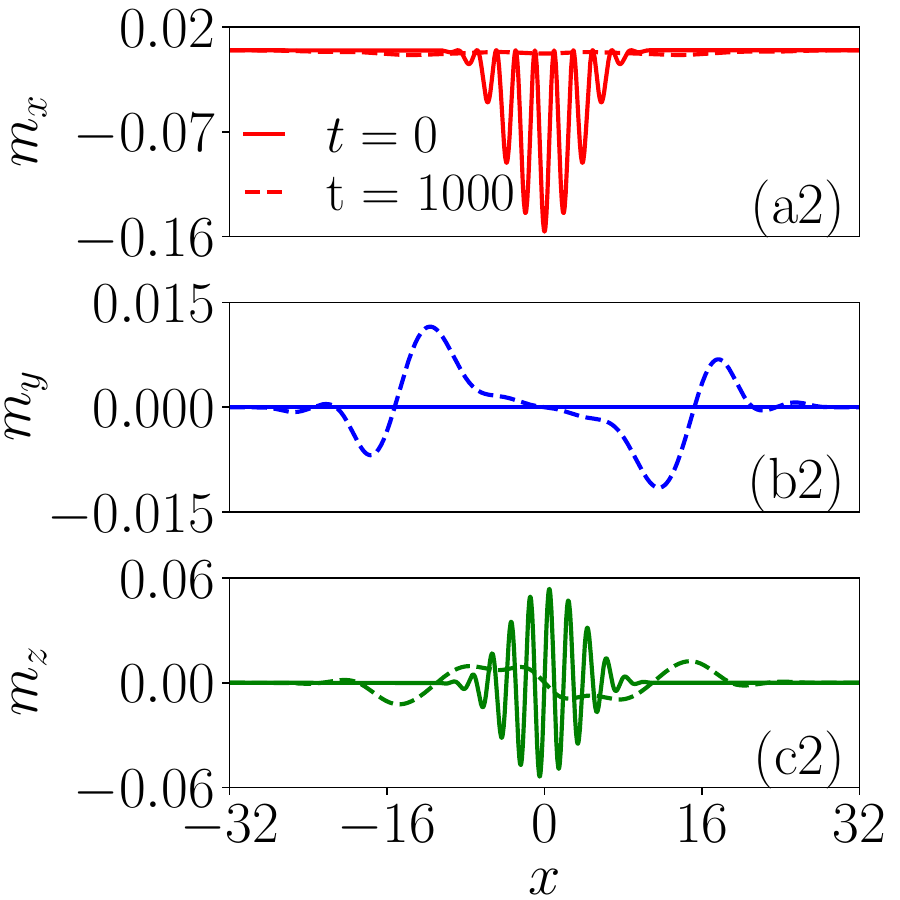}
\caption{Spin magnetization density components $(m_x,m_y,m_z)$ for the ground state at initial time $t=0$ (solid line),  and later time $t=1000$ (dashed line)  in the dynamically unstable region IIa in the $k_L - \Omega$ plane for different set of coupling parameters of the condensate with ferromagnetic interactions $c_{0} = 0.5$, and $c_{2} = -0.1$. (a1-c1) for $k_{L} = 2.0$, $\Omega = 2.0$ and (a2-c2) for $k_{L} = 2.35$, $\Omega = 4.0$. Owing to the dynamical instability, the spin texture deviates from its initial state during the time evolution.}
\label{fig65}
\end{figure*}%

\begin{figure*}[!ht]
\centering\includegraphics[width=0.425\linewidth]{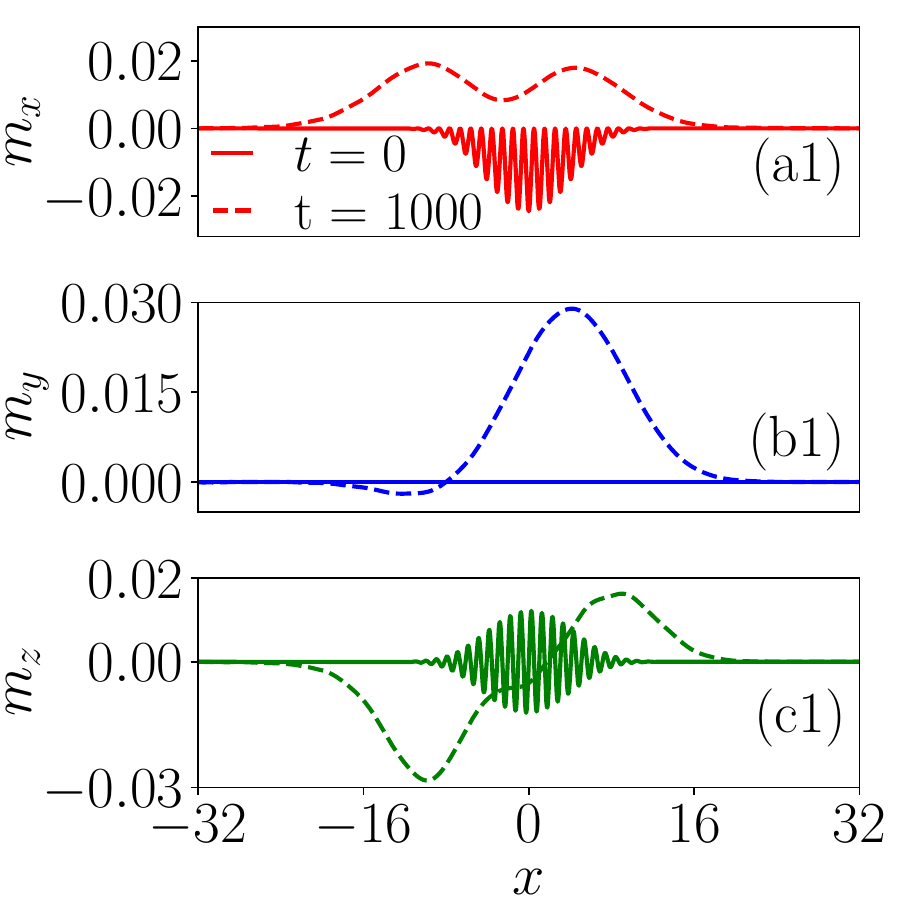}
\centering\includegraphics[width=0.425\linewidth]{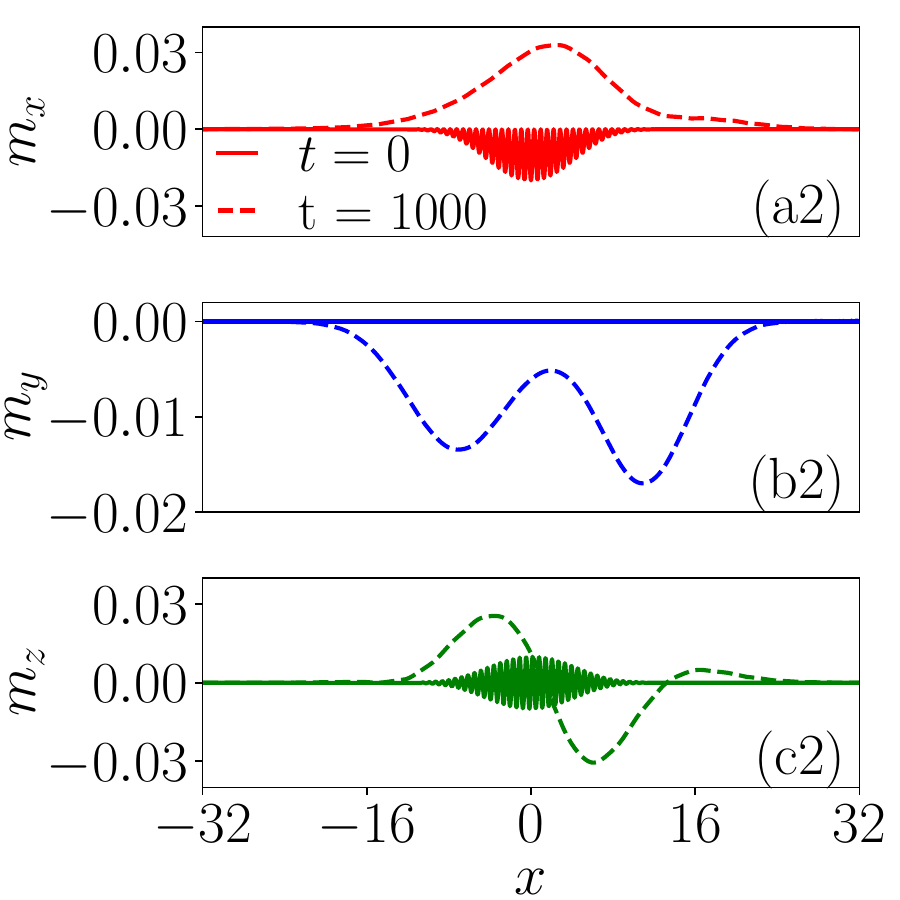}
\caption{Spin magnetization density components $(m_x,m_y,m_z)$ spatial profile for the ground state(solid line) at initial time $t=0$ and later time $t=1000$ (dashed line) in the dynamically unstable region IIb in the $k_L - \Omega$ plane for different set of coupling parameters of the condensate with ferromagnetic interactions $c_{0} = 0.5$, and $c_{2} = -0.1$. (a1-c1) for $k_{L} = 3.1$ and $\Omega = 1.17$ and (a2-c2) for $k_{L} = 5.0$ and $\Omega = 2.5$. Owing to the dynamical instability, spin-texture shows deviation from the initial behavior during time evolution.}
\label{fig66}
\end{figure*}%
 In Figs.~\ref{fig65} we show the components of the magnetization vector computed using Eqs.~\ref{eq:maga}-~\ref{eq:magc} for two sets of coupling parameters of the dynamically unstable region IIa of the $k_L - \Omega$ plane as depicted in the Fig.~\ref{fig8}. We show the magnetization profile at two instant of time t=0 (for the ground state) and at later instant t=1000 for $k_{L} = 2.0$, $\Omega = 2.0$ (see panel (a1-c1)) and same for other parameter $k_{L} = 2.35$, $\Omega = 4.0$ (see panel (a2-c2)). For both cases, all the magnetic components profile exhibits significant changes at a later time ($t=1000$) compared to those for the ground state confirming the dynamical instability of the condensate for these sets of the coupling parameters in the region IIa. This trend continues for other set of points in the region IIa. Similar trend of the magnetization vector component profile has been observed for the two sets of the coupling parameters, $k_L=3.1$ and $\Omega=1.17$ and $k_L=5.0$ and $\Omega=2.5$ of region IIb as depicted in the  Figs.~\ref{fig66} (a1-c1) and (a2-c2) respectively. 
 
\bibliography{cexandins.bib}

\begin{thebibliography}{67}%
\makeatletter
\providecommand \@ifxundefined [1]{%
 \@ifx{#1\undefined}
}%
\providecommand \@ifnum [1]{%
 \ifnum #1\expandafter \@firstoftwo
 \else \expandafter \@secondoftwo
 \fi
}%
\providecommand \@ifx [1]{%
 \ifx #1\expandafter \@firstoftwo
 \else \expandafter \@secondoftwo
 \fi
}%
\providecommand \natexlab [1]{#1}%
\providecommand \enquote  [1]{``#1''}%
\providecommand \bibnamefont  [1]{#1}%
\providecommand \bibfnamefont [1]{#1}%
\providecommand \citenamefont [1]{#1}%
\providecommand \href@noop [0]{\@secondoftwo}%
\providecommand \href [0]{\begingroup \@sanitize@url \@href}%
\providecommand \@href[1]{\@@startlink{#1}\@@href}%
\providecommand \@@href[1]{\endgroup#1\@@endlink}%
\providecommand \@sanitize@url [0]{\catcode `\\12\catcode `\$12\catcode
  `\&12\catcode `\#12\catcode `\^12\catcode `\_12\catcode `\%12\relax}%
\providecommand \@@startlink[1]{}%
\providecommand \@@endlink[0]{}%
\providecommand \url  [0]{\begingroup\@sanitize@url \@url }%
\providecommand \@url [1]{\endgroup\@href {#1}{\urlprefix }}%
\providecommand \urlprefix  [0]{URL }%
\providecommand \Eprint [0]{\href }%
\providecommand \doibase [0]{http://dx.doi.org/}%
\providecommand \selectlanguage [0]{\@gobble}%
\providecommand \bibinfo  [0]{\@secondoftwo}%
\providecommand \bibfield  [0]{\@secondoftwo}%
\providecommand \translation [1]{[#1]}%
\providecommand \BibitemOpen [0]{}%
\providecommand \bibitemStop [0]{}%
\providecommand \bibitemNoStop [0]{.\EOS\space}%
\providecommand \EOS [0]{\spacefactor3000\relax}%
\providecommand \BibitemShut  [1]{\csname bibitem#1\endcsname}%
\let\auto@bib@innerbib\@empty
\bibitem [{\citenamefont {Zhang}\ \emph {et~al.}(2016)\citenamefont {Zhang},
  \citenamefont {Mossman}, \citenamefont {Busch}, \citenamefont {Engels},\ and\
  \citenamefont {Zhang}}]{zhang2016properties}%
  \BibitemOpen
  \bibfield  {author} {\bibinfo {author} {\bibfnamefont {Y.}~\bibnamefont
  {Zhang}}, \bibinfo {author} {\bibfnamefont {M.~E.}\ \bibnamefont {Mossman}},
  \bibinfo {author} {\bibfnamefont {T.}~\bibnamefont {Busch}}, \bibinfo
  {author} {\bibfnamefont {P.}~\bibnamefont {Engels}}, \ and\ \bibinfo {author}
  {\bibfnamefont {C.}~\bibnamefont {Zhang}},\ }\href@noop {} {\bibfield
  {journal} {\bibinfo  {journal} {Front. Phys.}\ }\textbf {\bibinfo {volume}
  {11}},\ \bibinfo {pages} {1} (\bibinfo {year} {2016})}\BibitemShut {NoStop}%
\bibitem [{\citenamefont {Jacob}\ \emph {et~al.}(2012)\citenamefont {Jacob},
  \citenamefont {Shao}, \citenamefont {Corre}, \citenamefont {Zibold},
  \citenamefont {De~Sarlo}, \citenamefont {Mimoun}, \citenamefont {Dalibard},\
  and\ \citenamefont {Gerbier}}]{Jacob2012}%
  \BibitemOpen
  \bibfield  {author} {\bibinfo {author} {\bibfnamefont {D.}~\bibnamefont
  {Jacob}}, \bibinfo {author} {\bibfnamefont {L.}~\bibnamefont {Shao}},
  \bibinfo {author} {\bibfnamefont {V.}~\bibnamefont {Corre}}, \bibinfo
  {author} {\bibfnamefont {T.}~\bibnamefont {Zibold}}, \bibinfo {author}
  {\bibfnamefont {L.}~\bibnamefont {De~Sarlo}}, \bibinfo {author}
  {\bibfnamefont {E.}~\bibnamefont {Mimoun}}, \bibinfo {author} {\bibfnamefont
  {J.}~\bibnamefont {Dalibard}}, \ and\ \bibinfo {author} {\bibfnamefont
  {F.}~\bibnamefont {Gerbier}},\ }\href {\doibase 10.1103/PhysRevA.86.061601}
  {\bibfield  {journal} {\bibinfo  {journal} {Phys. Rev. A}\ }\textbf {\bibinfo
  {volume} {86}},\ \bibinfo {pages} {061601} (\bibinfo {year}
  {2012})}\BibitemShut {NoStop}%
\bibitem [{\citenamefont {Campbell}\ \emph {et~al.}(2016)\citenamefont
  {Campbell}, \citenamefont {Price}, \citenamefont {Putra}, \citenamefont
  {Vald{\'e}s-Curiel}, \citenamefont {Trypogeorgos},\ and\ \citenamefont
  {Spielman}}]{Campbell2016}%
  \BibitemOpen
  \bibfield  {author} {\bibinfo {author} {\bibfnamefont {D.}~\bibnamefont
  {Campbell}}, \bibinfo {author} {\bibfnamefont {R.}~\bibnamefont {Price}},
  \bibinfo {author} {\bibfnamefont {A.}~\bibnamefont {Putra}}, \bibinfo
  {author} {\bibfnamefont {A.}~\bibnamefont {Vald{\'e}s-Curiel}}, \bibinfo
  {author} {\bibfnamefont {D.}~\bibnamefont {Trypogeorgos}}, \ and\ \bibinfo
  {author} {\bibfnamefont {I.}~\bibnamefont {Spielman}},\ }\href@noop {}
  {\bibfield  {journal} {\bibinfo  {journal} {Nat. commun.}\ }\textbf {\bibinfo
  {volume} {7}},\ \bibinfo {pages} {1} (\bibinfo {year} {2016})}\BibitemShut
  {NoStop}%
\bibitem [{\citenamefont {Kasamatsu}\ \emph {et~al.}(2005)\citenamefont
  {Kasamatsu}, \citenamefont {Tsubota},\ and\ \citenamefont
  {Ueda}}]{Kasamatsu2005}%
  \BibitemOpen
  \bibfield  {author} {\bibinfo {author} {\bibfnamefont {K.}~\bibnamefont
  {Kasamatsu}}, \bibinfo {author} {\bibfnamefont {M.}~\bibnamefont {Tsubota}},
  \ and\ \bibinfo {author} {\bibfnamefont {M.}~\bibnamefont {Ueda}},\
  }\href@noop {} {\bibfield  {journal} {\bibinfo  {journal} {Int. J. Mod. Phys.
  B}\ }\textbf {\bibinfo {volume} {19}},\ \bibinfo {pages} {1835} (\bibinfo
  {year} {2005})}\BibitemShut {NoStop}%
\bibitem [{\citenamefont {Fujimoto}\ and\ \citenamefont
  {Tsubota}(2012)}]{Tsubota2012}%
  \BibitemOpen
  \bibfield  {author} {\bibinfo {author} {\bibfnamefont {K.}~\bibnamefont
  {Fujimoto}}\ and\ \bibinfo {author} {\bibfnamefont {M.}~\bibnamefont
  {Tsubota}},\ }\href {\doibase 10.1103/PhysRevA.85.033642} {\bibfield
  {journal} {\bibinfo  {journal} {Phys. Rev. A}\ }\textbf {\bibinfo {volume}
  {85}},\ \bibinfo {pages} {033642} (\bibinfo {year} {2012})}\BibitemShut
  {NoStop}%
\bibitem [{\citenamefont {Yu}(2016)}]{Yu2016}%
  \BibitemOpen
  \bibfield  {author} {\bibinfo {author} {\bibfnamefont {Z.-Q.}\ \bibnamefont
  {Yu}},\ }\href {\doibase 10.1103/PhysRevA.93.033648} {\bibfield  {journal}
  {\bibinfo  {journal} {Phys. Rev. A}\ }\textbf {\bibinfo {volume} {93}},\
  \bibinfo {pages} {033648} (\bibinfo {year} {2016})}\BibitemShut {NoStop}%
\bibitem [{\citenamefont {Martone}\ \emph {et~al.}(2016)\citenamefont
  {Martone}, \citenamefont {Pepe}, \citenamefont {Facchi}, \citenamefont
  {Pascazio},\ and\ \citenamefont {Stringari}}]{Martone2016}%
  \BibitemOpen
  \bibfield  {author} {\bibinfo {author} {\bibfnamefont {G.~I.}\ \bibnamefont
  {Martone}}, \bibinfo {author} {\bibfnamefont {F.~V.}\ \bibnamefont {Pepe}},
  \bibinfo {author} {\bibfnamefont {P.}~\bibnamefont {Facchi}}, \bibinfo
  {author} {\bibfnamefont {S.}~\bibnamefont {Pascazio}}, \ and\ \bibinfo
  {author} {\bibfnamefont {S.}~\bibnamefont {Stringari}},\ }\href {\doibase
  10.1103/PhysRevLett.117.125301} {\bibfield  {journal} {\bibinfo  {journal}
  {Phys. Rev. Lett.}\ }\textbf {\bibinfo {volume} {117}},\ \bibinfo {pages}
  {125301} (\bibinfo {year} {2016})}\BibitemShut {NoStop}%
\bibitem [{\citenamefont {Li}\ \emph {et~al.}(2017)\citenamefont {Li},
  \citenamefont {Chen}, \citenamefont {Peng}, \citenamefont {Li},\ and\
  \citenamefont {Bai}}]{Li2017}%
  \BibitemOpen
  \bibfield  {author} {\bibinfo {author} {\bibfnamefont {G.-Q.}\ \bibnamefont
  {Li}}, \bibinfo {author} {\bibfnamefont {G.-D.}\ \bibnamefont {Chen}},
  \bibinfo {author} {\bibfnamefont {P.}~\bibnamefont {Peng}}, \bibinfo {author}
  {\bibfnamefont {Z.}~\bibnamefont {Li}}, \ and\ \bibinfo {author}
  {\bibfnamefont {X.-D.}\ \bibnamefont {Bai}},\ }\href@noop {} {\bibfield
  {journal} {\bibinfo  {journal} {J. Phys. B: At. Mol. Opt. Phys.}\ }\textbf
  {\bibinfo {volume} {50}},\ \bibinfo {pages} {235302} (\bibinfo {year}
  {2017})}\BibitemShut {NoStop}%
\bibitem [{\citenamefont {Malomed}(2018)}]{Malomed2018}%
  \BibitemOpen
  \bibfield  {author} {\bibinfo {author} {\bibfnamefont {B.~A.}\ \bibnamefont
  {Malomed}},\ }\href@noop {} {\bibfield  {journal} {\bibinfo  {journal} {EPL}\
  }\textbf {\bibinfo {volume} {122}},\ \bibinfo {pages} {36001} (\bibinfo
  {year} {2018})}\BibitemShut {NoStop}%
\bibitem [{\citenamefont {Peng}\ \emph {et~al.}(2019)\citenamefont {Peng},
  \citenamefont {Li}, \citenamefont {Zhao}, \citenamefont {Yang},\ and\
  \citenamefont {Yang}}]{Peng2019}%
  \BibitemOpen
  \bibfield  {author} {\bibinfo {author} {\bibfnamefont {P.}~\bibnamefont
  {Peng}}, \bibinfo {author} {\bibfnamefont {G.-Q.}\ \bibnamefont {Li}},
  \bibinfo {author} {\bibfnamefont {L.-C.}\ \bibnamefont {Zhao}}, \bibinfo
  {author} {\bibfnamefont {W.-L.}\ \bibnamefont {Yang}}, \ and\ \bibinfo
  {author} {\bibfnamefont {Z.-Y.}\ \bibnamefont {Yang}},\ }\href@noop {}
  {\bibfield  {journal} {\bibinfo  {journal} {Phys. Lett. A}\ }\textbf
  {\bibinfo {volume} {383}},\ \bibinfo {pages} {2883} (\bibinfo {year}
  {2019})}\BibitemShut {NoStop}%
\bibitem [{\citenamefont {Adhikari}(2021{\natexlab{a}})}]{Adhikari2021}%
  \BibitemOpen
  \bibfield  {author} {\bibinfo {author} {\bibfnamefont {S.}~\bibnamefont
  {Adhikari}},\ }\href@noop {} {\bibfield  {journal} {\bibinfo  {journal} {J.
  Phys.: Condens. Matter}\ }\textbf {\bibinfo {volume} {33}},\ \bibinfo {pages}
  {265402} (\bibinfo {year} {2021}{\natexlab{a}})}\BibitemShut {NoStop}%
\bibitem [{\citenamefont {Gautam}\ and\ \citenamefont
  {Adhikari}(2014)}]{Gautam2014}%
  \BibitemOpen
  \bibfield  {author} {\bibinfo {author} {\bibfnamefont {S.}~\bibnamefont
  {Gautam}}\ and\ \bibinfo {author} {\bibfnamefont {S.~K.}\ \bibnamefont
  {Adhikari}},\ }\href {\doibase 10.1103/PhysRevA.90.043619} {\bibfield
  {journal} {\bibinfo  {journal} {Phys. Rev. A}\ }\textbf {\bibinfo {volume}
  {90}},\ \bibinfo {pages} {043619} (\bibinfo {year} {2014})}\BibitemShut
  {NoStop}%
\bibitem [{\citenamefont {Adhikari}(2019)}]{Adhikari2019}%
  \BibitemOpen
  \bibfield  {author} {\bibinfo {author} {\bibfnamefont {S.~K.}\ \bibnamefont
  {Adhikari}},\ }\href {\doibase 10.1103/PhysRevA.100.063618} {\bibfield
  {journal} {\bibinfo  {journal} {Phys. Rev. A}\ }\textbf {\bibinfo {volume}
  {100}},\ \bibinfo {pages} {063618} (\bibinfo {year} {2019})}\BibitemShut
  {NoStop}%
\bibitem [{\citenamefont {Gautam}\ and\ \citenamefont
  {Adhikari}(2015)}]{Gautam2015}%
  \BibitemOpen
  \bibfield  {author} {\bibinfo {author} {\bibfnamefont {S.}~\bibnamefont
  {Gautam}}\ and\ \bibinfo {author} {\bibfnamefont {S.~K.}\ \bibnamefont
  {Adhikari}},\ }\href@noop {} {\bibfield  {journal} {\bibinfo  {journal}
  {Laser Phys. Lett.}\ }\textbf {\bibinfo {volume} {12}},\ \bibinfo {pages}
  {045501} (\bibinfo {year} {2015})}\BibitemShut {NoStop}%
\bibitem [{\citenamefont {Adhikari}(2020)}]{Adhikari2020}%
  \BibitemOpen
  \bibfield  {author} {\bibinfo {author} {\bibfnamefont {S.~K.}\ \bibnamefont
  {Adhikari}},\ }\href@noop {} {\bibfield  {journal} {\bibinfo  {journal}
  {Physica E Low Dimens. Syst. Nanostruct.}\ }\textbf {\bibinfo {volume}
  {118}},\ \bibinfo {pages} {113892} (\bibinfo {year} {2020})}\BibitemShut
  {NoStop}%
\bibitem [{\citenamefont {Gautam}\ and\ \citenamefont
  {Adhikari}(2021)}]{Gautam2021}%
  \BibitemOpen
  \bibfield  {author} {\bibinfo {author} {\bibfnamefont {S.}~\bibnamefont
  {Gautam}}\ and\ \bibinfo {author} {\bibfnamefont {S.}~\bibnamefont
  {Adhikari}},\ }\href@noop {} {\bibfield  {journal} {\bibinfo  {journal}
  {Braz. J. Phys.}\ }\textbf {\bibinfo {volume} {51}},\ \bibinfo {pages} {298}
  (\bibinfo {year} {2021})}\BibitemShut {NoStop}%
\bibitem [{\citenamefont {Mardonov}\ \emph {et~al.}(2015)\citenamefont
  {Mardonov}, \citenamefont {Modugno},\ and\ \citenamefont
  {Sherman}}]{Mardonov2015}%
  \BibitemOpen
  \bibfield  {author} {\bibinfo {author} {\bibfnamefont {S.}~\bibnamefont
  {Mardonov}}, \bibinfo {author} {\bibfnamefont {M.}~\bibnamefont {Modugno}}, \
  and\ \bibinfo {author} {\bibfnamefont {E.~Y.}\ \bibnamefont {Sherman}},\
  }\href {\doibase 10.1103/PhysRevLett.115.180402} {\bibfield  {journal}
  {\bibinfo  {journal} {Phys. Rev. Lett.}\ }\textbf {\bibinfo {volume} {115}},\
  \bibinfo {pages} {180402} (\bibinfo {year} {2015})}\BibitemShut {NoStop}%
\bibitem [{\citenamefont {Mardonov}\ \emph {et~al.}(2018)\citenamefont
  {Mardonov}, \citenamefont {Konotop}, \citenamefont {Malomed}, \citenamefont
  {Modugno},\ and\ \citenamefont {Sherman}}]{Mardonov2018}%
  \BibitemOpen
  \bibfield  {author} {\bibinfo {author} {\bibfnamefont {S.}~\bibnamefont
  {Mardonov}}, \bibinfo {author} {\bibfnamefont {V.~V.}\ \bibnamefont
  {Konotop}}, \bibinfo {author} {\bibfnamefont {B.~A.}\ \bibnamefont
  {Malomed}}, \bibinfo {author} {\bibfnamefont {M.}~\bibnamefont {Modugno}}, \
  and\ \bibinfo {author} {\bibfnamefont {E.~Y.}\ \bibnamefont {Sherman}},\
  }\href {\doibase 10.1103/PhysRevA.98.023604} {\bibfield  {journal} {\bibinfo
  {journal} {Phys. Rev. A}\ }\textbf {\bibinfo {volume} {98}},\ \bibinfo
  {pages} {023604} (\bibinfo {year} {2018})}\BibitemShut {NoStop}%
\bibitem [{\citenamefont {Mardonov}\ \emph {et~al.}(2019)\citenamefont
  {Mardonov}, \citenamefont {Modugno}, \citenamefont {Sherman},\ and\
  \citenamefont {Malomed}}]{Mardonov2019}%
  \BibitemOpen
  \bibfield  {author} {\bibinfo {author} {\bibfnamefont {S.}~\bibnamefont
  {Mardonov}}, \bibinfo {author} {\bibfnamefont {M.}~\bibnamefont {Modugno}},
  \bibinfo {author} {\bibfnamefont {E.~Y.}\ \bibnamefont {Sherman}}, \ and\
  \bibinfo {author} {\bibfnamefont {B.~A.}\ \bibnamefont {Malomed}},\ }\href
  {\doibase 10.1103/PhysRevA.99.013611} {\bibfield  {journal} {\bibinfo
  {journal} {Phys. Rev. A}\ }\textbf {\bibinfo {volume} {99}},\ \bibinfo
  {pages} {013611} (\bibinfo {year} {2019})}\BibitemShut {NoStop}%
\bibitem [{\citenamefont {Mithun}\ and\ \citenamefont
  {Kasamatsu}(2019)}]{Mithun2019}%
  \BibitemOpen
  \bibfield  {author} {\bibinfo {author} {\bibfnamefont {T.}~\bibnamefont
  {Mithun}}\ and\ \bibinfo {author} {\bibfnamefont {K.}~\bibnamefont
  {Kasamatsu}},\ }\href@noop {} {\bibfield  {journal} {\bibinfo  {journal} {J.
  Phys. B: At. Mol. Opt. Phys.}\ }\textbf {\bibinfo {volume} {52}},\ \bibinfo
  {pages} {045301} (\bibinfo {year} {2019})}\BibitemShut {NoStop}%
\bibitem [{\citenamefont {Cabedo}\ \emph {et~al.}(2021)\citenamefont {Cabedo},
  \citenamefont {Claramunt},\ and\ \citenamefont {Celi}}]{Cabedo2021}%
  \BibitemOpen
  \bibfield  {author} {\bibinfo {author} {\bibfnamefont {J.}~\bibnamefont
  {Cabedo}}, \bibinfo {author} {\bibfnamefont {J.}~\bibnamefont {Claramunt}}, \
  and\ \bibinfo {author} {\bibfnamefont {A.}~\bibnamefont {Celi}},\ }\href
  {\doibase 10.1103/PhysRevA.104.L031305} {\bibfield  {journal} {\bibinfo
  {journal} {Phys. Rev. A}\ }\textbf {\bibinfo {volume} {104}},\ \bibinfo
  {pages} {L031305} (\bibinfo {year} {2021})}\BibitemShut {NoStop}%
\bibitem [{\citenamefont {Zhu}\ \emph {et~al.}(2020)\citenamefont {Zhu},
  \citenamefont {Pan},\ and\ \citenamefont {An}}]{Zhu2020}%
  \BibitemOpen
  \bibfield  {author} {\bibinfo {author} {\bibfnamefont {Q.-L.}\ \bibnamefont
  {Zhu}}, \bibinfo {author} {\bibfnamefont {L.}~\bibnamefont {Pan}}, \ and\
  \bibinfo {author} {\bibfnamefont {J.}~\bibnamefont {An}},\ }\href {\doibase
  10.1103/PhysRevA.102.053320} {\bibfield  {journal} {\bibinfo  {journal}
  {Phys. Rev. A}\ }\textbf {\bibinfo {volume} {102}},\ \bibinfo {pages}
  {053320} (\bibinfo {year} {2020})}\BibitemShut {NoStop}%
\bibitem [{\citenamefont {Lin}\ \emph {et~al.}(2011)\citenamefont {Lin},
  \citenamefont {Jim{\'e}nez-Garc{\'\i}a},\ and\ \citenamefont
  {Spielman}}]{Lin2011}%
  \BibitemOpen
  \bibfield  {author} {\bibinfo {author} {\bibfnamefont {Y.-J.}\ \bibnamefont
  {Lin}}, \bibinfo {author} {\bibfnamefont {K.}~\bibnamefont
  {Jim{\'e}nez-Garc{\'\i}a}}, \ and\ \bibinfo {author} {\bibfnamefont {I.~B.}\
  \bibnamefont {Spielman}},\ }\href@noop {} {\bibfield  {journal} {\bibinfo
  {journal} {Nature (London)}\ }\textbf {\bibinfo {volume} {471}},\ \bibinfo
  {pages} {83} (\bibinfo {year} {2011})}\BibitemShut {NoStop}%
\bibitem [{\citenamefont {Ravisankar}\ \emph
  {et~al.}(2021{\natexlab{a}})\citenamefont {Ravisankar}, \citenamefont
  {Sriraman}, \citenamefont {Kumar}, \citenamefont {Muruganandam},\ and\
  \citenamefont {Mishra}}]{Ravisankar2021influence}%
  \BibitemOpen
  \bibfield  {author} {\bibinfo {author} {\bibfnamefont {R.}~\bibnamefont
  {Ravisankar}}, \bibinfo {author} {\bibfnamefont {T.}~\bibnamefont
  {Sriraman}}, \bibinfo {author} {\bibfnamefont {R.~K.}\ \bibnamefont {Kumar}},
  \bibinfo {author} {\bibfnamefont {P.}~\bibnamefont {Muruganandam}}, \ and\
  \bibinfo {author} {\bibfnamefont {P.~K.}\ \bibnamefont {Mishra}},\
  }\href@noop {} {\bibfield  {journal} {\bibinfo  {journal} {J. Phys. B: At.
  Mol. Opt. Phys.}\ }\textbf {\bibinfo {volume} {54}},\ \bibinfo {pages}
  {225301} (\bibinfo {year} {2021}{\natexlab{a}})}\BibitemShut {NoStop}%
\bibitem [{\citenamefont {Goldstein}\ and\ \citenamefont
  {Meystre}(1997)}]{Goldstein1997}%
  \BibitemOpen
  \bibfield  {author} {\bibinfo {author} {\bibfnamefont {E.~V.}\ \bibnamefont
  {Goldstein}}\ and\ \bibinfo {author} {\bibfnamefont {P.}~\bibnamefont
  {Meystre}},\ }\href@noop {} {\bibfield  {journal} {\bibinfo  {journal} {Phys.
  Rev. A}\ }\textbf {\bibinfo {volume} {55}},\ \bibinfo {pages} {2935}
  (\bibinfo {year} {1997})}\BibitemShut {NoStop}%
\bibitem [{\citenamefont {Zhang}\ \emph {et~al.}(2005)\citenamefont {Zhang},
  \citenamefont {Zhou}, \citenamefont {Chang}, \citenamefont {Chapman},\ and\
  \citenamefont {You}}]{Zhang2005}%
  \BibitemOpen
  \bibfield  {author} {\bibinfo {author} {\bibfnamefont {W.}~\bibnamefont
  {Zhang}}, \bibinfo {author} {\bibfnamefont {D.~L.}\ \bibnamefont {Zhou}},
  \bibinfo {author} {\bibfnamefont {M.-S.}\ \bibnamefont {Chang}}, \bibinfo
  {author} {\bibfnamefont {M.~S.}\ \bibnamefont {Chapman}}, \ and\ \bibinfo
  {author} {\bibfnamefont {L.}~\bibnamefont {You}},\ }\href {\doibase
  10.1103/PhysRevLett.95.180403} {\bibfield  {journal} {\bibinfo  {journal}
  {Phys. Rev. Lett.}\ }\textbf {\bibinfo {volume} {95}},\ \bibinfo {pages}
  {180403} (\bibinfo {year} {2005})}\BibitemShut {NoStop}%
\bibitem [{\citenamefont {Pethick}\ and\ \citenamefont
  {Smith}(2008)}]{Pethick2008}%
  \BibitemOpen
  \bibfield  {author} {\bibinfo {author} {\bibfnamefont {C.~J.}\ \bibnamefont
  {Pethick}}\ and\ \bibinfo {author} {\bibfnamefont {H.}~\bibnamefont
  {Smith}},\ }\href@noop {} {\emph {\bibinfo {title} {Bose--Einstein
  condensation in dilute gases}}}\ (\bibinfo  {publisher} {Cambridge university
  press},\ \bibinfo {year} {2008})\BibitemShut {NoStop}%
\bibitem [{\citenamefont {Pitaevskii}\ and\ \citenamefont
  {Stringari}(2016)}]{Pitaevskii2016}%
  \BibitemOpen
  \bibfield  {author} {\bibinfo {author} {\bibfnamefont {L.}~\bibnamefont
  {Pitaevskii}}\ and\ \bibinfo {author} {\bibfnamefont {S.}~\bibnamefont
  {Stringari}},\ }\href@noop {} {\emph {\bibinfo {title} {Bose-Einstein
  condensation and superfluidity}}},\ Vol.\ \bibinfo {volume} {164}\ (\bibinfo
  {publisher} {Oxford University Press},\ \bibinfo {year} {2016})\BibitemShut
  {NoStop}%
\bibitem [{\citenamefont {Bogoliubov}(1947)}]{Bogoliubov1947}%
  \BibitemOpen
  \bibfield  {author} {\bibinfo {author} {\bibfnamefont {N.}~\bibnamefont
  {Bogoliubov}},\ }\href@noop {} {\bibfield  {journal} {\bibinfo  {journal} {J.
  Phys}\ }\textbf {\bibinfo {volume} {11}},\ \bibinfo {pages} {23} (\bibinfo
  {year} {1947})}\BibitemShut {NoStop}%
\bibitem [{\citenamefont {Jin}\ \emph {et~al.}(1996)\citenamefont {Jin},
  \citenamefont {Ensher}, \citenamefont {Matthews}, \citenamefont {Wieman},\
  and\ \citenamefont {Cornell}}]{Jin1996}%
  \BibitemOpen
  \bibfield  {author} {\bibinfo {author} {\bibfnamefont {D.~S.}\ \bibnamefont
  {Jin}}, \bibinfo {author} {\bibfnamefont {J.~R.}\ \bibnamefont {Ensher}},
  \bibinfo {author} {\bibfnamefont {M.~R.}\ \bibnamefont {Matthews}}, \bibinfo
  {author} {\bibfnamefont {C.~E.}\ \bibnamefont {Wieman}}, \ and\ \bibinfo
  {author} {\bibfnamefont {E.~A.}\ \bibnamefont {Cornell}},\ }\href {\doibase
  10.1103/PhysRevLett.77.420} {\bibfield  {journal} {\bibinfo  {journal} {Phys.
  Rev. Lett.}\ }\textbf {\bibinfo {volume} {77}},\ \bibinfo {pages} {420}
  (\bibinfo {year} {1996})}\BibitemShut {NoStop}%
\bibitem [{\citenamefont {Mewes}\ \emph {et~al.}(1996)\citenamefont {Mewes},
  \citenamefont {Andrews}, \citenamefont {van Druten}, \citenamefont {Kurn},
  \citenamefont {Durfee}, \citenamefont {Townsend},\ and\ \citenamefont
  {Ketterle}}]{Mewes1996}%
  \BibitemOpen
  \bibfield  {author} {\bibinfo {author} {\bibfnamefont {M.-O.}\ \bibnamefont
  {Mewes}}, \bibinfo {author} {\bibfnamefont {M.~R.}\ \bibnamefont {Andrews}},
  \bibinfo {author} {\bibfnamefont {N.~J.}\ \bibnamefont {van Druten}},
  \bibinfo {author} {\bibfnamefont {D.~M.}\ \bibnamefont {Kurn}}, \bibinfo
  {author} {\bibfnamefont {D.~S.}\ \bibnamefont {Durfee}}, \bibinfo {author}
  {\bibfnamefont {C.~G.}\ \bibnamefont {Townsend}}, \ and\ \bibinfo {author}
  {\bibfnamefont {W.}~\bibnamefont {Ketterle}},\ }\href {\doibase
  10.1103/PhysRevLett.77.988} {\bibfield  {journal} {\bibinfo  {journal} {Phys.
  Rev. Lett.}\ }\textbf {\bibinfo {volume} {77}},\ \bibinfo {pages} {988}
  (\bibinfo {year} {1996})}\BibitemShut {NoStop}%
\bibitem [{\citenamefont {Martone}\ \emph {et~al.}(2012)\citenamefont
  {Martone}, \citenamefont {Li}, \citenamefont {Pitaevskii},\ and\
  \citenamefont {Stringari}}]{Martone2012}%
  \BibitemOpen
  \bibfield  {author} {\bibinfo {author} {\bibfnamefont {G.~I.}\ \bibnamefont
  {Martone}}, \bibinfo {author} {\bibfnamefont {Y.}~\bibnamefont {Li}},
  \bibinfo {author} {\bibfnamefont {L.~P.}\ \bibnamefont {Pitaevskii}}, \ and\
  \bibinfo {author} {\bibfnamefont {S.}~\bibnamefont {Stringari}},\ }\href
  {\doibase 10.1103/PhysRevA.86.063621} {\bibfield  {journal} {\bibinfo
  {journal} {Phys. Rev. A}\ }\textbf {\bibinfo {volume} {86}},\ \bibinfo
  {pages} {063621} (\bibinfo {year} {2012})}\BibitemShut {NoStop}%
\bibitem [{\citenamefont {Li}\ \emph {et~al.}(2013)\citenamefont {Li},
  \citenamefont {Martone}, \citenamefont {Pitaevskii},\ and\ \citenamefont
  {Stringari}}]{Yun2013}%
  \BibitemOpen
  \bibfield  {author} {\bibinfo {author} {\bibfnamefont {Y.}~\bibnamefont
  {Li}}, \bibinfo {author} {\bibfnamefont {G.~I.}\ \bibnamefont {Martone}},
  \bibinfo {author} {\bibfnamefont {L.~P.}\ \bibnamefont {Pitaevskii}}, \ and\
  \bibinfo {author} {\bibfnamefont {S.}~\bibnamefont {Stringari}},\ }\href
  {\doibase 10.1103/PhysRevLett.110.235302} {\bibfield  {journal} {\bibinfo
  {journal} {Phys. Rev. Lett.}\ }\textbf {\bibinfo {volume} {110}},\ \bibinfo
  {pages} {235302} (\bibinfo {year} {2013})}\BibitemShut {NoStop}%
\bibitem [{\citenamefont {Khamehchi}\ \emph {et~al.}(2014)\citenamefont
  {Khamehchi}, \citenamefont {Zhang}, \citenamefont {Hamner}, \citenamefont
  {Busch},\ and\ \citenamefont {Engels}}]{Khamehchi2014}%
  \BibitemOpen
  \bibfield  {author} {\bibinfo {author} {\bibfnamefont {M.~A.}\ \bibnamefont
  {Khamehchi}}, \bibinfo {author} {\bibfnamefont {Y.}~\bibnamefont {Zhang}},
  \bibinfo {author} {\bibfnamefont {C.}~\bibnamefont {Hamner}}, \bibinfo
  {author} {\bibfnamefont {T.}~\bibnamefont {Busch}}, \ and\ \bibinfo {author}
  {\bibfnamefont {P.}~\bibnamefont {Engels}},\ }\href {\doibase
  10.1103/PhysRevA.90.063624} {\bibfield  {journal} {\bibinfo  {journal} {Phys.
  Rev. A}\ }\textbf {\bibinfo {volume} {90}},\ \bibinfo {pages} {063624}
  (\bibinfo {year} {2014})}\BibitemShut {NoStop}%
\bibitem [{\citenamefont {Chen}\ \emph {et~al.}(2017)\citenamefont {Chen},
  \citenamefont {Pu}, \citenamefont {Yu},\ and\ \citenamefont
  {Zhang}}]{Chen2017}%
  \BibitemOpen
  \bibfield  {author} {\bibinfo {author} {\bibfnamefont {L.}~\bibnamefont
  {Chen}}, \bibinfo {author} {\bibfnamefont {H.}~\bibnamefont {Pu}}, \bibinfo
  {author} {\bibfnamefont {Z.-Q.}\ \bibnamefont {Yu}}, \ and\ \bibinfo {author}
  {\bibfnamefont {Y.}~\bibnamefont {Zhang}},\ }\href {\doibase
  10.1103/PhysRevA.95.033616} {\bibfield  {journal} {\bibinfo  {journal} {Phys.
  Rev. A}\ }\textbf {\bibinfo {volume} {95}},\ \bibinfo {pages} {033616}
  (\bibinfo {year} {2017})}\BibitemShut {NoStop}%
\bibitem [{\citenamefont {Chen}\ \emph {et~al.}(2022)\citenamefont {Chen},
  \citenamefont {Lyu}, \citenamefont {Xu},\ and\ \citenamefont
  {Zhang}}]{chen2022elementary}%
  \BibitemOpen
  \bibfield  {author} {\bibinfo {author} {\bibfnamefont {Y.}~\bibnamefont
  {Chen}}, \bibinfo {author} {\bibfnamefont {H.}~\bibnamefont {Lyu}}, \bibinfo
  {author} {\bibfnamefont {Y.}~\bibnamefont {Xu}}, \ and\ \bibinfo {author}
  {\bibfnamefont {Y.}~\bibnamefont {Zhang}},\ }\href@noop {} {\bibfield
  {journal} {\bibinfo  {journal} {New J. Phys.}\ }\textbf {\bibinfo {volume}
  {24}},\ \bibinfo {pages} {073041} (\bibinfo {year} {2022})}\BibitemShut
  {NoStop}%
\bibitem [{\citenamefont {He}\ and\ \citenamefont
  {Lin}(2023)}]{he2023stationary}%
  \BibitemOpen
  \bibfield  {author} {\bibinfo {author} {\bibfnamefont {J.}~\bibnamefont
  {He}}\ and\ \bibinfo {author} {\bibfnamefont {J.}~\bibnamefont {Lin}},\
  }\href {\doibase 10.1088/1367-2630/acf8eb} {\bibfield  {journal} {\bibinfo
  {journal} {New J. Phys.}\ }\textbf {\bibinfo {volume} {25}},\ \bibinfo
  {pages} {093041} (\bibinfo {year} {2023})}\BibitemShut {NoStop}%
\bibitem [{\citenamefont {Ozawa}\ \emph {et~al.}(2013)\citenamefont {Ozawa},
  \citenamefont {Pitaevskii},\ and\ \citenamefont {Stringari}}]{Ozawa2013}%
  \BibitemOpen
  \bibfield  {author} {\bibinfo {author} {\bibfnamefont {T.}~\bibnamefont
  {Ozawa}}, \bibinfo {author} {\bibfnamefont {L.~P.}\ \bibnamefont
  {Pitaevskii}}, \ and\ \bibinfo {author} {\bibfnamefont {S.}~\bibnamefont
  {Stringari}},\ }\href {\doibase 10.1103/PhysRevA.87.063610} {\bibfield
  {journal} {\bibinfo  {journal} {Phys. Rev. A}\ }\textbf {\bibinfo {volume}
  {87}},\ \bibinfo {pages} {063610} (\bibinfo {year} {2013})}\BibitemShut
  {NoStop}%
\bibitem [{\citenamefont {Ravisankar}\ \emph
  {et~al.}(2021{\natexlab{b}})\citenamefont {Ravisankar}, \citenamefont
  {Fabrelli}, \citenamefont {Gammal}, \citenamefont {Muruganandam},\ and\
  \citenamefont {Mishra}}]{Mishra2021}%
  \BibitemOpen
  \bibfield  {author} {\bibinfo {author} {\bibfnamefont {R.}~\bibnamefont
  {Ravisankar}}, \bibinfo {author} {\bibfnamefont {H.}~\bibnamefont
  {Fabrelli}}, \bibinfo {author} {\bibfnamefont {A.}~\bibnamefont {Gammal}},
  \bibinfo {author} {\bibfnamefont {P.}~\bibnamefont {Muruganandam}}, \ and\
  \bibinfo {author} {\bibfnamefont {P.~K.}\ \bibnamefont {Mishra}},\ }\href
  {\doibase 10.1103/PhysRevA.104.053315} {\bibfield  {journal} {\bibinfo
  {journal} {Phys. Rev. A}\ }\textbf {\bibinfo {volume} {104}},\ \bibinfo
  {pages} {053315} (\bibinfo {year} {2021}{\natexlab{b}})}\BibitemShut
  {NoStop}%
\bibitem [{\citenamefont {Katsimiga}\ \emph {et~al.}(2021)\citenamefont
  {Katsimiga}, \citenamefont {Mistakidis}, \citenamefont {Schmelcher},\ and\
  \citenamefont {Kevrekidis}}]{Katsimiga2021}%
  \BibitemOpen
  \bibfield  {author} {\bibinfo {author} {\bibfnamefont {G.}~\bibnamefont
  {Katsimiga}}, \bibinfo {author} {\bibfnamefont {S.}~\bibnamefont
  {Mistakidis}}, \bibinfo {author} {\bibfnamefont {P.}~\bibnamefont
  {Schmelcher}}, \ and\ \bibinfo {author} {\bibfnamefont {P.}~\bibnamefont
  {Kevrekidis}},\ }\href@noop {} {\bibfield  {journal} {\bibinfo  {journal}
  {New J. Phys.}\ }\textbf {\bibinfo {volume} {23}},\ \bibinfo {pages} {013015}
  (\bibinfo {year} {2021})}\BibitemShut {NoStop}%
\bibitem [{\citenamefont {Katsimiga}\ \emph {et~al.}(2023)\citenamefont
  {Katsimiga}, \citenamefont {Mistakidis}, \citenamefont {Mukherjee},
  \citenamefont {Kevrekidis},\ and\ \citenamefont
  {Schmelcher}}]{Katsimiga2023}%
  \BibitemOpen
  \bibfield  {author} {\bibinfo {author} {\bibfnamefont {G.~C.}\ \bibnamefont
  {Katsimiga}}, \bibinfo {author} {\bibfnamefont {S.~I.}\ \bibnamefont
  {Mistakidis}}, \bibinfo {author} {\bibfnamefont {K.}~\bibnamefont
  {Mukherjee}}, \bibinfo {author} {\bibfnamefont {P.~G.}\ \bibnamefont
  {Kevrekidis}}, \ and\ \bibinfo {author} {\bibfnamefont {P.}~\bibnamefont
  {Schmelcher}},\ }\href {\doibase 10.1103/PhysRevA.107.013313} {\bibfield
  {journal} {\bibinfo  {journal} {Phys. Rev. A}\ }\textbf {\bibinfo {volume}
  {107}},\ \bibinfo {pages} {013313} (\bibinfo {year} {2023})}\BibitemShut
  {NoStop}%
\bibitem [{\citenamefont {Rajat}\ \emph {et~al.}(2022)\citenamefont {Rajat},
  \citenamefont {Roy},\ and\ \citenamefont {Gautam}}]{roy2022collective}%
  \BibitemOpen
  \bibfield  {author} {\bibinfo {author} {\bibnamefont {Rajat}}, \bibinfo
  {author} {\bibfnamefont {A.}~\bibnamefont {Roy}}, \ and\ \bibinfo {author}
  {\bibfnamefont {S.}~\bibnamefont {Gautam}},\ }\href {\doibase
  10.1103/PhysRevA.106.013304} {\bibfield  {journal} {\bibinfo  {journal}
  {Phys. Rev. A}\ }\textbf {\bibinfo {volume} {106}},\ \bibinfo {pages}
  {013304} (\bibinfo {year} {2022})}\BibitemShut {NoStop}%
\bibitem [{\citenamefont {Cross}\ and\ \citenamefont
  {Hohenberg}(1993)}]{Cross1993}%
  \BibitemOpen
  \bibfield  {author} {\bibinfo {author} {\bibfnamefont {M.~C.}\ \bibnamefont
  {Cross}}\ and\ \bibinfo {author} {\bibfnamefont {P.~C.}\ \bibnamefont
  {Hohenberg}},\ }\href {\doibase 10.1103/RevModPhys.65.851} {\bibfield
  {journal} {\bibinfo  {journal} {Rev. Mod. Phys.}\ }\textbf {\bibinfo {volume}
  {65}},\ \bibinfo {pages} {851} (\bibinfo {year} {1993})}\BibitemShut
  {NoStop}%
\bibitem [{\citenamefont {Bernier}\ \emph {et~al.}(2014)\citenamefont
  {Bernier}, \citenamefont {Dalla~Torre},\ and\ \citenamefont
  {Demler}}]{Bernier2014}%
  \BibitemOpen
  \bibfield  {author} {\bibinfo {author} {\bibfnamefont {N.~R.}\ \bibnamefont
  {Bernier}}, \bibinfo {author} {\bibfnamefont {E.~G.}\ \bibnamefont
  {Dalla~Torre}}, \ and\ \bibinfo {author} {\bibfnamefont {E.}~\bibnamefont
  {Demler}},\ }\href {\doibase 10.1103/PhysRevLett.113.065303} {\bibfield
  {journal} {\bibinfo  {journal} {Phys. Rev. Lett.}\ }\textbf {\bibinfo
  {volume} {113}},\ \bibinfo {pages} {065303} (\bibinfo {year}
  {2014})}\BibitemShut {NoStop}%
\bibitem [{\citenamefont {Salasnich}\ \emph {et~al.}(2002)\citenamefont
  {Salasnich}, \citenamefont {Parola},\ and\ \citenamefont
  {Reatto}}]{salas2002}%
  \BibitemOpen
  \bibfield  {author} {\bibinfo {author} {\bibfnamefont {L.}~\bibnamefont
  {Salasnich}}, \bibinfo {author} {\bibfnamefont {A.}~\bibnamefont {Parola}}, \
  and\ \bibinfo {author} {\bibfnamefont {L.}~\bibnamefont {Reatto}},\ }\href
  {\doibase 10.1103/PhysRevA.65.043614} {\bibfield  {journal} {\bibinfo
  {journal} {Phys. Rev. A}\ }\textbf {\bibinfo {volume} {65}},\ \bibinfo
  {pages} {043614} (\bibinfo {year} {2002})}\BibitemShut {NoStop}%
\bibitem [{\citenamefont {Adhikari}(2021{\natexlab{b}})}]{Adhikariss2021}%
  \BibitemOpen
  \bibfield  {author} {\bibinfo {author} {\bibfnamefont {S.~K.}\ \bibnamefont
  {Adhikari}},\ }\href {\doibase 10.1103/PhysRevE.104.024207} {\bibfield
  {journal} {\bibinfo  {journal} {Phys. Rev. E}\ }\textbf {\bibinfo {volume}
  {104}},\ \bibinfo {pages} {024207} (\bibinfo {year}
  {2021}{\natexlab{b}})}\BibitemShut {NoStop}%
\bibitem [{\citenamefont {Kawaguchi}\ and\ \citenamefont
  {Ueda}(2012)}]{Ueda2012}%
  \BibitemOpen
  \bibfield  {author} {\bibinfo {author} {\bibfnamefont {Y.}~\bibnamefont
  {Kawaguchi}}\ and\ \bibinfo {author} {\bibfnamefont {M.}~\bibnamefont
  {Ueda}},\ }\href@noop {} {\bibfield  {journal} {\bibinfo  {journal} {Phys.
  Rep.}\ }\textbf {\bibinfo {volume} {520}},\ \bibinfo {pages} {253} (\bibinfo
  {year} {2012})}\BibitemShut {NoStop}%
\bibitem [{\citenamefont {Stamper-Kurn}\ and\ \citenamefont
  {Ueda}(2013)}]{Stamper2013}%
  \BibitemOpen
  \bibfield  {author} {\bibinfo {author} {\bibfnamefont {D.~M.}\ \bibnamefont
  {Stamper-Kurn}}\ and\ \bibinfo {author} {\bibfnamefont {M.}~\bibnamefont
  {Ueda}},\ }\href {\doibase 10.1103/RevModPhys.85.1191} {\bibfield  {journal}
  {\bibinfo  {journal} {Rev. Mod. Phys.}\ }\textbf {\bibinfo {volume} {85}},\
  \bibinfo {pages} {1191} (\bibinfo {year} {2013})}\BibitemShut {NoStop}%
\bibitem [{\citenamefont {Ravisankar}\ \emph
  {et~al.}(2021{\natexlab{c}})\citenamefont {Ravisankar}, \citenamefont
  {Vudragovi{\'c}}, \citenamefont {Muruganandam}, \citenamefont {Bala{\v{z}}},\
  and\ \citenamefont {Adhikari}}]{Ravisankar2021}%
  \BibitemOpen
  \bibfield  {author} {\bibinfo {author} {\bibfnamefont {R.}~\bibnamefont
  {Ravisankar}}, \bibinfo {author} {\bibfnamefont {D.}~\bibnamefont
  {Vudragovi{\'c}}}, \bibinfo {author} {\bibfnamefont {P.}~\bibnamefont
  {Muruganandam}}, \bibinfo {author} {\bibfnamefont {A.}~\bibnamefont
  {Bala{\v{z}}}}, \ and\ \bibinfo {author} {\bibfnamefont {S.~K.}\ \bibnamefont
  {Adhikari}},\ }\href@noop {} {\bibfield  {journal} {\bibinfo  {journal}
  {Comput. Phys. Commun.}\ }\textbf {\bibinfo {volume} {259}},\ \bibinfo
  {pages} {107657} (\bibinfo {year} {2021}{\natexlab{c}})}\BibitemShut
  {NoStop}%
\bibitem [{\citenamefont {Inouye}\ \emph {et~al.}(1998)\citenamefont {Inouye},
  \citenamefont {Andrews}, \citenamefont {Stenger}, \citenamefont {Miesner},
  \citenamefont {Stamper-Kurn},\ and\ \citenamefont {Ketterle}}]{Inoye1998}%
  \BibitemOpen
  \bibfield  {author} {\bibinfo {author} {\bibfnamefont {S.}~\bibnamefont
  {Inouye}}, \bibinfo {author} {\bibfnamefont {M.}~\bibnamefont {Andrews}},
  \bibinfo {author} {\bibfnamefont {J.}~\bibnamefont {Stenger}}, \bibinfo
  {author} {\bibfnamefont {H.-J.}\ \bibnamefont {Miesner}}, \bibinfo {author}
  {\bibfnamefont {D.~M.}\ \bibnamefont {Stamper-Kurn}}, \ and\ \bibinfo
  {author} {\bibfnamefont {W.}~\bibnamefont {Ketterle}},\ }\href@noop {}
  {\bibfield  {journal} {\bibinfo  {journal} {Nature (London)}\ }\textbf
  {\bibinfo {volume} {392}},\ \bibinfo {pages} {151} (\bibinfo {year}
  {1998})}\BibitemShut {NoStop}%
\bibitem [{\citenamefont {Marte}\ \emph {et~al.}(2002)\citenamefont {Marte},
  \citenamefont {Volz}, \citenamefont {Schuster}, \citenamefont {D{\"u}rr},
  \citenamefont {Rempe}, \citenamefont {Van~Kempen},\ and\ \citenamefont
  {Verhaar}}]{Marte2002}%
  \BibitemOpen
  \bibfield  {author} {\bibinfo {author} {\bibfnamefont {A.}~\bibnamefont
  {Marte}}, \bibinfo {author} {\bibfnamefont {T.}~\bibnamefont {Volz}},
  \bibinfo {author} {\bibfnamefont {J.}~\bibnamefont {Schuster}}, \bibinfo
  {author} {\bibfnamefont {S.}~\bibnamefont {D{\"u}rr}}, \bibinfo {author}
  {\bibfnamefont {G.}~\bibnamefont {Rempe}}, \bibinfo {author} {\bibfnamefont
  {E.}~\bibnamefont {Van~Kempen}}, \ and\ \bibinfo {author} {\bibfnamefont
  {B.}~\bibnamefont {Verhaar}},\ }\href@noop {} {\bibfield  {journal} {\bibinfo
   {journal} {Phys. Rev. Lett.}\ }\textbf {\bibinfo {volume} {89}},\ \bibinfo
  {pages} {283202} (\bibinfo {year} {2002})}\BibitemShut {NoStop}%
\bibitem [{\citenamefont {Chin}\ \emph {et~al.}(2010)\citenamefont {Chin},
  \citenamefont {Grimm}, \citenamefont {Julienne},\ and\ \citenamefont
  {Tiesinga}}]{Chin2010}%
  \BibitemOpen
  \bibfield  {author} {\bibinfo {author} {\bibfnamefont {C.}~\bibnamefont
  {Chin}}, \bibinfo {author} {\bibfnamefont {R.}~\bibnamefont {Grimm}},
  \bibinfo {author} {\bibfnamefont {P.}~\bibnamefont {Julienne}}, \ and\
  \bibinfo {author} {\bibfnamefont {E.}~\bibnamefont {Tiesinga}},\ }\href
  {\doibase 10.1103/RevModPhys.82.1225} {\bibfield  {journal} {\bibinfo
  {journal} {Rev. Mod. Phys.}\ }\textbf {\bibinfo {volume} {82}},\ \bibinfo
  {pages} {1225} (\bibinfo {year} {2010})}\BibitemShut {NoStop}%
\bibitem [{\citenamefont {Wen}\ \emph {et~al.}(2012)\citenamefont {Wen},
  \citenamefont {Sun}, \citenamefont {Wang}, \citenamefont {Ji},\ and\
  \citenamefont {Liu}}]{Wen2012}%
  \BibitemOpen
  \bibfield  {author} {\bibinfo {author} {\bibfnamefont {L.}~\bibnamefont
  {Wen}}, \bibinfo {author} {\bibfnamefont {Q.}~\bibnamefont {Sun}}, \bibinfo
  {author} {\bibfnamefont {H.~Q.}\ \bibnamefont {Wang}}, \bibinfo {author}
  {\bibfnamefont {A.~C.}\ \bibnamefont {Ji}}, \ and\ \bibinfo {author}
  {\bibfnamefont {W.~M.}\ \bibnamefont {Liu}},\ }\href {\doibase
  10.1103/PhysRevA.86.043602} {\bibfield  {journal} {\bibinfo  {journal} {Phys.
  Rev. A}\ }\textbf {\bibinfo {volume} {86}},\ \bibinfo {pages} {043602}
  (\bibinfo {year} {2012})}\BibitemShut {NoStop}%
\bibitem [{\citenamefont {Rajaswathi}\ \emph {et~al.}(2023)\citenamefont
  {Rajaswathi}, \citenamefont {Bhuvaneswari}, \citenamefont {Radha},\ and\
  \citenamefont {Muruganandam}}]{Rajaswathi2023}%
  \BibitemOpen
  \bibfield  {author} {\bibinfo {author} {\bibfnamefont {K.}~\bibnamefont
  {Rajaswathi}}, \bibinfo {author} {\bibfnamefont {S.}~\bibnamefont
  {Bhuvaneswari}}, \bibinfo {author} {\bibfnamefont {R.}~\bibnamefont {Radha}},
  \ and\ \bibinfo {author} {\bibfnamefont {P.}~\bibnamefont {Muruganandam}},\
  }\href {\doibase 10.1103/PhysRevA.108.033317} {\bibfield  {journal} {\bibinfo
   {journal} {Phys. Rev. A}\ }\textbf {\bibinfo {volume} {108}},\ \bibinfo
  {pages} {033317} (\bibinfo {year} {2023})}\BibitemShut {NoStop}%
\bibitem [{\citenamefont {Wang}\ \emph {et~al.}(2010)\citenamefont {Wang},
  \citenamefont {Gao}, \citenamefont {Jian},\ and\ \citenamefont
  {Zhai}}]{Wang2010}%
  \BibitemOpen
  \bibfield  {author} {\bibinfo {author} {\bibfnamefont {C.}~\bibnamefont
  {Wang}}, \bibinfo {author} {\bibfnamefont {C.}~\bibnamefont {Gao}}, \bibinfo
  {author} {\bibfnamefont {C.-M.}\ \bibnamefont {Jian}}, \ and\ \bibinfo
  {author} {\bibfnamefont {H.}~\bibnamefont {Zhai}},\ }\href@noop {} {\bibfield
   {journal} {\bibinfo  {journal} {Phys. Rev. Lett.}\ }\textbf {\bibinfo
  {volume} {105}},\ \bibinfo {pages} {160403} (\bibinfo {year}
  {2010})}\BibitemShut {NoStop}%
\bibitem [{\citenamefont {Zhu}\ \emph {et~al.}(2012)\citenamefont {Zhu},
  \citenamefont {Zhang},\ and\ \citenamefont {Wu}}]{Zhu2012}%
  \BibitemOpen
  \bibfield  {author} {\bibinfo {author} {\bibfnamefont {Q.}~\bibnamefont
  {Zhu}}, \bibinfo {author} {\bibfnamefont {C.}~\bibnamefont {Zhang}}, \ and\
  \bibinfo {author} {\bibfnamefont {B.}~\bibnamefont {Wu}},\ }\href@noop {}
  {\bibfield  {journal} {\bibinfo  {journal} {EPL}\ }\textbf {\bibinfo {volume}
  {100}},\ \bibinfo {pages} {50003} (\bibinfo {year} {2012})}\BibitemShut
  {NoStop}%
\bibitem [{\citenamefont {Anderson}\ \emph {et~al.}(1999)\citenamefont
  {Anderson}, \citenamefont {Bai}, \citenamefont {Bischof}, \citenamefont
  {Blackford}, \citenamefont {Demmel}, \citenamefont {Dongarra}, \citenamefont
  {Du~Croz}, \citenamefont {Greenbaum}, \citenamefont {Hammarling},
  \citenamefont {McKenney} \emph {et~al.}}]{Anderson1999}%
  \BibitemOpen
  \bibfield  {author} {\bibinfo {author} {\bibfnamefont {E.}~\bibnamefont
  {Anderson}}, \bibinfo {author} {\bibfnamefont {Z.}~\bibnamefont {Bai}},
  \bibinfo {author} {\bibfnamefont {C.}~\bibnamefont {Bischof}}, \bibinfo
  {author} {\bibfnamefont {L.~S.}\ \bibnamefont {Blackford}}, \bibinfo {author}
  {\bibfnamefont {J.}~\bibnamefont {Demmel}}, \bibinfo {author} {\bibfnamefont
  {J.}~\bibnamefont {Dongarra}}, \bibinfo {author} {\bibfnamefont
  {J.}~\bibnamefont {Du~Croz}}, \bibinfo {author} {\bibfnamefont
  {A.}~\bibnamefont {Greenbaum}}, \bibinfo {author} {\bibfnamefont
  {S.}~\bibnamefont {Hammarling}}, \bibinfo {author} {\bibfnamefont
  {A.}~\bibnamefont {McKenney}},  \emph {et~al.},\ }\href@noop {} {\emph
  {\bibinfo {title} {LAPACK users' guide}}}\ (\bibinfo  {publisher} {SIAM},\
  \bibinfo {year} {1999})\BibitemShut {NoStop}%
\bibitem [{\citenamefont {Muruganandam}\ and\ \citenamefont
  {Adhikari}(2009)}]{Muruganandam2009}%
  \BibitemOpen
  \bibfield  {author} {\bibinfo {author} {\bibfnamefont {P.}~\bibnamefont
  {Muruganandam}}\ and\ \bibinfo {author} {\bibfnamefont {S.~K.}\ \bibnamefont
  {Adhikari}},\ }\href@noop {} {\bibfield  {journal} {\bibinfo  {journal}
  {Comput. Phys. Commun.}\ }\textbf {\bibinfo {volume} {180}},\ \bibinfo
  {pages} {1888} (\bibinfo {year} {2009})}\BibitemShut {NoStop}%
\bibitem [{\citenamefont {Young-S}\ \emph {et~al.}(2016)\citenamefont
  {Young-S}, \citenamefont {Vudragovi{\'c}}, \citenamefont {Muruganandam},
  \citenamefont {Adhikari},\ and\ \citenamefont {Bala{\v{z}}}}]{Luis2016}%
  \BibitemOpen
  \bibfield  {author} {\bibinfo {author} {\bibfnamefont {L.~E.}\ \bibnamefont
  {Young-S}}, \bibinfo {author} {\bibfnamefont {D.}~\bibnamefont
  {Vudragovi{\'c}}}, \bibinfo {author} {\bibfnamefont {P.}~\bibnamefont
  {Muruganandam}}, \bibinfo {author} {\bibfnamefont {S.~K.}\ \bibnamefont
  {Adhikari}}, \ and\ \bibinfo {author} {\bibfnamefont {A.}~\bibnamefont
  {Bala{\v{z}}}},\ }\href@noop {} {\bibfield  {journal} {\bibinfo  {journal}
  {Comput. Phys. Commun.}\ }\textbf {\bibinfo {volume} {204}},\ \bibinfo
  {pages} {209} (\bibinfo {year} {2016})}\BibitemShut {NoStop}%
\bibitem [{\citenamefont {Tasgal}\ and\ \citenamefont
  {Band}(2015)}]{Tasgal2015}%
  \BibitemOpen
  \bibfield  {author} {\bibinfo {author} {\bibfnamefont {R.~S.}\ \bibnamefont
  {Tasgal}}\ and\ \bibinfo {author} {\bibfnamefont {Y.~B.}\ \bibnamefont
  {Band}},\ }\href {\doibase 10.1103/PhysRevA.91.013615} {\bibfield  {journal}
  {\bibinfo  {journal} {Phys. Rev. A}\ }\textbf {\bibinfo {volume} {91}},\
  \bibinfo {pages} {013615} (\bibinfo {year} {2015})}\BibitemShut {NoStop}%
\bibitem [{\citenamefont {Abad}\ and\ \citenamefont {Recati}(2013)}]{Abad2013}%
  \BibitemOpen
  \bibfield  {author} {\bibinfo {author} {\bibfnamefont {M.}~\bibnamefont
  {Abad}}\ and\ \bibinfo {author} {\bibfnamefont {A.}~\bibnamefont {Recati}},\
  }\href {\doibase 10.1140/epjd/e2013-40053-2} {\bibfield  {journal} {\bibinfo
  {journal} {Eur. Phys. J. D}\ }\textbf {\bibinfo {volume} {67}},\ \bibinfo
  {pages} {148} (\bibinfo {year} {2013})}\BibitemShut {NoStop}%
\bibitem [{\citenamefont {Lyu}\ and\ \citenamefont {Zhang}(2020)}]{Lyu2020}%
  \BibitemOpen
  \bibfield  {author} {\bibinfo {author} {\bibfnamefont {H.}~\bibnamefont
  {Lyu}}\ and\ \bibinfo {author} {\bibfnamefont {Y.}~\bibnamefont {Zhang}},\
  }\href {\doibase 10.1103/PhysRevA.102.023327} {\bibfield  {journal} {\bibinfo
   {journal} {Phys. Rev. A}\ }\textbf {\bibinfo {volume} {102}},\ \bibinfo
  {pages} {023327} (\bibinfo {year} {2020})}\BibitemShut {NoStop}%
\bibitem [{\citenamefont {Cabrera}\ \emph {et~al.}(2018)\citenamefont
  {Cabrera}, \citenamefont {Tanzi}, \citenamefont {Sanz}, \citenamefont
  {Naylor}, \citenamefont {Thomas}, \citenamefont {Cheiney},\ and\
  \citenamefont {Tarruell}}]{Cabrera2018}%
  \BibitemOpen
  \bibfield  {author} {\bibinfo {author} {\bibfnamefont {C.~R.}\ \bibnamefont
  {Cabrera}}, \bibinfo {author} {\bibfnamefont {L.}~\bibnamefont {Tanzi}},
  \bibinfo {author} {\bibfnamefont {J.}~\bibnamefont {Sanz}}, \bibinfo {author}
  {\bibfnamefont {B.}~\bibnamefont {Naylor}}, \bibinfo {author} {\bibfnamefont
  {P.}~\bibnamefont {Thomas}}, \bibinfo {author} {\bibfnamefont
  {P.}~\bibnamefont {Cheiney}}, \ and\ \bibinfo {author} {\bibfnamefont
  {L.}~\bibnamefont {Tarruell}},\ }\href {\doibase 10.1126/science.aao5686}
  {\bibfield  {journal} {\bibinfo  {journal} {Science}\ }\textbf {\bibinfo
  {volume} {359}},\ \bibinfo {pages} {301} (\bibinfo {year}
  {2018})}\BibitemShut {NoStop}%
\bibitem [{\citenamefont {Petrov}(2015)}]{Petrov2015}%
  \BibitemOpen
  \bibfield  {author} {\bibinfo {author} {\bibfnamefont {D.~S.}\ \bibnamefont
  {Petrov}},\ }\href {\doibase 10.1103/PhysRevLett.115.155302} {\bibfield
  {journal} {\bibinfo  {journal} {Phys. Rev. Lett.}\ }\textbf {\bibinfo
  {volume} {115}},\ \bibinfo {pages} {155302} (\bibinfo {year}
  {2015})}\BibitemShut {NoStop}%
\bibitem [{\citenamefont {Gangwar}\ \emph {et~al.}(2022)\citenamefont
  {Gangwar}, \citenamefont {Ravisankar}, \citenamefont {Muruganandam},\ and\
  \citenamefont {Mishra}}]{Gangwar2022}%
  \BibitemOpen
  \bibfield  {author} {\bibinfo {author} {\bibfnamefont {S.}~\bibnamefont
  {Gangwar}}, \bibinfo {author} {\bibfnamefont {R.}~\bibnamefont {Ravisankar}},
  \bibinfo {author} {\bibfnamefont {P.}~\bibnamefont {Muruganandam}}, \ and\
  \bibinfo {author} {\bibfnamefont {P.~K.}\ \bibnamefont {Mishra}},\
  }\href@noop {} {\bibfield  {journal} {\bibinfo  {journal} {Phys. Rev. A}\
  }\textbf {\bibinfo {volume} {106}},\ \bibinfo {pages} {063315} (\bibinfo
  {year} {2022})}\BibitemShut {NoStop}%
\bibitem [{\citenamefont {Gangwar}\ \emph {et~al.}(2023)\citenamefont
  {Gangwar}, \citenamefont {Ravisankar}, \citenamefont {Mistakidis},
  \citenamefont {Muruganandam},\ and\ \citenamefont {Mishra}}]{Gangwar2023}%
  \BibitemOpen
  \bibfield  {author} {\bibinfo {author} {\bibfnamefont {S.}~\bibnamefont
  {Gangwar}}, \bibinfo {author} {\bibfnamefont {R.}~\bibnamefont {Ravisankar}},
  \bibinfo {author} {\bibfnamefont {S.}~\bibnamefont {Mistakidis}}, \bibinfo
  {author} {\bibfnamefont {P.}~\bibnamefont {Muruganandam}}, \ and\ \bibinfo
  {author} {\bibfnamefont {P.~K.}\ \bibnamefont {Mishra}},\ }\href@noop {}
  {\bibfield  {journal} {\bibinfo  {journal} {arXiv preprint arXiv:2307.16742}\
  } (\bibinfo {year} {2023})}\BibitemShut {NoStop}%
\bibitem [{\citenamefont {Sadler}\ \emph {et~al.}(2006)\citenamefont {Sadler},
  \citenamefont {Higbie}, \citenamefont {Leslie}, \citenamefont
  {Vengalattore},\ and\ \citenamefont {Stamper-Kurn}}]{Sadler2006}%
  \BibitemOpen
  \bibfield  {author} {\bibinfo {author} {\bibfnamefont {L.}~\bibnamefont
  {Sadler}}, \bibinfo {author} {\bibfnamefont {J.}~\bibnamefont {Higbie}},
  \bibinfo {author} {\bibfnamefont {S.}~\bibnamefont {Leslie}}, \bibinfo
  {author} {\bibfnamefont {M.}~\bibnamefont {Vengalattore}}, \ and\ \bibinfo
  {author} {\bibfnamefont {D.}~\bibnamefont {Stamper-Kurn}},\ }\href@noop {}
  {\bibfield  {journal} {\bibinfo  {journal} {Nature}\ }\textbf {\bibinfo
  {volume} {443}},\ \bibinfo {pages} {312} (\bibinfo {year}
  {2006})}\BibitemShut {NoStop}%
\end{thebibliography}%
\end{document}